\documentclass{article}
\usepackage{pdfpages}
\usepackage{amssymb}
\usepackage{subfigure}   
\usepackage{ws-rv-van}   
\usepackage{color}

\usepackage{fancyhdr}
\makeindex
\newcommand{\apjl}{Astrophysical Journal, Letters}

\newcommand{\apjs}{Astrophysical Journal Supplements}

\newcommand{\aap}{Astronomy \& Astrophysics}

\newcommand{\blue}{\textcolor{blue}}

\def\simgreat{\lower2pt\hbox{$\buildrel {\scriptstyle >}
   \over {\scriptstyle\sim}$}}
\def\simless{\lower2pt\hbox{$\buildrel {\scriptstyle <}
   \over {\scriptstyle\sim}$}}
\def\msun{\,{\rm M_\odot}}
\def\ergs{\,{\rm erg\,s^{-1}}}
\def\kms{\,{\rm km\,s^{-1}}}

\begin{document}

            



\begin{titlepage}
   \vspace*{\stretch{1.0}}
   \begin{center}
      \LARGE\textbf{Gravitational wave sources in the era of multi-frequency gravitational wave astronomy}\\
   \vspace*{\stretch{0.1}}

   \Large{Monica Colpi$^{1,2,\star}$, Alberto  Sesana$^{3,\dagger}$}

   \vspace*{\stretch{0.05}}
   \small{\it $^1$ Dipartimento di Fisica G. Occhialini, Universit\`a degli Studi di Milano Bicocca, Piazza della Scienza 3, I-20126 Milano, Italy\\
   $^2$ INFN, Sezione di Milano-Bicocca, Piazza della Scienza 3, I-20126 Milano, Italy\\
   $^3$ School of Physics and Astronomy, The University of Birmingham, Edgbaston, Birmingham B15 2TT, UK\\}
   \vskip  1.0 truecm
   \large{{to appear in the book} }
   
   \Large{{\sl An Overview of Gravitational Waves: \\Theory, Sources and Detection}}
   
   \vskip 0.1 truecm
   \large{edited by G. Auger and E. Plagnol (World Scientific, 2016)}
   \vskip 0.5 truecm
   
   \end{center}
   \vspace*{\stretch{2.0}}

\begin{flushleft}   
   $^\star$ email: Monica.Colpi@mib.infn.it\\
   $^\dagger$ email: asesana@star.sr.bham.ac.uk\\ 
\end{flushleft}   
   

\end{titlepage}


\def\simgreat{\lower2pt\hbox{$\buildrel {\scriptstyle >}
   \over {\scriptstyle\sim}$}}
\def\simless{\lower2pt\hbox{$\buildrel {\scriptstyle <}
   \over {\scriptstyle\sim}$}}
\def\msun{\,{\rm M_\odot}}
\def\ergs{\,{\rm erg\,s^{-1}}}
\def\kms{\,{\rm km\,s^{-1}}}

\tableofcontents    

\newpage
\begin{abstract}
\textit{The focus of this Chapter is on describing the prospective sources of the gravitational wave universe accessible to present and future observations,  from kHz, to mHz down to nano-Hz frequencies. The multi-frequency gravitational wave universe gives a deep view into the cosmos, inaccessible otherwise. It has as main actors core-collapsing massive stars, neutron stars, coalescing compact object binaries of different flavours and stellar origin, coalescing massive black hole binaries, extreme mass ratio inspirals, and possibly the very early universe itself. Here, we highlight the science aims and  describe the gravitational wave signals expected from the sources and the information gathered in it. We show that the observation of gravitational wave sources will play a transformative role in our understanding of the processes ruling the formation and evolution of stars and black holes, galaxy clustering and evolution, the nature of the strong forces in neutron star interiors, and the most mysterious interaction of Nature: gravity. The discovery, by the LIGO Scientific Collaboration and Virgo Collaboration, of the first source of gravitational waves from the cosmos GW150914, and the superb technological achievement of the space mission LISA Pathfinder herald the beginning of the new phase of exploration of the universe.}
\end{abstract}

\section{Key science objectives of the multi-band gravitational wave astronomy}\label{key-questions}

Gravitational wave sources have been anticipated and studied in the literature quite extensively during the last twenty years.  These studies flourished in parallel to 
the building on Earth of the interferometric detectors Advanced LIGO and Virgo designed to 
explore the high frequency gravitational wave universe, and the proposal
to construct in space a Laser Interferometer Space Antenna (LISA) to investigate lower frequency sources.  The Pulsar Timing Array experiment (PTA) has been joining this world-wide effort by exploiting millisecond pulsars as high precision  clocks to investigate the very low
frequency domain.

With time, it has become clear that exploring the universe with gravitational waves from
kHz to nano Hz makes it possible to discover new sources never anticipated before, and to provide complementary 
avenues for expanding our knowledge on the  {\it  Laws of  Nature}, on {\it Cosmology}, and  on the processes ruling
the formation and evolution of compact objects as black holes 
within the realm of  {\it Relativistic Astrophysics} and {\it Galaxy Structure Formation}.
Exploring the universe with gravitational waves will help answering a number of fundamental questions in all these domains:
\vskip 1 truecm

\noindent
{\bf Laws of Nature}\\
\begin{itemize}
\item {\sl Is gravity in the strong field regime and dynamical sector as predicted by Einstein's theory?}
\item  {\sl Are the properties of gravitational radiation as predicated by Einstein's theory?}

\item  {\sl Does gravity couple to other dynamical fields, such as,
massless or massive scalars?}
\item  {\sl Are (astrophysical) black holes described by the Kerr metric?}
\item  {\sl Are black holes hairless?}
\item {\sl Are there naked singularities?}
\item {\sl What is the behaviour of the short-range interaction at supra-nuclear densities?}
\item {\sl What is the lowest energy state of baryonic matter at supra-nuclear densities?}
\item {\sl Is gravitational collapse to a Kerr black hole unescapable?}
\item {\sl Is the signal of coalescence from close pairs of neutron stars or/and black holes as predicted from Einstein's theory?}
\end{itemize}

\medskip
\noindent
{\bf Relativistic Astrophysics and Galaxy Structure Formation}\\

\begin{itemize}

\item  {\sl What is the maximum mass of a neutron star and the minimum and maximum mass of a stellar origin black hole?} 
\item   {\sl What is the mass function and redshift distribution of stellar origin  neutron stars and black holes?}
\item   {\sl How do neutron star and black hole masses and spins evolve in relation to the environment and with cosmic epoch?}
\item  {\sl What is the physical mechanism behind supernovae and how asymmetric is their collapse?}
\item   {\sl How do  stellar origin compact  binaries form? Do they form in binary stars, 
or dynamically in dense star clusters or both?} 
\item  {\sl Do black hole coalescence events of any flavour have an associated electromagnetic counterpart?}
\item  {\sl How can we identify the counterparts of neutron star binary mergers and of black hole-neutron star mergers? Are they related to known sources as short gamma-ray bursts (GRBs)
and kilo-novae?}
\item   {\sl How many  compact  binaries of all flavours  exist in the Milky Way and what do they tell us about the star formation history of our own galaxy?}
\item {\sl Are ultra-compact  white dwarf binaries the progenitors of Type Ia supernovae?}
\item {\sl How do massive black hole form? Via accretion or/and aggregation of stellar origin black holes, or via the direct collapse of supermassive stars?}
\item  {\sl How do seed black holes grow to become giant through accretion and mergers, and how fast do they grow over cosmic time?}
\item {\sl How often compact object binaries of the different  flavours coalesce in galaxies and how does their coalescence rate evolve with redshift?}
\item  {\sl When did the first black holes form in pre-galactic halos, and what is their initial mass and spin distribution?}
\item {\sl How do massive black holes pair in galaxy mergers and how fast to they coalesce?}
\item {\sl What is the role of black hole mergers in galaxy formation?}
\item {\sl Are massive black holes as light as $10^{3-5}\msun$ inhabiting the cores of all dwarf galaxies?}
\item  {\sl What is the mass distribution of stellar remnants at the galactic centres and what is the role of mass 
segregation and relaxation in determining the nature of the stellar populations around the nuclear black holes in galaxies?}
\item {\sl What is the merger rate of extreme mass ratio inspirals in galactic nuclei?}

\end{itemize}

\medskip
\noindent
{\bf Cosmography-Cosmology}\\ 
\begin{itemize}

\item {\sl What is the architecture of the universe?}
\item {\sl Using precise gravitationally calibrated distances and redshift measurements of coalescence events to what precision can we
measure the Hubble flow?}
\item {\sl What is the equation of state of dark energy, and to what precision can it be inferred from gravitational wave sources?}
\item {\sl What can gravitational waves tell us about the physics beyond the Standard Model?}
\item {\sl Can we measure or set bounds on cosmological gravitational wave backgrounds from the very early
universe?}

\end{itemize}

\section{Prologue: GW150914 - the first cosmic source of gravitational waves}\label{prologue}

In the last years the detection of gravitational waves was perceived as imminent, following the rebuild of the LIGO 
interferometers at Hanford and Livingston \cite{Abbott-2}, and of the interferometer Virgo
in Pisa.  Operating in the frequency interval between 10 Hz and 1000 Hz, Advanced LIGO and Virgo are designed to detect gravitational waves emitted by 
highly perturbed/deformed neutron stars,  core collapsing massive stars and by the merger of pairs of neutron stars 
and stellar origin black holes.  The rates of these events were so uncertain to prevent any definite prediction on the nature of the first signal, whether coming from neutron stars or black holes, or a combination of the two. But, on February 11th 2016, during the drafting of this chapter, from  the LIGO Scientific Collaboration and The Virgo Collaboration came the announcement of \cite{Abbott-1}
\begin{figure}[!t]
\begin{center}
\includegraphics[width=.99\textwidth]{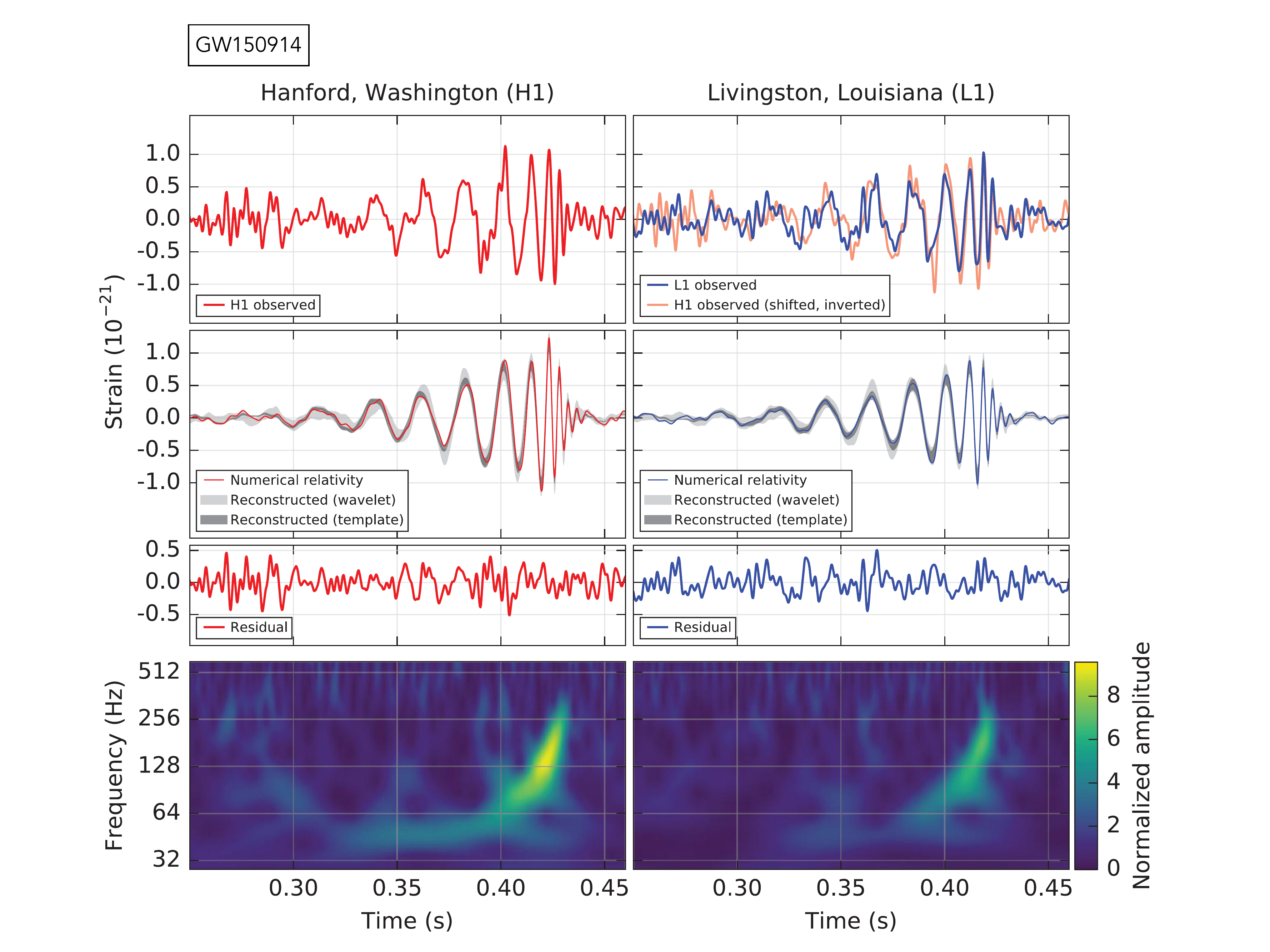}
\caption{The gravitational wave signal from the first cosmic source detected in gravitational waves: GW150914 observed by the LIGO Hanford (H1, left panel) and Livingston (L1, right 
 panel) interferometers \cite{Abbott-1}. The first row gives the strain with time in the two detectors. The second row 
 gives the gravitational wave strain projected onto each
detector in the 35-350 Hz band, and the solid lines show the numerical relativity waveform for a source with parameters consistent 
with the inspiral, merger and ringdown of two coalescing black holes of $36^{+5}_{-4} \msun$ and $29^{+4}_{-4}\msun$,
forming a new black hole of  $62^{+4}_{-4}\msun$. The bottom row gives a time-frequency representation of the
strain data, showing the signal frequency increasing over time. Courtesy of the LIGO Scientific Collaboration and Virgo Collaboration \cite{Abbott-1}.}
\label{GW150914} 
\end{center} 

\end{figure}
{\color{blue}
\begin{itemize}
\item \blue{\sl the first direct detection of gravitational waves from a cosmic source;}
\item \blue{\sl the discovery of the most  powerful  astronomical event ever observed since the Big Bang;}
\item \blue{\sl the first detection of two stellar origin black holes coalescing in a single black hole,
  according to general relativity, and observed through the inspiral, merging and ringdown phases. }
\end{itemize}
}

\noindent
The discovery of this event, named GW150914, confirms, within statistical uncertainties and current precision\footnote{We remark that the current signal from GW150914 is 
not sufficient to exclude more exotic configurations than black holes as described in general relativity. 
 Only precision observations of the late-time ringdown signal, where the differences in the quasinormal-mode spectrum eventually emerge, can be used to rule out exotic alternatives to black holes and to test quantum effects at the horizon scale. \cite{Cardoso16a,Cardoso16c}.
},
{\color{blue}
\begin{itemize}
\item  {\sl Einstein's theory of gravity in the dynamical strong field regime, never tested before;} 
\item  {\sl Einstein's theory on the generation of gravitational radiation;}
\item {\sl the unimpeded propagation of gravitational waves  across the universe;}
\item {\sl the existence of highly dynamical space-times that can form in the cosmos;}
\item {\sl the existence and overall simplicity of black holes.}
\end{itemize}
}
\noindent
From an astrophysical perspective, the discovery of  GW150914 provides
{\color{blue}
\begin{itemize}
\item {\sl the first measure of the mass and spin of stellar black holes through gravity's own messenger: the 
gravitational waves;}
\item {\sl the first identification of "heavy" stellar origin black holes with mass $\sim 30\msun$;}
\item {\sl the first definite proof of the existence of "binary black holes";}
\item {\sl the most massive stellar origin black hole known to date, resulting from a merger product:  $62\msun.$}
\end{itemize}
}

The  signal from GW150914, lasting less than 0.45 seconds, is extraordinary {\it simple} \cite{Abbott-1}.  As shown in Figure \ref{GW150914}, the signal
sweeps upwards in amplitude and frequency of oscillation from 30 Hz to 250 Hz, with a peak gravitational wave strain $h$ of $1.0\times 10^{-21}$ 
and significance of $5.1\sigma$.  The signal matches the 
waveform predicted by general relativity of the {\it inspiral, merger,} and {\it ringdown} of two black holes of $36^{+5}_{-4} \msun$ and $29^{+4}_{-4}\msun$ in the source-frame. 
The new black hole that formed has rest-frame mass $62^{+4}_{-4}\msun$ 
and spin $s_{\rm spin}=0.67^{+0.05}_{-0.07}$.  The energy radiated in gravitational waves
corresponds to $3.0^{+0.5}_{-0.5}\msun c^2$, and to a peak luminosity of $3.6^{+0.5}_{-0.4}\times 10^{56}\ergs$ equivalent to $200^{+30}_{-20}\msun c^2\,\rm s^{-1}$. 
The source was observed with a matched filter signal-to-noise ratio of 23.7, and lies at a luminosity distance of $410^{+160}_{-180}$ Mpc, 
corresponding to a redshift $z=0.09^{+0.03}_{-0.04}$ 
(assuming the current  $\Lambda$-CDM cosmological model).

During the editing of this Chapter, the LIGO Scientific Collaboration and the Virgo Collaboration announced the discovery of a second signal:
 GW151226, detected at a significance greater than $5\sigma$ \cite{Abbott-GW151226}. Again a remarkable finding. The signal 
 lasted in the LIGO frequency band for approximately 1 second, increasing in frequency and
amplitude over about 55 cycles from 35 to 450 Hz (higher than that of the first event), and reached a peak gravitational strain of $3.4^{+0.7}_{-0.9}\times 10^{-22}$.
The event detected with a signal-to-noise ratio of 13 is consistent again with the coalescence of two stellar origin black holes of $14.2^{+8.3}_{-2.3}\msun$ and $7.5^{+2.3}_{-2.3}\msun$ and a final mass of
$20.8^{+6.2}_{-1.7}\msun$, in the rest frame of the source.
The long lasting signal allowed the measure of the spin of one of the component black holes, which has spin parameter greater than 0.2.
GW151226 is located at a distance of $440^{+180}_{-190}$ Mpc, corresponding to a redshift $z=0.09^{+0.03}_{-0.04}$, similar to that of GW150914.
While the high black hole masses in GW150914 lie in an almost unexplored interval, the masses in GW151226 are in line with those inferred in X-ray binaries \cite{Ozel10}.
In $\S 4.1$ we will discuss the repercussions of these findings on the origin and nature of stellar mass black holes.
 
In general, the signal expected from {\it coalescing binaries} of all flavours carries exquisite, and unique information on the {\it masses} and {\it spins} of the sources and of their {\it internal} structure: in the case of GW150914 and GW151226 the simplicity of two colliding black holes.
Masses and spins will be the leitmotif of this Chapter on gravitational wave sources, as they are key parameters with which we will be describing 
the gravitational wave universe in a way complementary  to that offered by electromagnetic observations, and often unaccessible otherwise.
Key sources of the gravitational wave universe have been already presented in \cite{Sathya09,Andersson13,GWnote13,SathyaET12} and will be described ahead in this Chapter. \footnote{Regretfully, the references have been limited to a minimum, due to the vastness of the topic. We will be mentioning main key and recent papers
supplemented by specific reviews, which do include more extensive
references.}

\section{A new cosmic landscape}\label{settingthestage}\label{stage}

The theory of general relativity by Einstein makes five key predictions: the existence of 

\smallskip
\noindent
\begin{itemize} 
\item black holes; 
\item an upper limit on the compactness of any self-gravitating object endowed of a surface (neutron stars being a chief example);
\item a maximum mass for stable, degenerate matter at nuclear densities, known as Oppenheimer Volkoff limit;
\item a maximum  mass for (idealised) supermassive stars dominated by radiation pressure; 
\item gravitational waves emitted by accelerated masses in non spherical motion.
\end{itemize}
\medskip
\noindent
The electromagnetic observations of the universe reveal the existence of 

\smallskip
\noindent
\begin{itemize}
\item neutron stars, endowed by strong gravity, rapid rotation, intense magnetic fields,  and high temperatures. Powered either by rotation or accretion or magnetic field dissipation, neutron stars are living in isolation or in binaries with stars 
or neutron stars as companions. They are ubiquitous and widespread in all the galaxies;
\item  stellar origin black holes,  powered by accretion and observed in a variety of X-ray binary systems with stars as companions. They are ubiquitous and widespread in all the  galaxies;
\item active and quiescent supermassive black holes at the centres of galaxies; 
 \item galaxies with central supermassive black holes on their way to collide and merge;
\item an expanding universe changing and evolving on all scales, with gas fragmenting into stars inside dark matter halos (on the smallest scales), and 
 galaxies embedded in dark matter halos assembling in galaxy's clusters (on the largest scales);
\item an expanding universe at large, dominated by a dark energy component of unknown origin.
\end{itemize}

\noindent
The {\it combination} of these two items set the frame for constructing a new cosmic landscape, that of the {\it gravitational wave universe}. By observing the universe with gravitational waves as messengers we will be able to answer to the deep questions outlined in $\S 1$. We now explore the content of each item, 
establishing connections among the different voices.

\subsection{Black holes, neutron stars and supermassive stars as basic equilibrium objects}\label{bh,ns,sm}

\medskip
\noindent
$\bullet$ \underline {Black Holes } \\ 
The simplest object to describe in nature is a  {\it black hole}, representing 
the exact solution
for the metric tensor of a {\it point mass}  $M,$ in otherwise empty space \cite{Gravitation73}. 
The black hole solution found by Schwarzschild in 1916 describes the
static, isotropic gravitational field generated by an uncharged point mass $M$.
The mass $M$ confined into a null volume, at $r=0$, is surrounded by an {\it event horizon}, i.e. a boundary in spacetime 
 defined with respect to the external universe inside of which events cannot affect any external observer.  $M$ is not
 the baryonic mass only, but  it includes the gravitational energy.
 Within the event horizon of a black hole all paths that electromagnetic waves could take 
are warped so as to fall farther into the black hole.
In Schwarzschild coordinates, the event horizon, i.e. the surface of no return,  appears as a critical
spherical surface of radius  $R_{\rm S}=2GM/c^2$, the Schwarzschild's radius of the point mass $M$.
For the Sun the Schwarzschild radius of 2.95 km is deep in the solar interior, where Einstein's equations 
in matter space exhibit no singularity.  A key property of the Schwarzschild metric, which has no Newtonian analogue, is that below a radius, $R_{\rm isco}=3R_{\rm S},$ all 
circular orbits of massive particles are unstable ("isco" is acronym of innermost stable circular orbit).  Massive particles moving on these geodesics are fated to cross the horizon, if  subjected to an infinitesimal perturbation.

The Schwarzschild solution is a limiting case of a more general solution of the Einstein's field equations 
found by Roy Kerr \cite{Kerr63}, which describes the spacetime metric of an {\it axially-symmetric} point mass  $M$ 
surrounded by an event horizon, and describes an uncharged, {\it rotating black hole} with mass $M$: here $M$ includes the negative
contribution from the gravitational energy and the positive contribution by rotation, besides the matter load.
The finding by Kerr is remarkable as it shows, contrary to Newtonian gravitation, that a mass endowed with rotation warps spacetime 
 as the energy content from rotation becomes  source
of gravity itself, due to the non linearity of Einstein's equation. Rotation is described by the spin vector ${\bf S}= {\bf s}_{\rm spin}GM^2/c, $ 
where the norm of  the vector  ${\bf s}_{\rm spin}$ is the spin parameter taking values between 0 and 1. 
Counterintuitively, the horizon of a Kerr black hole is smaller than $R_{\rm S}$ and takes a simple expression:
$R_{\rm horizon}=(G/c^2)[M+(M^2-s_{\rm spin}^2M^2)^{1/2}]$. 
Also $R_{\rm isco}$ depends on $s_{\rm spin}$, and for $s_{\rm spin}=1$, a test particle in co-rotation (counter-rotation) has $R_{\rm isco}$ placed at $GM/c^2$ ($9GM/c^2$).

Kerr black holes have become central for understanding the nature of singularities in general relativity, and a conjecture has been posed, known as  
 {\it cosmic censorship conjecture} which asserts that no {\it naked} singularities form in Nature.  In other words, it asserts 
that singularities (present in the classical description of gravity) are enclosed by a horizon so that information does not propagate into the rest of the universe, hidden from any observer at infinity by the event horizon of a black hole. 
 For this reason the spin of a Kerr black hole is limited to values $s_{\rm spin}\leq 1$, with $s_{\rm spin}=1$ corresponding to a  maximally rotating Kerr black hole (for $s_{\rm spin}>1$ a naked singularity would appear).  
 
Black holes are fundamentally geometrical objects and a theorem, known as  {\it uniqueness} theorem, states that  Kerr black holes are the unique end-state of gravitational collapse \cite{Carter68,Hawking72,Robinson75}.  This created the belief that all "astrophysical" black holes
(that form in Nature) are Kerr black holes, being the Kerr solution the only stationary solution of Einstein's equation.  The uniqueness theorem paved  the way to a further conjecture, known as  {\it no-hair theorem.}
The no-hair theorem postulates that all black hole solutions of the Einstein-Maxwell equations of gravitation and electromagnetism in
general relativity  can be completely characterised by only three externally observable parameters: the mass $M$,
angular momentum ${\bf S}$, and electric charge. All other information (for which "hair" is a metaphor) about the matter which formed a black hole or fall into it, disappears behind the black hole event horizon and is therefore permanently 
not accessible to external observers.
 A corollary of the no-hair conjecture asserts that the only deformations  that black holes admit  
are those obtained by a change of mass and angular momentum as it occurs during the merger of two black holes, or during their 
substantial growth by accretion of in-falling matter.  Any residual deformation is then radiated away
by gravitational waves. \footnote{We do not consider here charged black holes and they are of no interest in astrophysics.
Astrophysical black holes do not live in isolation and thus do not carry a charge. If charged, matter of opposite charge would fall in  to obliterate any charge excess, 
making the black hole neutral. }

While  the theory of general relativity poses a limit on
the angular momentum, {\it no upper limit} exists  on the {\it mass} of a classical Kerr black hole. Only a lower bound exists, imposed by 
quantum mechanics natural units: the Planck mass $M_{\rm Pl}=(\hbar c/8\pi G)^{1/2}=4.34\, \mu\rm g$ below which a quantum
description of gravity is desired \cite{Hawking74-evap}. 
Astrophysical black holes are  grouped in three classes or flavours possibly because of their different origin: the stellar black holes with
masses in the interval  [$3 \msun,100\msun$], the (super-)massive black holes of [$10^5\msun,10^{10}\msun$], and the
middleweight or intermediate mass black holes of [$100\msun,10^5\msun$].  The boundaries of each interval are loosely defined and still arbitrary,
as the physical mechanisms leading to the formation of massive black holes are uncertain. Detecting black holes of all flavours  as gravitational wave sources will shed light into these mechanisms, and their potential connections.

\medskip

\noindent
$\bullet$ \underline {Compactness} \\
Let us now proceed on considering the limit on the compactness of any astrophysical object endowed by a {\it surface}.
A consequence of the Oppenheimer Volkoff equation for the equilibrium of a self-gravitating spherical body, of mass $M$ and radius $R,$ is that its {\it compactness} defined as 
${\cal C}\equiv GM/Rc^2$ can not exceed a limiting value  ${\cal C}<{\cal C}_{\rm max}=4/9,$  which holds for all stars, incompressible or not.
{\it An equilibrium object with  ${\cal C}>{\cal C}_{\rm max}$ can not exist with a finite surface.}  Note that the Schwarzschild radius $R_{\rm S}$ (and in general
$R_{\rm horizon}$) violates this condition,  implying the presence of a singularity in the interior solution of the Einstein's equations.

Not only this condition poses a {\it lower limit} on the radius $R$ of any star,  but paves the way to the idea that Kerr black holes inevitably form in nature, as soon as the condition ${\cal C}<{\cal C}_{\rm max}$ is violated. Instability to collapse occurs when the pressure support against gravity, determined by
the {\it microphysical} properties of matter, drops to the point that the total energy $E$ of the configuration 
is no longer a minimum \cite{ShapiroTeukolskybook86}. \footnote{The minimum here is computed with respect to all variations in the density profile $\rho(r)$ that leave 
the number of particles $N$ unchanged, and unchanged and uniform 
the entropy per nucleon and chemical composition.  }
The consequent loss of stability  and evolution toward gravitational collapse occurs for a variety of reasons and here we highlight the most important.

\medskip
\noindent
$\bullet$ \underline{Neutron stars} \\
In Newtonian gravity, gravitational collapse occurs when a stellar core supported by the degeneracy pressure
of cold electrons  becomes massive enough that electrons in their quantum states become ultra-relativistic, i.e.
when their Fermi energy exceeds the electron rest mass energy $m_{\rm e}c^2$.
The reduced pressure support implied by this microphysical state transition occurs at the {\it  Chandrasekhar mass limit}  of $M_{\rm Ch}\sim 1.4\msun$
(whose exact value depends on the chemical composition of the stellar core). 
Stellar evolution models show that when the iron core of a massive evolving star increases above $M_{\rm Ch},$ core collapse ensues promptly. 
The dynamical contraction comes to a halt when the entire 
iron core at the Chandrasekhar mass limit has been transformed in a core of neutrons plus few exotic nuclei,  at around or even above nuclear density $ \rho_{\rm nuc}\equiv 1.4\times 10^{14}\,\rm g\,cm^{-3}$.  A new equilibrium, i.e. a {\it neutron star} endowed with a surface (of radius $R\sim 10$ km) forms,  supported by neutron degeneracy.
The transformation of nuclear matter occurs  following photodissociation of iron group nuclei and helium present in the core, and deleptonization of matter 
via weak interactions (mostly electron captures by protons) with the concomitant emission of neutrinos of all flavours. 
The universality of the process leads to the prediction that neutron stars at birth carry a mass $M_{\rm NS}$ close to the Chandrasekhar mass limit $M_{\rm Ch}.$

The Oppenheimer Volkoff equation, which describes non-rotating neutron stars, does not admit stable equilibria above a {\it maximum mass}, $M_{\rm max}^{\rm NS}$, whose exact value depends on the details of the equation of state  (EoS)
of matter above nuclear density \cite{Lattimer16}.  
If the Chandrasekhar mass limit refers to a Newtonian instability driven by microphysical processes (change in the degree of degeneracy) inside the star, the instability of neutron stars above the maximum mass limit is induced by general relativity only \cite{ShapiroTeukolskybook86}. 
 What triggers the instability in a neutron star above the maximum mass is the pressure source term in the right hand side of the Oppenheimer Volkoff  equation.
 The huge pressure (from microphysical processes) requested to counteract relativistic gravity acts as gravity source (before the degenerate neutrons become ultra-relativistic).

 Rotation can contrast gravity, and uniformly rotating neutron stars can carry
 a mass higher than the corresponding static limit.  The maximum mass of a uniformly rotating star is determined by the spin rate at which a fluid element at the equator moves on a geodesic so that any further speed-up would lead to mass shedding.  This maximum mass can be determined numerically and is found to be at most $\sim 20\%$ larger than the non-rotating value \cite{Cook94}.  Only differentially rotating neutron stars can support significantly more mass than their non-rotating or uniformly rotating counterparts \cite{Baumgarte2000}.  
  {\it Hyper-massive neutron stars}  are 
 differentially rotating neutron stars with masses exceeding the maximum mass of a uniformly rotating neutron star.
 Because of the large angular momentum and shear, the hyper-massive neutron star is dynamically unstable to nonlinear instabilities leading to a bar mode deformation \cite{Baiotti-review16}.
 In general, the collapse of the hyper-massive neutron star to a rotating black hole is temporarily prevented by its differential rotation, but a number of dissipative processes, such as magnetic fields, viscosity, or gravitational wave
emission, will act so as to remove differential rotation. The hyper-massive neutron star can collapse  directly to a black  hole on the dynamical timescale. Alternatively, the star by loosing its differential rotation 
may evolve into a so-called {\it supra-massive} neutron star, i.e., an axisymmetric and uniformly
rotating neutron star with mass exceeding the limit for non-rotating neutron stars (defined in the static limit). 
Magneto-rotational energy losses ultimately drive the star to collapse secularly  to a black hole. In Nature, hyper-massive neutron 
stars likely form in the aftermath of a neutron star-neutron star merger that we will describe in $\S 7.3$.

\begin{figure}[!t]
\begin{center}
\includegraphics[width=1.00\textwidth]{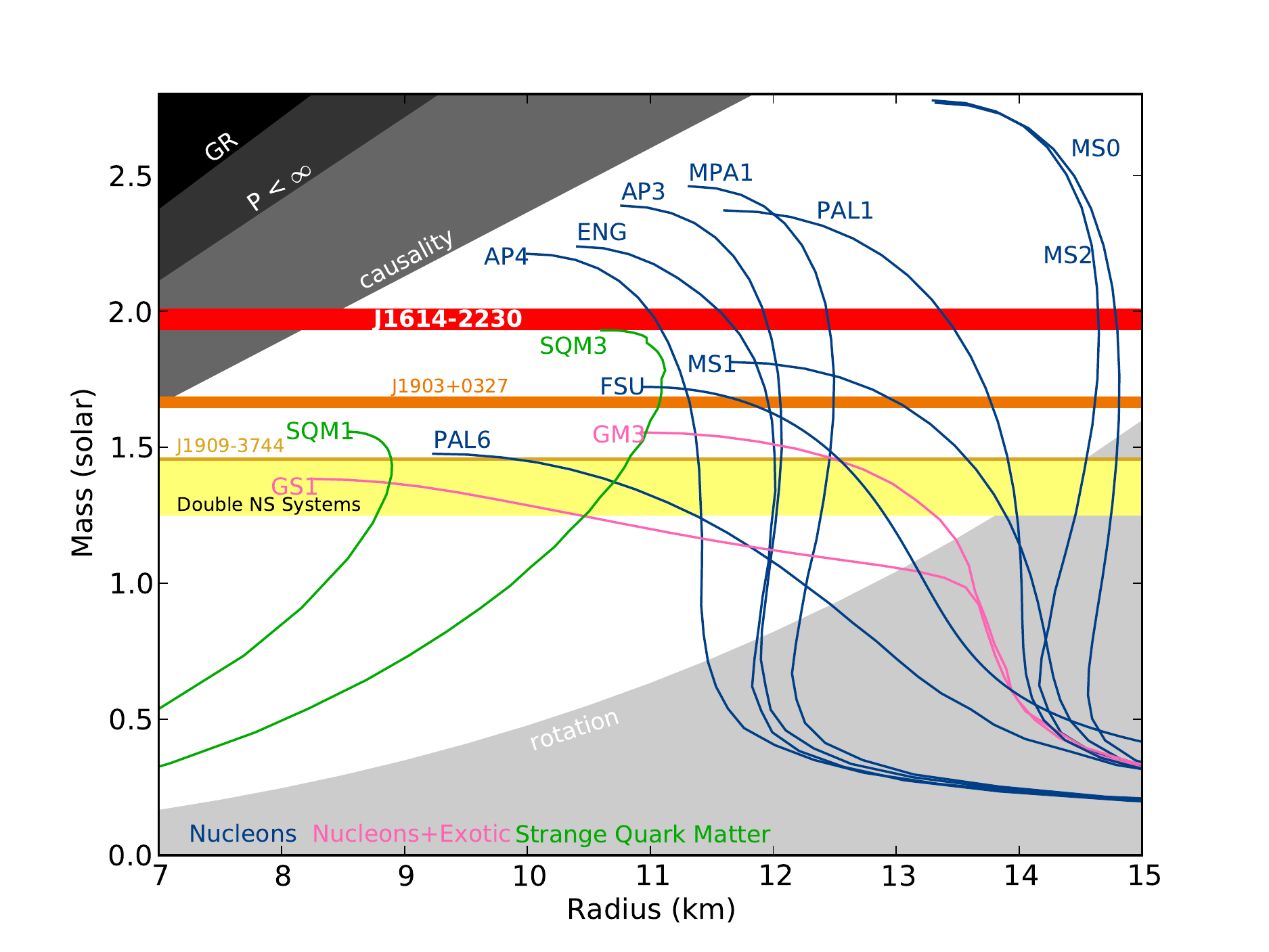}
\caption{Sequences of neutron star equilibria for selected Equations of State  (EoS) \cite{Lattimer16}. The
  figure shows the mass $M$ versus radius $R$ relation, for non-spinning neutron star models. 
  The yellow band shows the interval of masses inferred from the sample of double neutron star binaries.
  The orange (red) band indicates the mass of the neutron star in J1903+0327  (J1614-2230).
  The largest mass currently measured is that in J1614-2230, of 1.97$\pm$0.04$\,\msun$.  Thus, any EoS line that does
  not intersect the J1614-2230 band is ruled out by this measurement \cite{Demorest10}. In
  particular, most EoS involving exotic matter, such as kaon
  condensates or hyperons, tend to predict maximum neutron star masses  below
  $2.0\msun$, and are therefore ruled out.  Green lines refer to strange star models. The upper left 
  grey areas of different intensity refer to regions excluded by general relativity and causality, and the lower grey area
  is excluded by observations of the fastest pulsars.  Courtesy of P. Demorest.}
  \label{MR-neutron-stars}
\end{center}  
  \label{fig:1RM}
\end{figure}

According to many theoretical studies, there exists a range of values for  the mass $M_{\rm max}^{\rm NS}$ in the static limit, between $1.5\msun$ and $2.7\msun$, due to current uncertainties in the behaviour of matter at supra-nuclear densities as shown in Figure \ref{MR-neutron-stars}.  Quantum chromodynamics is expected to give a complete description of matter at the energy-densities  of neutron stars. At present no unique model exists for describing 
the many-body ($10^{57}$ baryons) nuclear interaction, understood as a residual coupling of the more fundamental interactions among quarks, and a phase transition to a free quark state may occur inside the star.  A strange star made of strange quarks, representing the lowest energy state of matter at zero temperature, may also form in Nature \cite{Lattimer16}.

\medskip

\noindent
$\bullet$ 
\underline{ Supermassive stars } \\
 A second example of a relativistic instability which conducts to the black hole concept is that of an equilibrium configuration 
dominated by {\it radiation} \cite{ShapiroTeukolskybook86}.   This is called {\it supermassive star}, with hypothetical masses clustering around $10^5\msun$, but  unknown dispersion.

Consider a supermassive star dominated by radiation, 
 in convective equilibrium and with uniform chemical composition. Its mass $M_{\rm sms}$ is determined uniquely
 by   the value of the photon entropy per baryon $s_{\rm rad}=(4/3) aT^3/n$, where $T$ is the temperature of radiation and matter, $n$ the baryon number density and $a=\pi^2k_{\rm B}/(15c^3\hbar^3)$ the radiation constant (with $k_{\rm B}$ and $\hbar$ the Boltzmann and Planck constants).  Since a supermassive star has the structure of a Newtonian polytrope of index $\gamma=4/3$,  supermassive stars can cover a wide mass spectrum, depending only on $s_{\rm rad}$, and their mass is 
 \begin{equation}
M_{\rm sms} = 1.127\left ( {s_{\rm rad}\over k_{\rm B}} \right )^2 \msun,
 \end{equation}
  independent on the central density \cite{ShapiroTeukolskybook86}. The interesting fact is that one can relate $s_{\rm rad}$ to the baryon load of the star.
  A simple calculation shows that $s_{\rm rad}/k_{\rm B}=8\beta^{-1}$, where $\beta$ is the matter to radiation pressure ratio. Arbitrarily large masses
  can be assembled for arbitrarily small values of $\beta$, since $M_{\rm sms}\propto \beta^{-2}$.  Since a polytropic relation connects the mass, radius and central density, one
  can compute the compactness parameter ${\cal C}$ which turns out to be very small $\ll 0.39$ since supermassive stars are rarefied, loosely bound objects (for $M_{\rm sms}=10^4\msun$  
  the radius is $R\gg 10^4$ km) 
  with  low surface gravity owing to radiation pressure.
    
 Although general relativity does not intervene in determining the overall structure of the star, it affects instead its stability, as a polytrope of $\gamma=4/3$ is, in stellar structure, the trembling limit between stability and instability of the star, so it is necessary to take into account the small
 effect of matter pressure and general relativity which play little or no role in their structure calculation.
 A further important fact is that a supermassive star radiates energy at a rate very close to the Eddington limit
$ L_{\rm Edd}= 4\pi GM_{\rm sms} mc/\sigma_T=1.3\times 10^{42}(M_{\rm sms}/10^4\msun)\,\rm erg\,s^{-1}$ (where $m$ is the mean baryon mass and $\sigma_T$ the Thomson cross section), and therefore its total energy $E$ continues  to decrease. Thus, the star evolves over time  toward states or lower energy (more bound)  and higher compaction.
 When plasma and general relativity effects are included in the stability analysis, one can show that the supermassive star loses  its stability when 
 its central density has reached a limiting value 
 $\rho_{\rm crit}=1.996\times 10^4(10^4\msun/M_{\rm sms})^{7/2}\,\rm gr\, cm^{-3}$  after having radiated away an energy $E_{\rm rad}=3.583\times 10^{54}$ erg, over its lifetime,
 a quite large value which is independent on the mass of the supermassive star.  At the critical point of instability, the core temperature  $T_{\rm crit}=4\times 10^9 (10^4\msun/M_{\rm sms}) \, \rm K$ is large enough to ignite nuclear reactions which can affect the final fate.
 Once the supermassive star has reached the instability line through a progression of quasi-static equilibrium states, it can either explode or collapse to a black hole.  
 
 General relativity calculations \cite{ShapiroTeukolskybook86} have shown that the nominal range
 of supermassive stars collapsing into a black hole lies between $\sim 5\times 10^{5}\msun$ and $\sim 10^8\msun$. 
 \footnote {The maximum mass of a supermassive star is  set by the comparison of two timescales: the thermal  timescale $\tau_{\rm thermal}$
and the timescale for the star to adjust to a new hydrostatic equilibrium, i.e. the dynamical timescale $\tau_{\rm dyn}$.  If  $\tau_{\rm thermal}$ is shorter than $\tau_{\rm dyn}$  the star can not any longer recover equilibrium, and rapid cooling leads the whole star to collapse. The thermal timescale at the boundary of stability $\tau_{\rm thermal}\sim E_{\rm crit}/L_{\rm E}$ equals  $\tau_{\rm dyn}\sim (G\rho_{\rm crit})^{-1/2}$ at a mass $M_{\rm max}^{\rm sms}\sim 10^8\msun$ \cite{ShapiroTeukolskybook86}. }  The collapse is homologous, with a velocity  essentially linear with radius, and density profiles self-similar, although increasing in magnitude. Due to the homologous nature of the collapse, the entire mass moves inward coherently, crossing the event horizon in only a few light travel times $GM_{\rm sms}/c^3$ \cite{ShapiroTeukolskybook86}.
The concept of supermassive star has now evolved  into the modern one of DCBH, acronym of a process which call for the Direct Collapse of a equilibrium structure as a supermassive star into a massive Black Hole \cite{Omukai13,Latif16review}.  Supermassive stars, should they exist, are believed to grow via accretion onto an embryo of solar mass, in pristine, non-fragmenting gas clouds. In the context of structure formation models, they
are viewed as equilibrium structures transiting through progressively more massive states (typically of $10^5\msun$) 
that are conducive to stable episodes of hydrogen and helium burning. Later, they collapse into a black hole when 
crossing the general relativity instability limit either after exhaustion of the fuel or during the nuclear- burning phase \cite{Umeda16}.
 Ahead in this chapter we will return on this issue.

\subsection {Gravitational wave sources: a first glimpse} \label{glimpse}
Neutron stars and black holes are  the most bound, lowest-energy states of self-gravitating matter known in the universe where gravity is in the  {\it strong field regime} and the field is {\it stationary.}  No processes can lower their energy state.
When do neutron stars and black holes  become sources of gravitational waves?

 Gravity is the weakest interaction in nature but when high compactness  combines with large scale, non-spherical coherent mass motions with velocity $v$  near the speed of light (as in a merger of two compact objects), 
 then an immense luminosity can be emitted in gravitational waves, the luminosity scaling as $L\sim (c^5/G)(R_{\rm S}/R)(v/c)^6$.
  GW150914 is the first extraordinary example of a merger of  two black holes moving at near 1/2 of the speed of light releasing a  luminosity in excess of $10^{56}\ergs$.

 When perturbed  out of equilibrium, during their formation or when colliding, neutron stars and black holes become often, and for a short time lapse, among the loudest sources of gravitational radiation in the universe. In particular, binary coalescences of compact objects are among  the most powerful emitters 
 that theory predicts. For binary coalescence we refer to
 the process of inspiral of two compact objects in a binary terminating with their merger into a new single unit, as is the case of GW150914 and GW151226.

 In essence, the trait of a powerful gravitational wave source stems in its exquisite high degree
of disequilibrium, leading to non-spherical dynamics under extreme conditions of compactness.
 Neutron stars, and black holes over a wide
spectrum of masses are the protagonists of most of the violent events detectable
by both current and next generations of interferometers, on Earth and in space.

\medskip
\noindent
$\bullet$ \underline {The frequency of gravitational waves}\\
Gravitational wave sources emit over a broad frequency range, and there is a close link between the 
frequency of the gravitational wave $f$ and mass $M$ and compactness ${\cal C}$ of the source.
The natural unit for  $f$ is 
\begin{equation}
f_{\rm o}\equiv {c^3\over GM}={c\over R_{\rm G}}=2 \times 10^5  {\msun\over M}\,{\rm  Hz}, 
\label{fnat}
\end{equation} 
where $R_{\rm G}\equiv GM/c^2$.
 In any self-bound system of mass $M$ and size $R$, the natural frequency of oscillation, rotation, orbital revolution and dynamical collapse is 
 of the order of 
\begin{equation}
f_{\rm source}
\sim  \left (  {GM\over R^3}\right )^{1/2}= f_{\rm o}\left ({R_G\over R}\right )^{3/2}= f_{\rm o}\,{\cal C}^{3/2}
\label{gw-n}\end{equation}
 Since gravitational waves are emitted by accelerated, asymmetric mass motions, the  frequency  $f$ of the gravitational wave 
 is expected to be close to the frequency $f_{\rm source}$ of the source's mass motions; $f\simeq f_{\rm source}$, and in general $f<f_{\rm o},$ as ${\cal C}< 1$.
 
 We focus now on the case of compact binary {\it coalescences} (CBCs).  
 For {\it  black holes}, whose horizon  $R_{\rm horizon}$ is between 2 and 1 $R_G$ (depending on the spin parameter $s_{\rm spin}$), the  characteristic frequency of the wave near coalescence  is $\sim f_{\rm o}$, so that the total mass $M$ of the binary system determines the highest frequency of a coalescence signal.
It is customary to introduce the frequency  $f_{\rm isco}\equiv f_{\rm o}/(\pi 6^{3/2})$,  equal to twice the Keplerian frequency of a test mass  at 
the innermost stable circular orbit $R_{\rm isco}$, as the characteristic frequency of a binary near coalescence. For stellar black holes with typical mass of $\sim 10 \msun$
 \begin{equation}
f^{\rm BH*}_{\rm isco} \approx 0.44 \times 10^3  \left ( {10 {\rm M}_\odot\over M}\right ) \,{\rm Hz},
\label{isco-bh*}
\end{equation}
whereas 
for  massive black holes 
\begin{equation}
f^{\rm BH,massive}_{\rm isco}\approx  4.4\times 10^{-3} \left ( {10^6 {\rm M}_\odot\over M}\right )\,{\rm Hz}. 
\end{equation}
During the inspiral and merger  $f$ sweeps upwards for $f\ll f_{\rm isco},$ up to  $f_{\rm isco}$. 
 As neutron stars carry masses $\sim 1.35\,\rm M_\odot$ they can extend their gravitational wave emission at slightly higher frequencies $f^{\rm NS}_{\rm isco}\simeq 1.6$ kHz than stellar black holes. 
 
 According to the above relations, coalescing {\it massive  black holes}  are intrinsically {\it low frequency} sources, whereas coalescing {\it stellar origin black holes} and {\it neutron stars} combined in different arrangements are {\it high frequency} sources.  In the case
 of a binary composed by a massive and a stellar black hole (denoted as extreme mass ratio inspiral, EMRI) the reference mass is that of the largest hole.
 Thus, {\it EMRIs} belong to the {\it  low frequency} universe as the mass of the big black hole  sets the frequency of emission.
  
In more detail, if  $f_{\rm min}$ is the minimum  ($f_{\rm max}$ is the maximum) frequency of operation of an interferometer, equation (\ref{isco-bh*}) sets an upper (lower)  limit on the mass of a binary that can be detected when nearing the final phase of plunge and coalescence.  In the case of high frequency sources, this leads to
\begin{equation}
2.20  \left ({2000 \,{\rm Hz}\over f_{\rm max}}\right ) \msun <M< 440 \left ({10 \,{\rm Hz}\over f_{\rm min}}\right ) \msun.
\label{bandLIGO}
\end {equation}
In the case of low-frequency sources 
\begin{equation}
4.4\times 10^4  \left ({0.1 \,{\rm Hz}\over f_{\rm max}}\right ) \msun <M< 4.4\times 10^7 \left ({10^{-4} \,{\rm Hz}\over f_{\rm min}}\right ) \msun.
\label{bandLISA}
\end {equation}

At nanoHz frequencies, the typical mass of a black hole binary near coalescence would be far in excess of $10^{10}\msun$. At these very low frequencies, its is possible to detect the signal from
supermassive black holes of $10^{8-9.5}\msun$ far from coalescence, i.e. at $f\ll f_{\circ}$.  The signal in this case is continuous and nearly monochromatic. 
By contrast, the signal from coalescing binaries is transient in nature and peaks during the latest phases of inspiral and plunge when the 
 two objects reach separations comparable to  their sizes:  $\sim 60$ km for neutron stars, $\sim 200$ km for stellar origin black holes,
and a few AU up to $\simless 10-100$ AU for massive black holes, depending in their mass.  How this variety of coalescing binaries form in Nature 
 is the subject of the following section. To this purpose  we here overview key observational properties and key notions in the realm of current astronomical 
 observations.

\section{The electromagnetic universe}\label{em}

\subsection{Neutron stars and stellar origin black holes in the realm of observations}\label{ns-bh-obs} 
Neutron stars are known to form in the aftermath of the gravitational collapse of massive stars ($\simgreat 10\msun$) whose degenerate iron core is driven above the Chandrasekhar mass limit. The collapse releases $\sim 10^{53}$ erg. Most of the energy (99\%) is emitted in neutrinos and only about $10^{51}$ erg into kinetic energy of
the supernova explosion which is associated to 
the propagation of a shock wave, emerging when the infalling star's envelope impacts on the dense neutronized core that settles into equilibrium: that of a young, hot neutron star. During shock break-out, the stellar envelope unbinds producing a luminous supernova \cite{Burrows13}. Crab with its remnants is a magnificent example
of a successful supernova explosion.  However, if the shock break-out is weaker, as for the case of heavier stars,  a stellar black hole forms by  {\it fall back} of part of the envelope onto the proto-neutron star driven above
its maximum mass. At the extremes, {\it direct collapse} to a black hole can occur.

The mass of a compact object can be measured when it is a member of a binary system.
At present, data from a variety of observations indicate that {\it neutron stars} likely show a bimodal, asymmetric distribution in their masses, with a low mass 
component centred around $1.393\msun$ and dispersion $0.064\msun$,  and a heavier component  with a mass mean of $1.807 \msun$ and dispersion $0.177\msun$ \cite{Antoniadis16}.  
The yellow strip in Figure \ref{MR-neutron-stars} shows the range of neutron star masses observed in double neutron star binaries, and the red line
shows the heaviest neutron star ever detected of $1.97\pm 0.04\msun$  \cite{Demorest10} consistent with the expectation that neutron stars in "binaries" experience {\it re-cycling}, i.e. a long-lived phase during which 
they accrete matter  from the companion star. In this case, the mass of the compact object may not represent the mass at birth and gives information on the interaction of the neutron star with its companion.  As shown in Figure \ref{MR-neutron-stars}, this finding already rules out the softest EoS for nuclear matter.

\begin{figure}[!t]
\begin{center}
\includegraphics[width=1.20\textwidth]{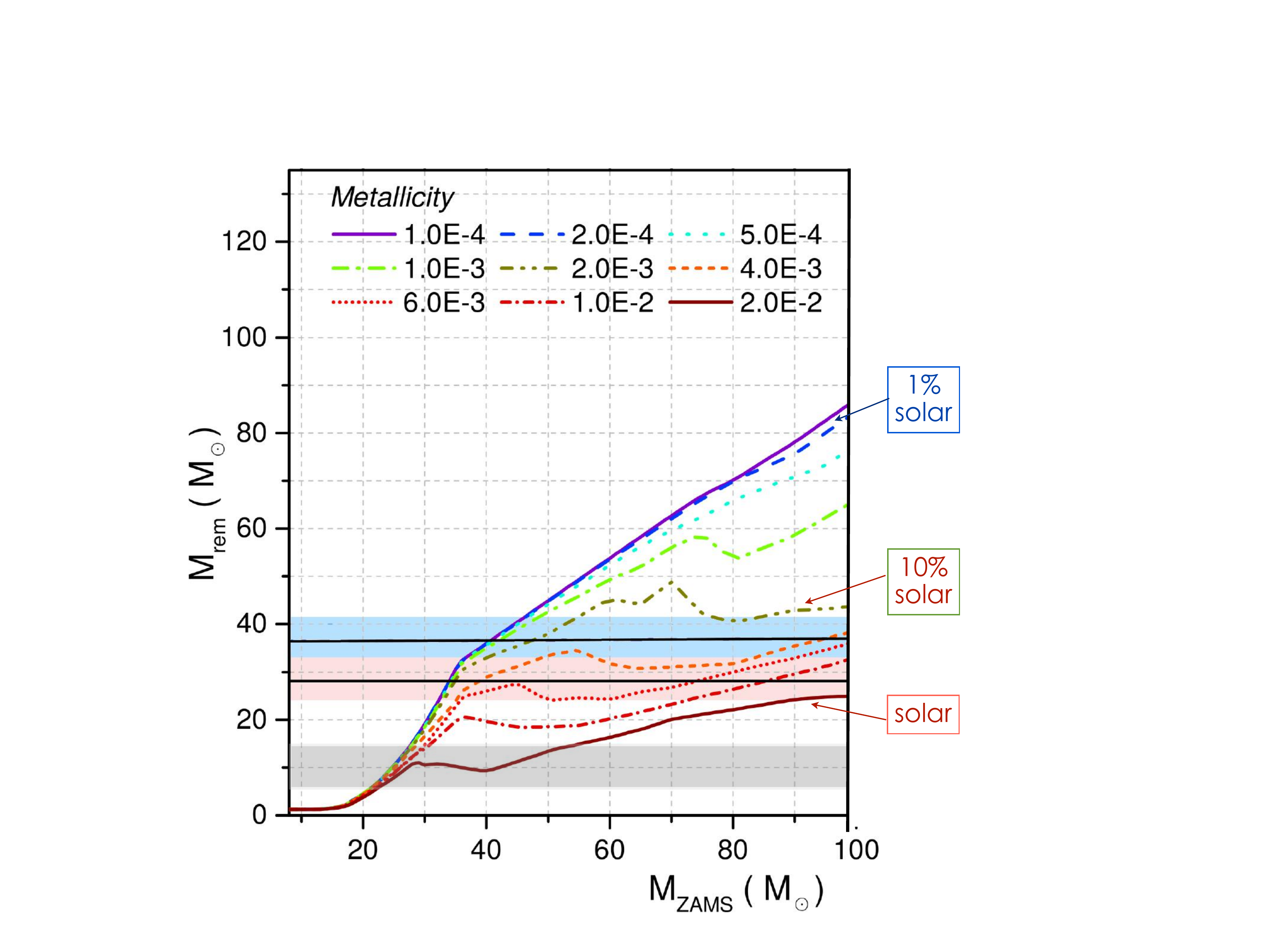}
\caption{The figure shows the mass of the remnant as a function of the mass of the progenitor star in the interval between [10$\msun$, 100$\msun$]
 for different values of  the stellar metallicity, from \cite{Spera15} . 
The lower (solid red) line refers to $Z=0.02$ (Sun's metallicity). The upper line refers to $Z=10^{-4}$. In between the metallicities (from bottom to top) are $0.01,0.006,0.004,0.002,0.0001,
5\times 10^{-4}, 2\times 10^{-4}$. (A black hole is assumed  to form above $M_{\rm max}^{\rm NS}=3\msun.$)  
The upper horizontal bands indicate the two values of the black hole masses of GW150914 (and uncertainty interval), prior to merger, and the 
lower pink strip indicates the interval of black hole masses measured electromagnetically in galactic X-ray binaries \cite{Ozel10}. 
The black holes in GW151226 fall in this lower strip. 
Courtesy of M. Mapelli.}
\label{BH-mass-metallicity} 
\end{center} 
\end{figure}

For {\it stellar origin black holes},  reliable dynamical mass measurements in low mass X-ray binaries are best described by a narrow mass 
distribution peaked around $7.8 \pm 1.2\msun$ \cite{Ozel10} with a clear divide between neutron stars and black holes, i.e. no 
remnants between
$\sim 3$ and $5\msun$ \cite{Ozel12}, often referred to as {\it gap}. Higher mass values are inferred in high mass X-ray binaries, and the mass of Cyg X-1, the first black hole discovered in X-rays is bound to values
$\sim 14-16\msun$. The currently observed range of black hole masses is indicated in Figure \ref{BH-mass-metallicity} as lower grey strip, and the two black holes
in GW151226 fall in the same range.
But, the discovery of the two "heavy" stellar black holes of $29$ and $36 \msun$ in GW150914 came as a surprise\cite{Abbott-3-science}, though 
hypothesised  by \cite{Belci10metallicity,Dominik12} in their studies on binaries. Heavy stellar black holes, resulting from low metallicity progenitor stars, were considered earlier by \cite{Mapelli09} in the context of a class of sources known as Ultra Luminous X-ray sources, which often inhabit low metallicity galaxies.

The fate of massive stars and nature of the relic is a complex process to model.
Studies by \cite{Heger02,Heger03,Zhang08} show that the nature of the remnant depends to a large extent:
(i) on the mass loss by stellar winds during the different evolutionary stages, driven by the opacity of the metals present in the star's envelope and measured by the metallicity $Z$ 
(defined as the logarithm
in power of ten of the iron to hydrogen abundance ratio, and often expressed in units of the solar metallicity $Z_\odot=0.02$); (ii) on rotation;  (iii) on the strength of the shock break-out through the stellar envelope after core collapse and bounce; 
(iv) on the interplay between neutrino cooling and heating 
at the interface between the neutrino-sphere and the (stalled) shock; 
and (v) on the amount of fall-back material accreted onto the newly born hot  neutron star, after formation of a reverse shock. These processes establish whether the collapse is delayed  (lasting longer than 0.5 sec) or prompt (lasting less than 250 msec) and determine the value of the mass of the relic star.

Figure \ref{BH-mass-metallicity} shows the mass of the compact remnants as a function of the star's initial mass, in the interval between [10$\msun$, 100$\msun$] for different values of the absolute 
metallicity $Z$, predicted by the models of \cite{Spera15} .  The figure indicates that 
lower metallicity progenitor stars leave heavier relic black holes.
The upper horizontal  bands in the figure indicate the masses of the two black holes in GW150914 (with their uncertainties). Metallicities
of the order of  $0.1-0.2\, Z_{\odot}$ are necessary to form stellar origin black holes as heavy as those
in GW150914, from stellar evolution models  \cite{Belci10metallicity,Spera15,Belci16}. 
We further note that black holes with masses $\simless 15\msun$ form in any metallicity environment. Thus the black holes in GW151226 do not pose constraints on the metallicity of the host galaxy \cite{Abbott-GW151226,Abbott-16-three-sources}.

Following the discovery of GW150914 \cite{Abbott-1,Abbott-3-science} and of GW151226 \cite{Abbott-GW151226}
new questions arise: \\ 
{\color{red} 
\begin{itemize}
\item  What is the shape and normalisation of the mass function of relic stars? 
\item How far does the black hole mass function extend at high masses?
\item How can pairs of relativistic objects as those in GW150914 and GW151226 form in binaries and coalesce within the age of the universe? 
\item Which are the astrophysical conditions for the rise of a substantial population of gravitational wave sources as GW150914?
\end{itemize}
 
}
\subsection{Forming stellar origin compact binaries}

Tutukov and Yungelson were the first to study the evolution of isolated massive binaries, and predicted the formation of merging binary compact objects of the different flavours  \cite{Tutukov93,Rasio12-review}.
The formation of neutron star binaries (NS,NS) became the subject of intense studies soon after the discovery of PSR1913+16, the first binary pulsar for which we had evidence, albeit indirect, of the existence of gravitational waves \cite{Hulse75,Taylor82}. Formation models of neutron star-black hole (NS,BH*) and 
black hole-black hole (BH*,BH*) binaries have been developed in parallel despite lacking
of any observational evidence.  Compact binaries can form  in the galactic field as outcome of stellar evolution in primordial binaries \cite{vandenheuvel91,
Belci10metallicity,Dominik12,Mandel16};  in dense star clusters via 
stellar dynamical exchanges involving stars and black holes \cite{Mapelli16,Rodriguez16,Benacquista13,Portegies00};  or  in more exotic environments as in the discs of active galactic nuclei \cite{Stone16}.

\begin{figure}[!t] 
\begin{center}
\includegraphics[width=1.2\textwidth]{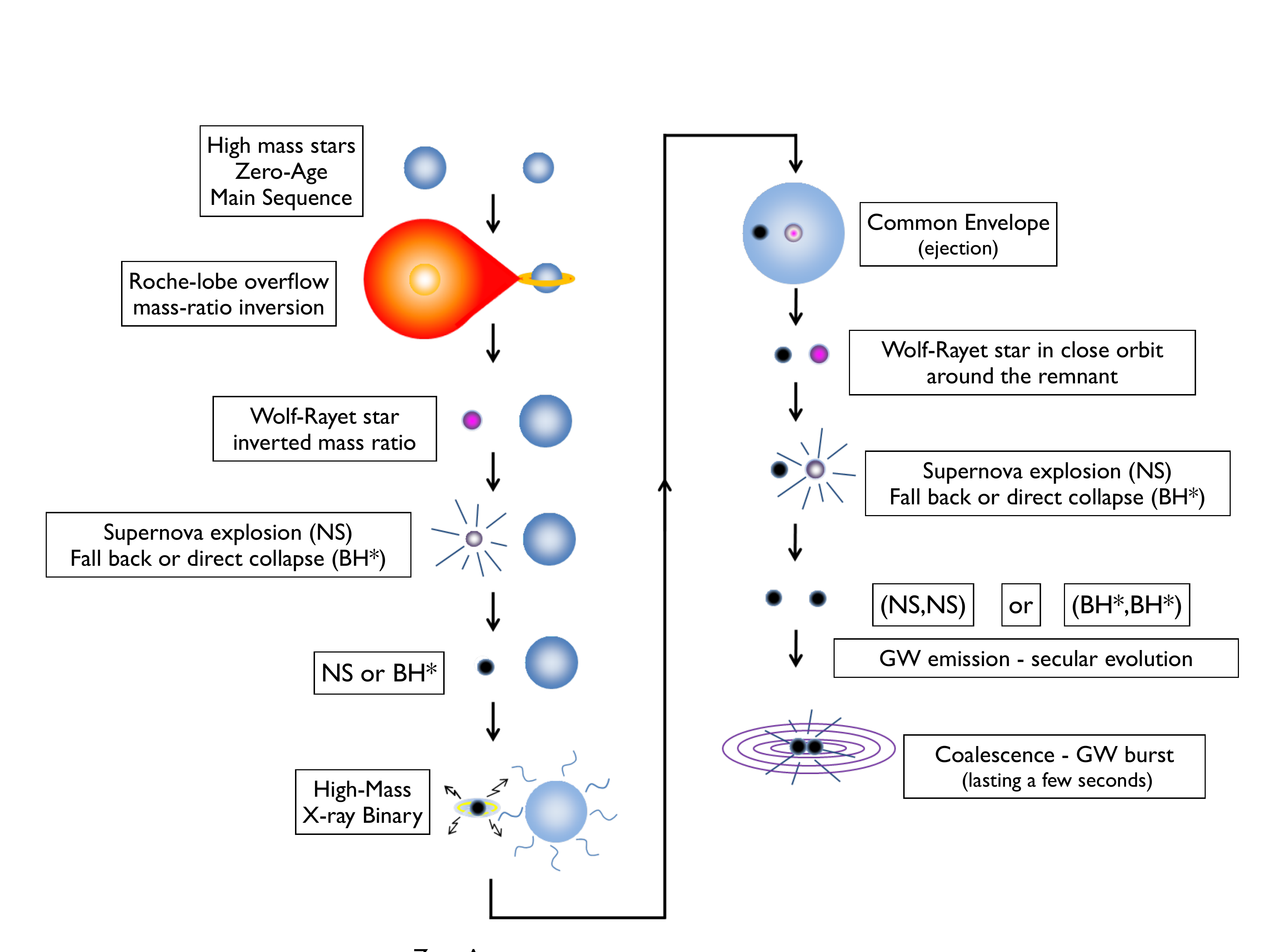}
\caption{The path of formation of compact binaries in the primordial binary scenario.  
The initial mass of the stars determine the nature of the compact binary. To reproduce the data of GW150914,
 Belczynski et al. \cite{Belci16} start with two stars of $96\msun$  and $60\msun, $ at a separation of 2463 R$_\odot$
 in a low
 metallicity environment, with $Z=0.03 Z_\odot$ (we defer to Figure 1 of \cite{Belci16}). The figure is rearranged from \cite{Marchant16}.  Courtesy of T. Tauris.}
\label{BH*formation} 
\end{center} 
\end{figure}
Stellar population synthesis models are a powerful tool to establish how and in which fraction close pairs of compact objects can form in primordial binaries and coalesce within a Hubble time.  
The input parameters for starting a simulation are: (i) the shape of the underlying gravitational potential,  
(ii)  the initial mass function (IMF) of massive stars on the zero-age main sequence, (iii)  the metallicity
of the parent gas cloud, (iv) the fraction of primordial binaries (and triplets), and (v) the distribution of the initial binary separation and eccentricity, which affect the degree of interaction of the two stars over their lifetime. 

Stars lose mass via winds, but in binaries they can also donate their mass to the companion via mass exchange, as illustrated in Figure \ref{BH*formation}, 
which depicts the evolution of a binary system in a simplified way. Mass transfer occurs when the most massive star, 
which evolves first away from the main sequence, fills its Roche lobe. The pouring of mass on the companion star leads to a re-equilibration of the mass ratio and in general makes the less massive star  the heaviest in the system as time evolves.
After mass exchange, the star that  evolves first becomes a Wolf-Rayet or a helium star (depending on the initial mass) that can go supernova.
The supernova explosion can unbind the binary due to mass loss  and recoil that accompany anisotropic core-collapse. 
Neutron stars are known to receive natal kicks at the time of their formation, with mean velocities of $\sim 400\kms$ \cite{Hobbs05}, so that the binary can break up.
In fact, as many as $\sim 90\%$ of potential binaries may end up 
being disrupted after the first supernova explosion. This makes (NS,star) binaries very rare objects.
Black holes which form either through fall back (with supernova display) or direct collapse,  likely receive lower kicks but the three-dimensional distribution of
their natal velocities is unknown \cite{Repetto12}. Thus, the weaker mass loss that may accompany their formation, and lower natal kicks may help a heavy binary to survive almost intact after the formation of the first compact object.  Thus the rate of formation of (NS,star), (BH*,star) systems is not directly set by the shape of the IMF, as disruption mechanisms that break lighter binaries can limit the number of double neutron star systems  that may form.

After birth of the first remnant,  evolution continues, through a phase of Common Envelope evolution (depicted in Figure \ref{BH*formation}) when the second star becomes a giant and starts engulfing the 
 companion remnant which spirals inwards via gas dynamical friction losing orbital angular momentum and energy, which is deposited as heat in the largely convective envelope. 
Then, the remnant star either merges plunging inside the dense core of the companion star, or lands on a close, very tight orbit after having ejected the entire envelope. In this last case,  the core of the star evolves into a relic object and may go supernova, fall-back or direct collapse depending on its mass, so that binaries of all the three flavours can form.  After common envelope and mass ejection, the tight binary that forms is less fragile against break up and can survive.  If the two relic stars that managed to remain bound are sufficiently close (a few solar radii in separation), gravitational waves will drive the binary toward coalescence (as described in $\S 6$),
on timescales that may vary between a few Myr to Gyr or more.
This avenue is affected by uncertainties on the common envelope evolution process, the kick distribution, and supernova modelling, so that the formation of compact binaries is a genuine statistical process and 
the rate of coalescences largely undetermined.

\begin{figure}[!t]
\begin{center}
\includegraphics[width=1.0\textwidth]{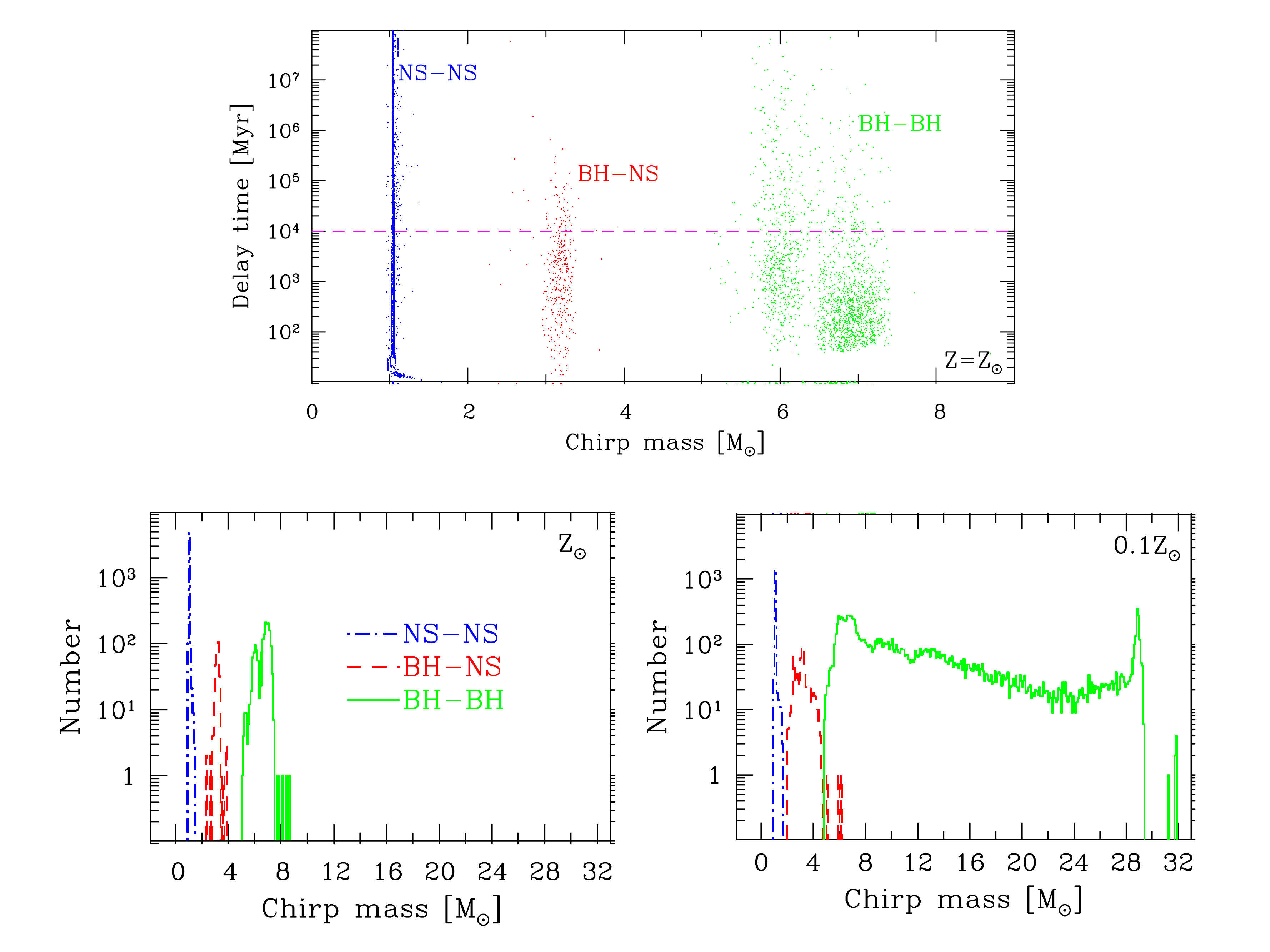}
\caption{Upper panel: Distribution of (NS,NS), (NS,BH*), and (BH*,BH*) delay times as a function of the chirp mass $M_{\rm c}=\nu^{3/5}M$ of the binary, where
$M$ is the total mass, $\mu$ the reduce mass and $\nu=\mu/M$ the symmetric mass ratio ($\nu=0.25$ for equal mass binaries). (For equal mass (NS,NS)
binaries the chirp mass is $\sim 1.2\msun$.)
The plot rearranged from \cite{Dominik12} refers to a simulated population of primordial binaries 
of solar metallicity. Note that the delay time to coalescence can by far exceed the age of the universe (horizontal dashed line). Lower panels depict the distribution of compact binaries that form in the three different flavours as a function
of the chirp mass $M_c$, for two different metallicities: solar (sub-solar) in the left (right) panel. The chirp mass of GW150914 is $27.9\msun$
and that of GW151226 is  $8.9\msun$. Courtesy of T. Bulik.}
\label{Binary-distributions} 
\end{center} 
\end{figure}

Within the scenario of primordial field binary formation, the effect of metallicity was anticipated by \cite{Belci10metallicity} who pointed out that heavy black hole binaries could form in low-metallicity environments. 
Figure \ref{Binary-distributions}  shows  the distribution of (NS,NS), (NS,BH*), and (BH*,BH*) mergers as a function of the chirp
   mass of the binary $M_{\rm c}$ (defined in the caption) from a population synthesis model and for two different metallicities. 
      Depending on $Z$, whether it is solar or sub-solar, the expected number of compact binaries in all their arrangements changes dramatically, with black holes
   filling the high end of the mass distribution  in the low metallicity channel. In the upper panel of Figure  \ref{Binary-distributions}, we show the broad distribution of delay times  as a function of the chirp mass. The  {\it delay time} is defined as the time it takes a binary to coalesce since its formation as primordial stellar system.  The distribution of delay times is  broad, going from a Myr up to $10^9$ Myr. In general, population synthesis models suggest 
  that the delay times follow a power-law distribution with slope $-1$, in the interval from 10 Myr up to $10^4$ Myr, i.e. a uniform distribution for logarithmic bin.
 In the context of primordial, isolated binaries, Belczynski et al. \cite{Belci16} find that the typical channel for the formation of GW150914 like binaries involves two very massive stars 40-100$\msun$ formed
 in a low metallicity environment with $Z<0.1\,Z_\odot $ that interacted once
through stable mass transfer and once through a phase of common envelope evolution,
with both black holes forming without no supernova display and low natal kick.
   
   Recently and again in the context of primordial binaries, an alternative channel as been proposed for the origin of GW150914, named MOB (massive over-contact binary), which involves two very massive low-metallicity ($Z\sim 0.1 Z_\odot $)
stars in a tight binary which remains fully mixed due to their high spins induced by orbit synchronism driven from tides \cite{Mandel16,deMink16,Marchant16}.  
Rotation and tides transport the products of hydrogen burning throughout the stellar envelopes, enriching the entire star with helium and preventing the build-up of an internal chemical gradient. In this scenario there is no giant phase: both stars remain in stable contact filling their Roche lobes and eventually form two massive black holes, because the cores that collapse are massive.
   
   Compact binaries can also form in dense, young star clusters or globular clusters and galactic nuclei, via dynamical processes \cite{Benacquista13}.
      The high stellar densities, of the order of $10^{3-6}$ stars pc$^{-3},$ present in star's clusters, favour the formation of black hole binaries via exchange interactions with other stars \cite{Ziosi14}. In particular  three body exchange interactions   (BH*,star)+BH*$\to$ (BH*,BH*)+star  can lead to the built up of a population of
   massive compact object binaries \cite{Portegies00}.  Being  the heaviest objects in the cluster, these binaries mass segregate at the cluster centre
   on a timescale shorter than the two-body relaxation time, and continue to experience exchange encounters with other black holes that can further rearrange
   them in progressively heavier binaries, that can be as massive as GW150914 \cite{Rodriguez16}. Hardening due to scattering off stars can drive these
   binaries to coalesce within  $\sim 1-10$ Gyr, and may also escape the parent cluster due to dynamical recoil \cite{Portegies00,Benacquista13}.

 \subsection{Massive black holes in the realm of observations}\label{MBH-em}
 
There is compelling evidence that besides stellar origin black holes, there exists a substantial population of 
{\it supermassive} black holes of $10^5\msun-10^{9}\msun$ that inhabit the centres of galaxies.
This "other flavour" is observed in two states: an {\it active} and a {\it dormant} state \cite{Merloni16,Kormendy13}.

Active supermassive black holes are {\it accreting} black holes at the centre of galaxies, which power
 the luminous, highly variable QSOs, and the less luminous Active Galactic Nuclei (AGN).
 The accretion paradigm  states that 
outside the
event horizon of a supermassive black hole, radiation is generated with high efficiency ($\varepsilon_{\rm acc} \sim10$\%, higher than nuclear reactions) through the  viscous dissipation of kinetic energy  from gas orbiting deep in the
gravitational potential of the hole. The energy escapes in the form of radiation, high velocity plasma outflows, and relativistic particles to produce
 luminosities of $10^{44}-10^{47} \ergs$  emitted over a wide spectrum and in 10\% of the cases  in the form of collimated  radiation. 

Dormant supermassive black holes appear ubiquitous in nearby bright galaxy spheroids.
When dormant, their presence in inactive galaxies is revealed, albeit indirectly, through the measure of Doppler displacements in the spectral lines of 
 stars and/or gas in the nuclear region of the galaxy.  Often line-of-sight velocities show a Keplerian rise attributed to 
the presence of a point like gravitational potential dominating that of stars in the centre-most region of the galaxy.
 The Galactic Centre provides the most compelling evidence of a supermassive black hole.
The Milky Way hosts a $4\times 10^6\msun$ "dark object" surrounded by a swarm of stars in Keplerian motion 
as close as $(1-10)\times 10^3R_{\rm S}.$  The distance of the nearest star to the central object  poses a lower limit on its
compactness, found of the order of $10^{15}\msun\, \rm pc^{-3}$ \cite{Genzel10}. No nuclear star cluster can remain in dynamical equilibrium at these densities, 
so that the black hole 
is the most simple and elegant hypothesis. 

 \begin{figure}[!t]
\begin{center}
\includegraphics[width=0.99\textwidth]{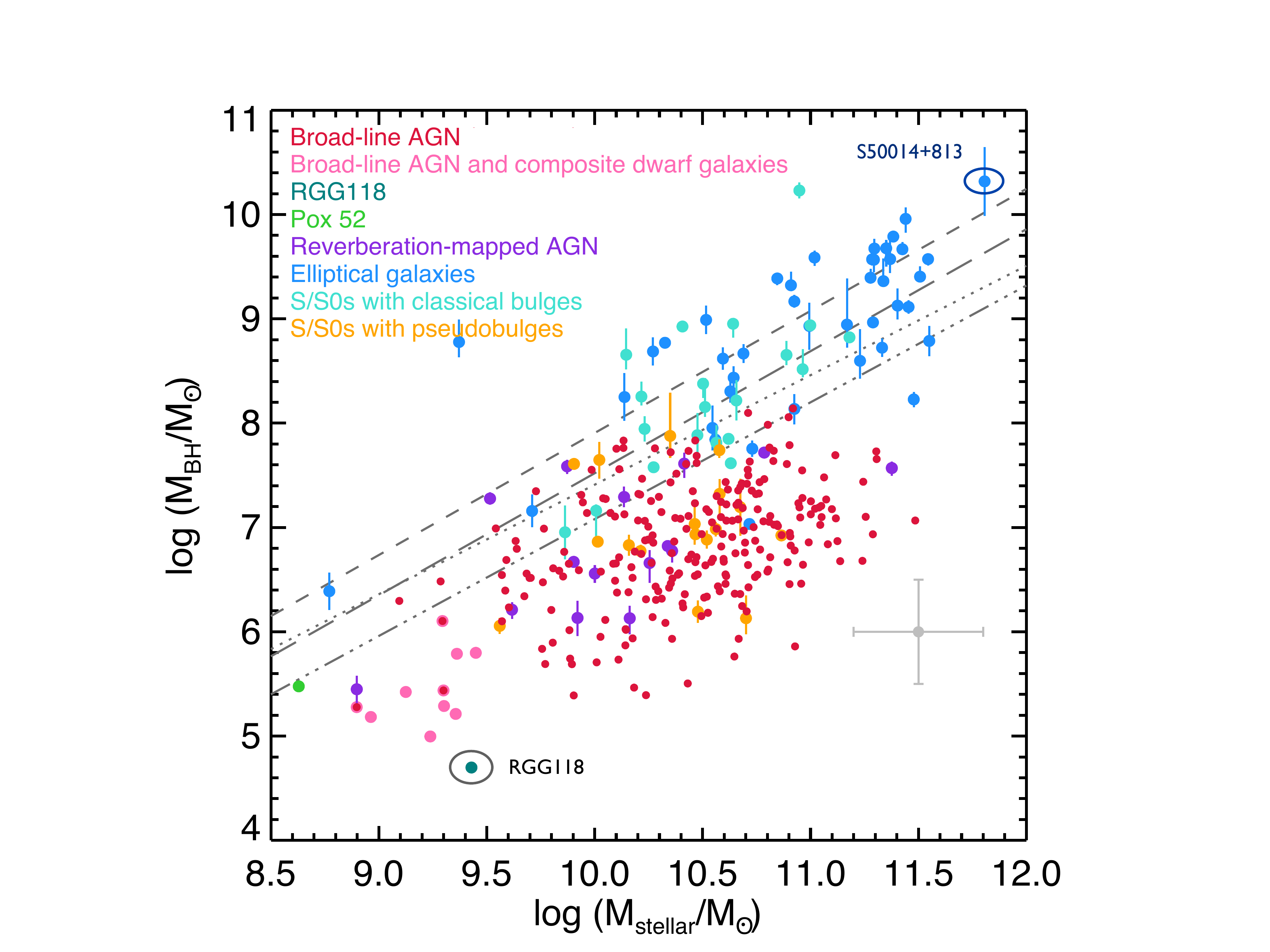}
\caption{Black hole mass versus total host galaxy's stellar mass.  The figure rearranged from \cite{Reines15} illustrates the vastness of the mass spectrum of supermassive black holes
residing at the centres of galaxies.  
The sample  comprises  244 broad-line AGN, 15 reverberation-mapped AGN, and
a sample of dormant supermassive black holes for which the mass is inferred from dynamical measurements.
Elliptical galaxies are shown in blue, S/S0 galaxies with classical bulges in  turquoise, and S/S0 galaxies with pseudo-bulges in orange points.  Dashed lines show the $M_{\rm BH}-M_*$ correlation from different authors.  
The two data points with a circle and associated name 
indicate the lightest and heaviest black holes observed as of today. 
Details on the plot can be found in \cite{Reines15}. Courtesy of M. Volonteri.}
\label{M-mass} 
\end{center} 

\end{figure}
Figure \ref{M-mass} illustrates how broad is the mass distribution inferred from a sample of both dormant and active close-by supermassive black holes in galaxies of different morphology, kinematics and stellar masses $M_*$. It extends from $5\times 10^4\msun$ (the lightest black hole discovered in RGG118)  to $4\times 10^{10}\msun$ (the giant black hole in S50014+831).  (Recently, the mass of  S50014+813 as been revised downwards to $7\times 10^9\msun$: see e.g. \cite{Sbarrato16}).\footnote{The mass of the black hole $M_{\rm BH}$  in bright, massive spheroids correlates with properties of the host galaxy in  ellipticals and S/S0s with classical bulges. 
Two are the correlations, the  $M_{\rm BH}-M_*$, where $M_*$ is the stellar mass of the host galaxy, and a second (tighter) 
between $M_{\rm BH}$ and the velocity dispersion $\sigma$ of the stars, 
measured far from the black hole \cite{Kormendy13}. 
These correlations (often referred to as $M_{\rm BH}-\sigma$, and $M_{\rm BH}-M_*$ relations, the last shown as dashed or dotted lines in Figure \ref{M-mass}:
see \cite{Reines15} for details) state that bright galaxy spheroids with higher stellar velocity
dispersions, i.e. with deeper gravitational potential wells, grow heavier black holes, and that 
brighter, more massive galaxies host more massive black holes.
Despite being black holes tiny objects, with an influence gravitational radius $R_{\rm inf}\sim GM_{\rm BH} /\sigma^2$
extending out to $\sim 1$ pc (much smaller than the galaxy's size of tens of kpc), 
black holes  "see" the galaxy they inhabit, and galaxies "see" the central black hole they host.  Consensus is rising that the $M_{\rm BH}-\sigma$ relation 
is fossil evidence of a symbiotic co-evolution of black
  holes and bright spheroids. Most likely, the relation was established during 
the course of galaxy formation and assembly in episodes of self-regulated accretion and mergers when the black holes were active, creating a balance between 
accretion flows with their radiated power and
gas at disposal for triggering  both/either star formation and accretion. The correlation is poor instead when extended to a larger sample of galaxies types and
galaxy masses, indicating  that particularly in lower mass systems co-evolution never got to completion, or never started.}

From the study of the kinematics of stars and gas in nearby galaxies, one can estimate  the black hole local mass
density: ${\rho_\bullet} \sim {(2 -5) \times 10^{5}}\msun\,\rm Mpc{^{-3}}.$
This mass density is remarkably close to the mass density increment $\Delta \rho_\bullet=3.5\times 10^{5}(\varepsilon_{\rm acc}/0.1)^{-1}$
that black holes experience over cosmic history (between $0.5<z<3$) due to efficient accretion \cite{GWnote13,Merloni16}.  
This last value is
inferred considering that active black holes in galaxies contribute to the rise of the cosmic X-ray background resulting mostly from unresolved, obscured AGN of mass $10^8\msun$ - $10^9\msun$.
As the contribution to the local black hole mass density
${\rho_\bullet}$ results from black holes in the same mass range, the close match
between the two independent measures, $\ensuremath{\rho_\bullet}$ and
$\Delta\ensuremath{\rho_\bullet}$, indicates that   radiatively efficient
accretion  ($\varepsilon_{\rm acc} \approx 0.1$) played a large part in the building of the mass of the 
supermassive black holes in galaxies, from redshift $z\sim3$ to now.  It
further indicates that information residing in the initial mass distribution of
the, albeit unknown, black hole seed population is erased during events of
copious accretion, along the course of cosmic evolution.
Thus, QSOs and AGN are believed to emerge from a population of {\it seed 
black holes}  with masses in a range largely unconstrained (from $100\msun$ up to $10^{4-6}\msun$).
This is because  the mass of black holes increases sizeably due to accretion, over a relatively short  $e$-folding  timescale $\tau_{\rm BH}\approx {4.7
  \times 10^{8}}\, \varepsilon_{\rm acc} f_{\rm E}^{-1}(1-\varepsilon_{\rm acc} )^{-1}$ yr compared to the age of the universe
 (where $f_{\rm E}=L/L_{\rm E}\sim 0.1$ gives the luminosity in units of the Eddington luminosity, and $(1-\varepsilon_{\rm acc})$ the fraction of mass accreted by the hole in order to radiate a luminosity  $f_{\rm E}$ with efficiency $\varepsilon_{\rm acc}$).

The {\it spin} of a black hole is also a key parameter in the context of gravitational wave astronomy, as together with the mass it can be inferred from the rich structure of the  waveform (as illustrated ahead in this Chapter).  Mass and spin are strongly coupled across the accretion history of a growing black hole.  Spins determine directly
the radiative efficiency, and thus also the rate at
which the black hole mass is increasing. In radiatively efficient accretion discs, the efficiency 
 $\varepsilon_{\rm acc}$ varies from 0.057 (for $s_{\rm spin}=0$) to 0.15 (for $s_{\rm spin}=0.9$)
and 0.43  ($s_{\rm spin}=1$, for a maximally rotating black hole).  Accretion on the other hand
determines black hole's spins since matter carries with it angular
momentum (the angular momentum at the innermost stable circular orbit $R_{\rm isco}$).  A non rotating black hole is spun-up to
$s_{\rm spin}=1$ after increasing its mass by a factor $\sqrt 6$, for
prograde accretion. \footnote{ Gas accretion from a
geometrically thin disc limits the black-hole spin to
$s^{\rm acc}_{\rm spin}=0.9980$ \cite{Thorne74}, as photons emitted with angular momentum anti-parallel to the black hole spin
are preferentially captured, having a larger cross section.  In a magnetised, turbulent thick disc, the spin attains an equilibrium value $s^{\rm acc,mag}_{\rm spin}\simeq 0.93$ \cite{Gammie04}.}
Conversely, a
maximally rotating black hole is spun-down by retrograde accretion to
$s_{\rm spin}\sim 0$, after growing by a factor $\sqrt {3/2}$ \cite{Bardeen70}.

The direction and norm of the black hole spin play a key role in the study of the spin-mass evolution of black holes. 
In a viscous accretion disc whose angular momentum ${\bf J}_{\rm disc}$ is
initially misaligned with the spin ${\bf S}$ of the black hole, Lense-Thirring precession 
of the orbital plane, acting on the fluid elements, warps the disc forcing the gas
close to the black hole to align (either parallel or anti-parallel)
with the spin vector of the black hole. The timescale for warp propagation is very rapid and the warp extends out to rather large radii  \cite{Bardeen75}.
Following
conservation of total angular momentum, the black hole responds changing its spin direction ${\bf  S}$.
The spin starts precessing and  
the system evolves into a configuration of minimum energy where ${\bf S}$ and ${\bf J}_{\rm disc}$ are aligned
and parallel, if  $S/J_{\rm disc}<1$.
Black hole precession and alignment occur on a timescale $\tau_{\rm align}\propto s_{\rm spin}^{5/7}$ shorter than the 
$e$- folding accretion time scale (typically $\tau_{\rm align}\sim 10^{5-6}$ yr)\cite{Perego09}.
If $S/J_{\rm disc}<1$ accretion tends to spin the black hole up  after re-orienting the black hole spin.
By contrast heavier black holes for which $S/J_{\rm disc}>1$ oppose more inertia  and the spin direction does not suffer
major re-orientations \cite{Perego09,Sesana14spin}.

Two limiting scenarios for the spin evolution have been proposed:
{\it Coherent accretion} refers to accretion episodes 
from a geometrically thin disc lasting for a time $\tau_{\rm episode}$ longer than the  black hole
mass growth $e$-folding time $\tau_{\rm BH}$, bringing its spin up to its limiting value $s_{\rm spin}\sim 1$, and with  ${\bf S}$ parallel to ${\bf J}_{\rm disc}$.   By contrast,
{\it chaotic accretion} refers to a succession of accretion episodes
that are incoherent, i.e. randomly oriented with $\tau_{\rm episode}<\tau_{\rm BH}$. The black hole can then
be spun-up or down, depending on the comparison between  ${\bf J}_{\rm disc}$ 
and ${\bf S}$.  If accretion proceeds via
uncorrelated episodes with co-rotating and
counter-rotating material equally probable,  the spin direction continues to change.
Counter-rotating material spins the black hole down
more than co-rotating material spins it up, as the innermost stable
orbit  of a counter-rotating test particle is located at a larger
radius ($R_{\rm isco}=9R_{\rm G}$ for $s_{\rm spin}=1$) than that of a co-rotating particle ($R_{\rm isco}=R_{\rm G}$ for $s_{\rm spin}=1$), and accordingly carries
a larger orbital angular momentum.  If $\tau_{\rm episode}<\tau_{\rm align}$  chaotic accretion results in low spins \cite{King05}. 

The two scenarios of coherent and chaotic are at the extremes of a wide distribution
of angular momenta for the accreting gas which is determined by the kinematic and dynamical properties of 
gas and stars
at the galaxy's centres which change over cosmic time as galaxies are not isolated systems.

  \begin{figure}[!t]
\begin{center}
\includegraphics[width=1.00\textwidth]{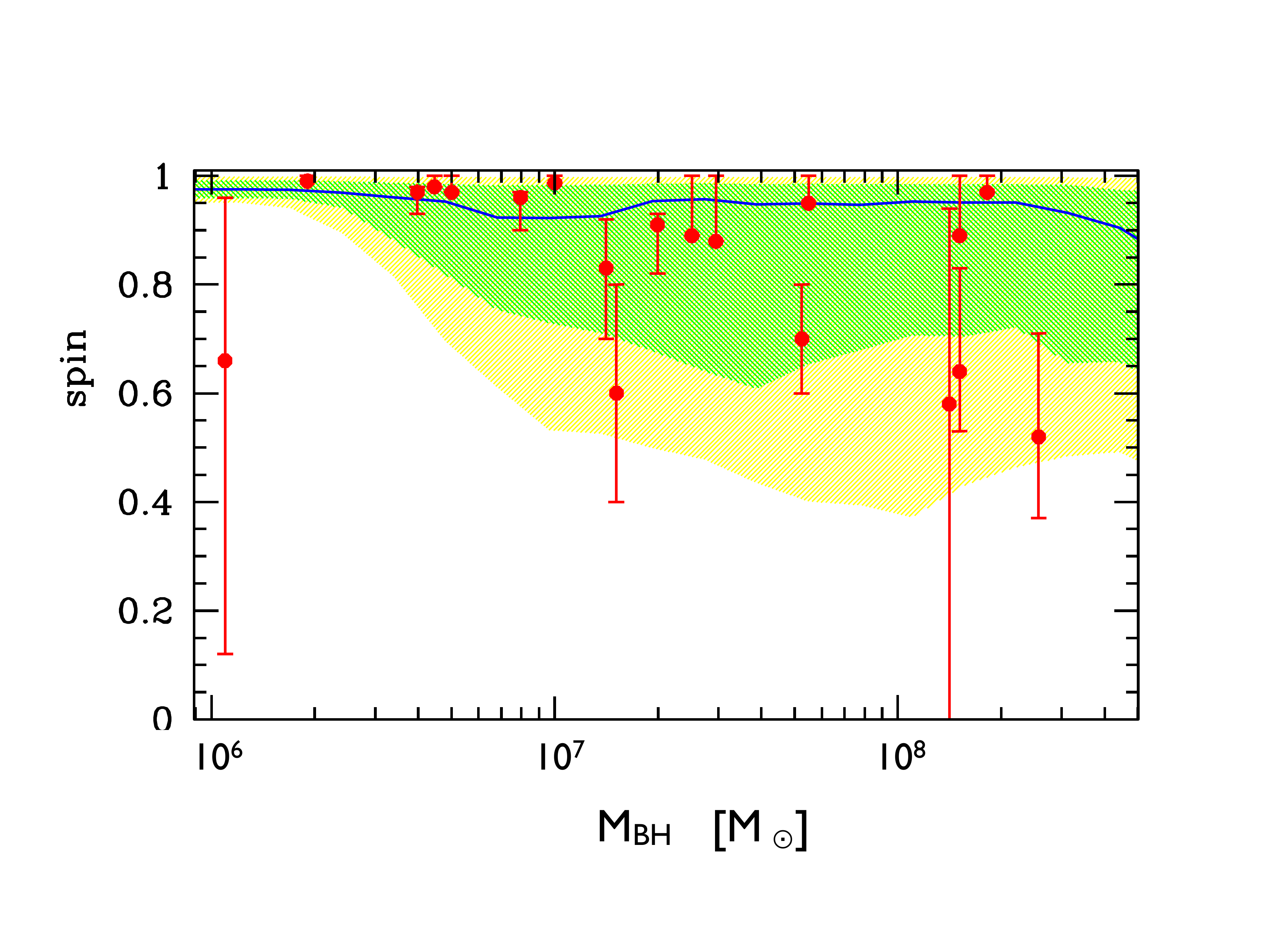}
\caption{Comparison between measured spins from a sample of AGN (red dots) and predictions of a model by \cite{Sesana14spin} from  a simulated sample of accreting
massive black holes (with Eddington ratio $f_{\rm E}\simgreat 0.01$) in spiral galaxies.  The observational data appear to disfavour both coherent accretion along a fixed direction and chaotic (isotropic) fuelling. When the properties of the accretion flow are anchored to the kinematics of the host galaxy,  a combination of coherent and incoherent accretion phases best describes the data.}
\label{MBH-spin} 
\end{center} 

\end{figure}
At present the spin moduli of a handful (20) of AGN, hosted in low redshift late type galaxies, has been measured  through 
the spectra of relativistically broadened $K\alpha$  iron lines, and  are reported in 
Figure \ref{MBH-spin}. The data points are then compared with a hybrid model by \cite{Sesana14spin} which follows the joint evolution of the mass $M_{\rm BH}$ and  the spin vector  ${\bf S}$ by precession and accretion, of a simulated population of growing black holes in late type (spiral) galaxies. 

Mass and spin
are directly encoded in the gravitational wave signal emitted during the merger of massive black holes, and mergers are detectable with space-borne
detectors
out to very large cosmological distances.  Therefore,  measuring the masses and spins of coalescing black holes over  cosmic time will offer unprecedented details on how they  have been evolving via repeated episodes of accretion and mergers.

\subsection{The black hole desert }

\begin{figure}[!t]
\begin{center}
\includegraphics[width=1.00\textwidth]{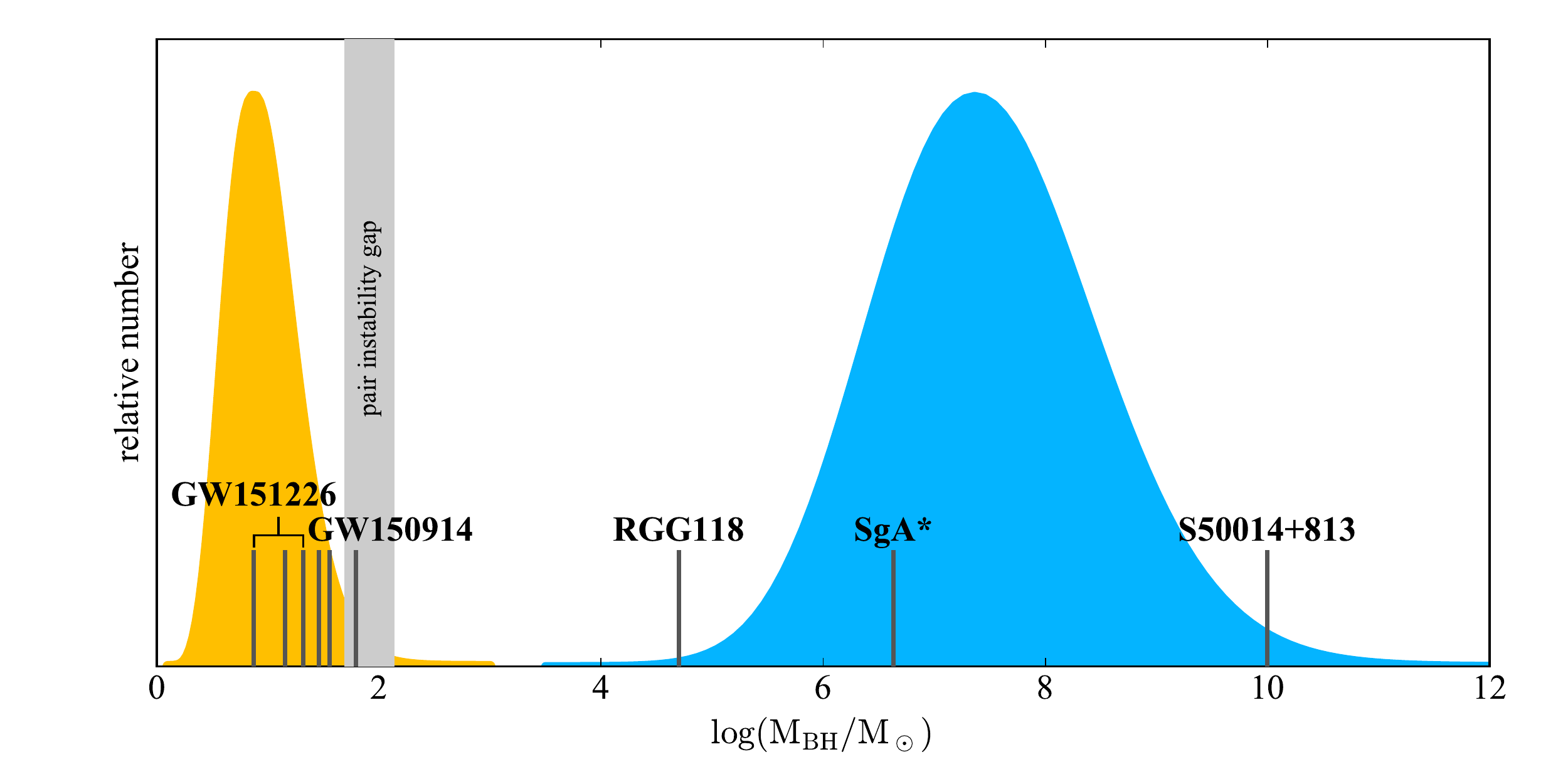}
\caption{A pictorial representation of the black hole mass function (modelled as a log-normal distribution), encompassing the mass intervals known, from stellar origin black holes to massive and giant black holes at the centre of galaxies,
to illustrate the presence of a {\it desert}  at intermediate mass scales. Vertical lines denote the black hole masses in GW150914 and GW151226 (including
the mass of the new black hole, result of the merger) and the smallest and largest mass of the two active black holes known as of today in galaxies. The grey vertical strip denotes the region of Pair instability Supernovae
that leave no remnant \cite{Heger02,Belci-Heger16}.
Advanced LIGO and Virgo jointly with LISA in space will shed light into the physical mechanisms that lead to the formation of heavy stellar black holes and black hole seeds covering the desert zone.}
\label{BHdesert} 
\end{center} 

\end{figure}

There is a black hole {\it desert} in the mass range between $\sim 65\msun$, the mass of the remnant black hole in GW150914 (the highest known as of today) 
and the mass of the lightest supermassive black hole known at the centre of the dwarf galaxy RGG118, of $5\times 10^4\msun$, as depicted in Figure \ref{BHdesert}.  Key questions arise that the new gravitational wave astronomy  can answer:\\
{\color{red}
\begin{itemize}
\item Is the desert real, i.e. empty of middle sized black holes, or is the desert inhabited by black holes which we still do not detect?
\item If inhabited, is the desert populated by transition objects, resulting from the clustering/accretion of stellar black holes viewed as single building blocks?
\item Or is there a "genetic divide"   between stellar origin black holes and massive black holes growing from seeds of unknown origin?
\item What is the maximum mass for a stellar origin black hole$^{\circ\circ}$?
\item Is there a gap in the mass function of stellar origin black holes induced by pair-instability supernovae?
\end{itemize}
} 
\noindent
$^{\circ \circ}$(Here a {\it maximum mass} for a stellar origin black hole $M_{\rm max}^{\rm BH*}$ is intended not as fundamental mass limit (as in the 
case of neutron stars)  but as a value related to the existence of a limit on the maximum mass $M^*_{\rm max}$
of stars on the zero-age main sequence \cite{Krumholz15,Belci10maxBHmass,Belci-Heger16}.)

To elaborate more on the above questions we notice that  there is a conceptual distinction between  stellar origin black holes and  supermassive black holes: the first are the relic of 
the very massive stars that experienced stable and long lived episodes of nuclear burning. The second are possibly the relic of rare supermassive stars 
that may never experienced long-lasting phases of nuclear burning and which formed in rather extreme, isolated environments.  There is also a "morphological" distinction: stellar black holes (typically more than several millions per galaxy) are spread everywhere in all the galaxies
of the universe, as stars are.  In addition, they continue to form as stars do, in the galaxies. Instead,  massive black holes (from the middleweight size to the giants) are found at the centres of galaxies (perhaps not in all), as single dark massive objects, and may have formed at early cosmic epochs or over a narrower interval of cosmic times, when the first galaxies were
forming and assembling \cite{Volonteri10}.  If the 
desert is devoid of objects, this would unambiguously indicate that the physical conditions leading to the two flavours are distinct.  Locating the dividing line is not easy.


Massive stars with sub-solar metallicities and masses between $\sim 100\msun$ and $\sim 260\msun$  explode as pair instability supernovae \cite{Heger02,Woosley07,Belci-Heger16}. The pair instability is encountered when, late in the star's life, a large amount of radiative energy  is converted into electron-positron pairs which reduce the pressure support against gravity, cause rapid contraction of the core and trigger explosive burning of the CO core, ultimately leading to the
disruption of the star.
Thus, if the IMF is devoid of stars with mass in excess of $\sim 260\msun$, pair instability supernovae
 may actually limit the value of the maximum mass of a stellar origin black hole  $M^{\rm BH*}_{\rm max}$ \cite{Belci-Heger16}.
 \begin{figure}[!t]
\begin{center}
\includegraphics[width=1.00\textwidth]{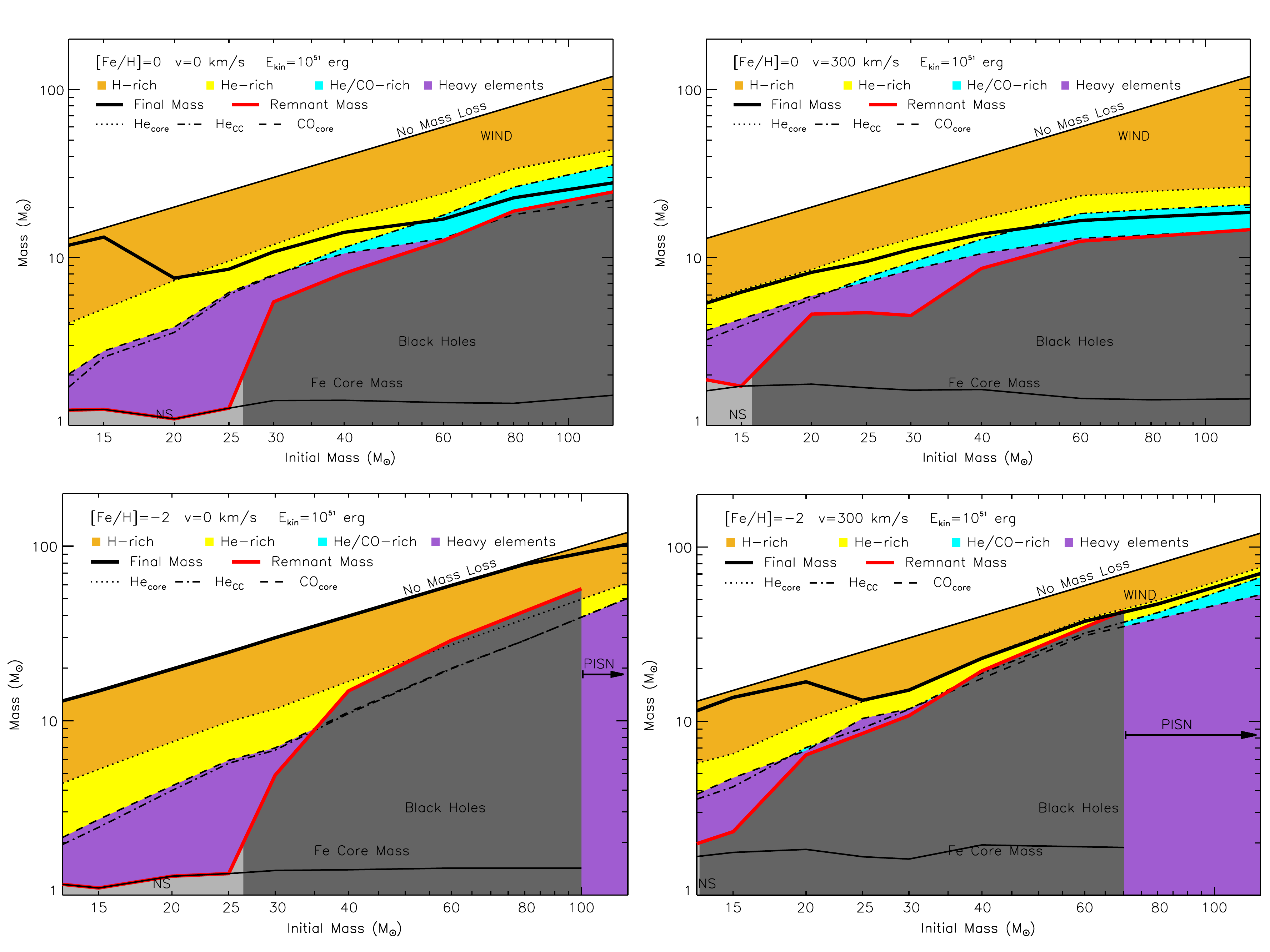}
\caption{Global properties of a generation of massive stars as a function of the metallicity\cite{Limongi12}. 
The red line in each panel gives the mass of the remnant versus the initial mass (in $\msun$). In the legenda
symbols and colours describe the mass of different elements from hydrogen to iron inside the relic star prior to core collapse.
Upper panels refer to stars  of 
solar metallicities  $Z=0.02$ ([Fe/H]=0), and rotation velocities of $V=0\kms$ (left), $V=300\kms$ (right).  At solar metallicities,
mass losses by winds inhibit the growth of a massive CO core and the pair instability is never encountered. The large mass losses also limit the mass of the relic black hole to values $<20\msun$. Lower panels refer to stars with 
$Z=3.236\times 10^{-4}$ ([Fe/H]=-3) which encounter the pair instability supernova (PISN).  The black hole maximum mass is close to $60\msun$, depending weakly on rotation, 
and it is attained at the threshold of PISN. 
The figures are adapted from the chapter "Supernovae from massive stars" by M. Limongi in "Handook of Supernovae", edited by Springer.  Courtesy of M. Limongi. 
\label{pair-instability} }
\end{center} 
\end{figure}
In the four panels of Figure~\ref{pair-instability},  the mass of the relic black holes (red line) is plotted versus
the initial mass of the progenitor star, for two selected values of 
the metallicity, $Z=Z_\odot$ and $Z=0.016 Z_\odot$ respectively, for both non-rotating stars and  rapidly rotating stars
(with velocity $300\kms$).  At sub-solar metallicities the maximum mass of the relic black hole
coasts around values $\sim 60\msun$ almost independently on rotation and on metallicity, provided is sub solar 
($Z<0.1\,Z_\odot$)\cite{Limongi12}.  As shown in the bottom right panel,
 rapid rotation limits again $M^{\rm BH*}_{\rm max}\sim 50\msun,$ and this occurs at lower 
progenitor masses $\sim70 \msun$. During core He burning, rotation driven mixing causes diffusion of 
matter from the He convective core into the surrounding radiative zones. 
Such an occurrence has the consequence of increasing the CO core mass at core He
depletion and therefore of reducing  the limiting initial mass that enters the pair instability
regime.
If the IMF is top heavy and contains stars in excess of $260\msun$, a gap between $50\msun$ and $135\msun$ (the He mass inside a 260$\msun$ star \cite{Woosley07}) should appear in the mass function of (single) stellar origin black holes, as anticipated in \cite{Belci-Heger16}.  Advanced LIGO and Virgo will likely help to shed light into this 
problem, if/when  "heavy" stellar  binary black hole mergers will be further discovered.

In the logic of a "genetic divide", supermassive stars likely play a key role.
The formation of  radiation dominated  equilibrium  states with masses $\simgreat 10^{5}\msun$\cite{Umeda16},  evolving into a DCBH, i.e. a configuration collapsing directly into a black hole, requires rather extreme conditions to form\cite{Loeb13-book}, (i) namely pristine, metal free gas clouds (site of the forming supermassive star)  irradiated by an intense flux of 
ultraviolet radiation to promote the dissociation of the main coolant, i.e. molecular hydrogen, and avoid fragmentation \cite{Omukai13,Latif16review,Umeda16}; or/and (ii) the coherent collapse
of massive gas clouds in major galaxy mergers\cite{Mayer15}. These scenarios would lead to a clean divide
between the two black hole flavours, but Nature appears to be more complex and continuous.

There is  indeed the possibility that the desert is filled of {\it transition} black holes\cite{Volonteri10}.
Processes of aggregation might have been in action, e.g. using as {\it single building blocks} stellar origin black holes which merge inside nuclear star clusters \cite{Lupi14}. 
Alternatively, episodes of supercritical accretion may drive stellar origin black holes to grow up to "seed" sizes
of $10^{3}\msun$  when
 residing  in gas rich disky galaxies\cite{Lupi16}.  Photon trapping ensures that 
the momentum from outgoing radiation does not feed back to halt accretion which can continue until 
fuel exhaustion. 
In these cases, the desert zone would be filled
by transition  black holes with a broad range of masses between 100 and 1000$\msun$ or more.

Runway collisions among massive stars, in young, dense nuclear star clusters at the centre of unstable proto-galactic discs \cite{Portegies02,Devecchi12,Mapelli16} may also lead to middleweight black holes. The resulting 
runaway super-massive star, product of the multiple merger of stars (typically of mass $\simless 10^3\msun$) may  evolve into a {\it quasi-star}, i.e. a black hole surrounded by a massive accretion envelope \cite{Begelman10}, ultimately forming a black hole of intermediate mass\cite{Fiacconi-Rossi16}. 
Detecting black hole coalescences of different flavours with both Earth and space-based detectors
will enable us to shed light into this complex problem.

 \subsection {Formation of gravitational wave sources: the cosmological perspective}  
 
In this short paragraph, we show that the existence of close pairs of massive black holes fated to coalesce is a key, unescapable prediction
of the process of clustering of cosmic structures, and that  the formation of close binary systems comprising stellar origin black holes and neutron stars 
has a natural connection with the overall star formation history in the universe. The progenitor stars of GW150914 may have formed in a binary  $\sim 2$ Gyr after the Big Bang (at $z\sim 3$) in a metal poor environment, according to \cite{Belci16}.  

A plethora of observations show that today the energy content of our expanding universe is dominated by dark energy (68.3\%), and by cold dark matter (CDM, 26.8\%), with baryons contributing only at  4.9\% level \cite{Planck13}, and that the present spectrum of primordial density fluctuations contains more power at lower masses. At the earliest epoch, the universe was dominated by small scale density fluctuations. Regions with higher density grow in time to the point where they decouple from the Hubble flow and collapse and virialize, forming self-gravitating halos. The first dark matter halos that form grow bigger through mergers with other halos and accretion of surrounding matter along cosmic filaments.   This is a bottom up process which leads to the hierarchical clustering of dark matter sub-structures and of the luminous components, the galaxies \cite{Loeb13-book}. 

 \begin{figure}[!t]
\begin{center}
\includegraphics[width=.9900\textwidth]{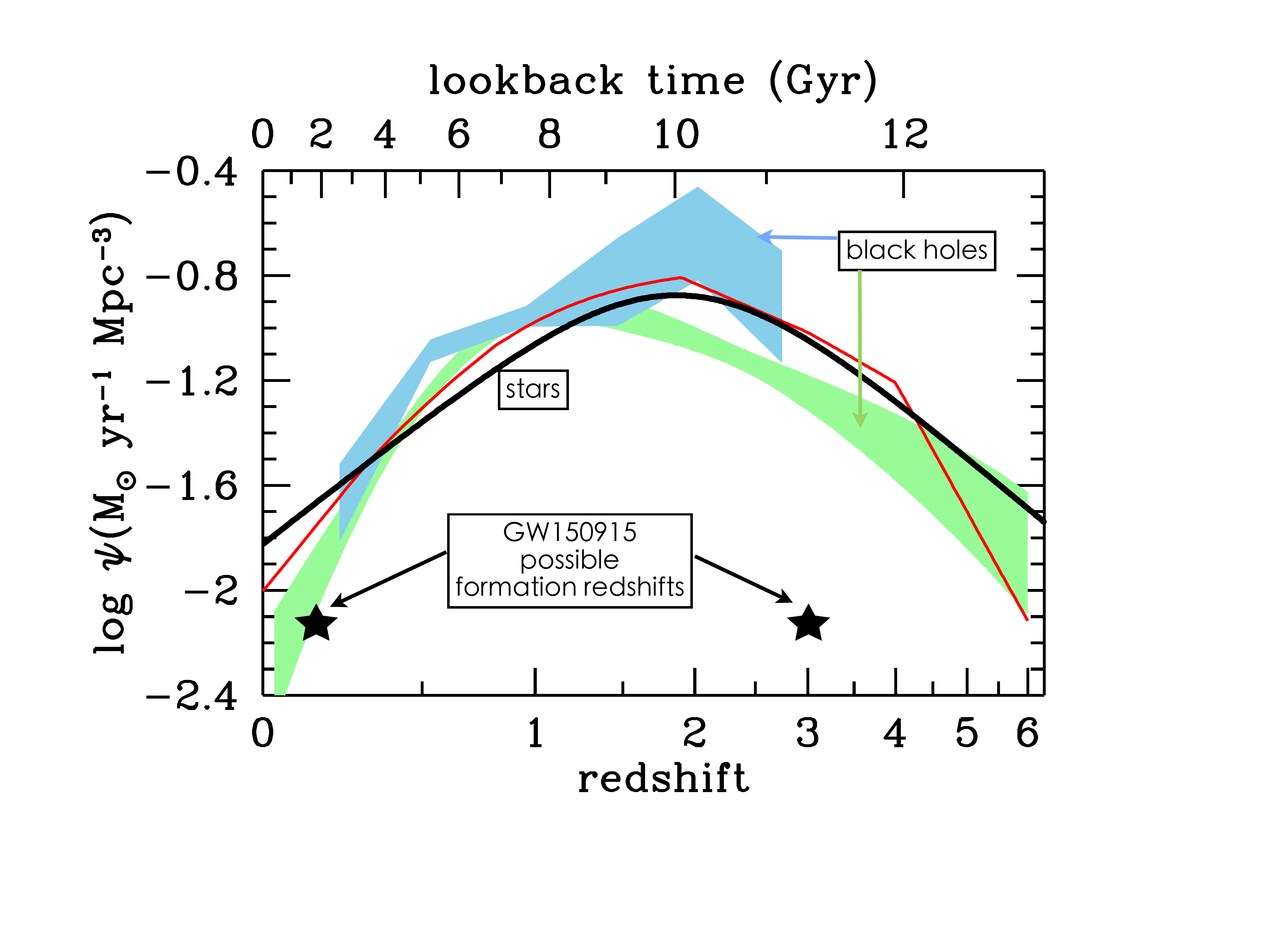}
\caption{Comparison of the best-fit star formation history (thick black solid curve) with the massive black
hole accretion history from different data (shaded green, light blu, and red line), adapted from \cite{Madau14-review}. The shading indicates the $\pm \sigma$
uncertainty range on the total bolometric luminosity density. The comoving rates of black hole accretion have been scaled up by a factor of 3,300 to facilitate
visual comparison to the star-formation history. The two black stars indicate the possible redshifts of formation of
the progenitor binary to the GW150914 event, as calculated in \cite{Belci16}. Courtesy of P. Madau.
}
\label{sfr-qso} 

\end{center} 
\end{figure}

At present, most of the investigations of galaxies and of QSOs in the electromagnetic universe 
feature the occurrence of three main epochs of evolution, along cosmic history \cite{Madau14-review}:\\
$\bullet$ The  {\underline {\sl cosmic dawn}} which is the epoch extending from cosmic redshift $z\sim 15$ when the universe was only a few $100$ Myr old to redshift  
$z\sim 6$,  corresponding to $\simless 1$ Gyr. During this epoch, baryons in dark matter halos of $\simgreat 10^6\msun$  begin to collapse and the first stars form
and first seed black holes. Planck data indicate that between 
$z=7.8$ and 8.8 the universe completed the phase of cosmic {\it re-ionisation} of gas turning intergalactic neutral hydrogen  into a 
 hot tenuous plasma \cite{Planck16}.  At the limits of current capabilities,  GRB 090423, the farthest  long GRB observed (which signals the formation of a stellar origin black hole), exploded at $z=9.4$, when the universe was 520 Myr old, and  the most distant galaxy MACS0647-JD 420  and the most distant QSO ULAS J1120+0641
 are found  at  $z=10.7$ and  $z=7.07$, 420 and 770 Myr after the big bang, respectively \cite{GWnote13}.  
 These  brightest sources are just probing the tip of an underlying distribution of fainter early objects for which little is known and which represent the building blocks of
 the largest structures.\\
 $\bullet$ The  {\underline  {\sl cosmic high noon}} follows, which is an epoch of critical transformations for galaxies, extending from $z\sim 6$ to $2$.
Around $z\sim 2$, the luminous QSOs and the cosmic-integrated star formation rate have their {\it peak}. This is illustrated in Figure \ref{sfr-qso} where
we show the cosmic-averaged star formation rate per unit comoving volume  (in units of $\msun\,\rm yr^{-1}\,Mpc^{-3}$) and the massive black hole accretion history
(in the same units but enhanced by a factor 3,300 to help the comparison) as  function of lookback time and redshift.
 Galaxies and seed black holes
are expected to grow fast in this epoch which erases information of their properties at birth. In between redshift $1$ and 2,  galaxies acquires about  50\% of their mass, and widespread star formation can lead to the build up of populations of (NS,NS), (NS,BH*) and (BH*,BH*) fated to coalesce over cosmic time, and accessible to forthcoming
and future observations.\\
$\bullet$ The last epoch of {\underline  {\sl cosmic fading} }
traces a phase where star formation in galaxies, and QSO's activity in galactic nuclei are both declining. It is a phase of slow evolution extending from $z\sim 1$ to the present. 
 Observations of galaxies and AGN give a description of a quieter universe where dormant supermassive black holes lurk at the centre of bright
 elliptical galaxies likely formed through galaxy mergers.  Less massive (dwarf)  galaxies in the near universe have undergone 
 a quieter merger and accretion history than their brighter analogues (which formed earlier).
They represent the closest analogue of lower mass high redshift dark matter halos from which galaxy assembly took off during cosmic dawn.
 Local, dwarf galaxies are  the
preferred site for the search of middleweight (or intermediate) black holes of $10^{3-6}\msun$ \citep{volonteri09}.    
NGC~4359, a close-by bulgeless disky dwarf, houses
in its centre a black hole of only ${3.6 \times
  10^{5}}\msun$.  This indicates that
nature provides a channel for the formation of middleweight  black holes  also in potential
wells much shallower than that of the massive spheroids, and these galaxies are expected to host 
a class of gravitational waves sources, known as Extreme Mass Ratio Inspirals (EMRIs) that we will discuss ahead in this Chapter.

A number of important questions can be posed in the context of galaxy formation and evolution that the gravitational universe will try to answer:\\
{\color{red} 
\begin{itemize}
\item When did the first black hole seeds form? Did they form only during cosmic dawn, i.e. over a limited interval of cosmic time?
\item How does the black hole  mass and spin distribution evolve with cosmic time?
\item To what extent mergers affect the cosmic evolution of massive black holes?
\end{itemize}
}

\subsection {Massive black hole binary mergers across cosmic ages}\label{MBH-ages}

During cosmic dawn and high noon, the bottom-up assembly of galactic halos through  galaxy mergers inevitably lead to the growth of an evolving population of
 binary black holes in a mass range between $10^{4-7}\msun$.  These are the target sources of the upcoming LISA-like observatory, in space.
When two galaxies with their dark matter halos merge, the time-varying gravitational field induced by the grand collision redistribute the 
orbital energy of stars  and gas discs in such a way that a new
 galaxy with new morphology forms. At the same time, the black holes nested at the centres of the interacting galaxies have a long journey to travel  before entering the phase of 
gravitational driven inspiral \cite{Colpi14}. They experience four critical phases covering more than 10 orders of magnitude in dynamic range: \\

 \begin{figure}[!t]
\begin{center}
\includegraphics[width=1.\textwidth]{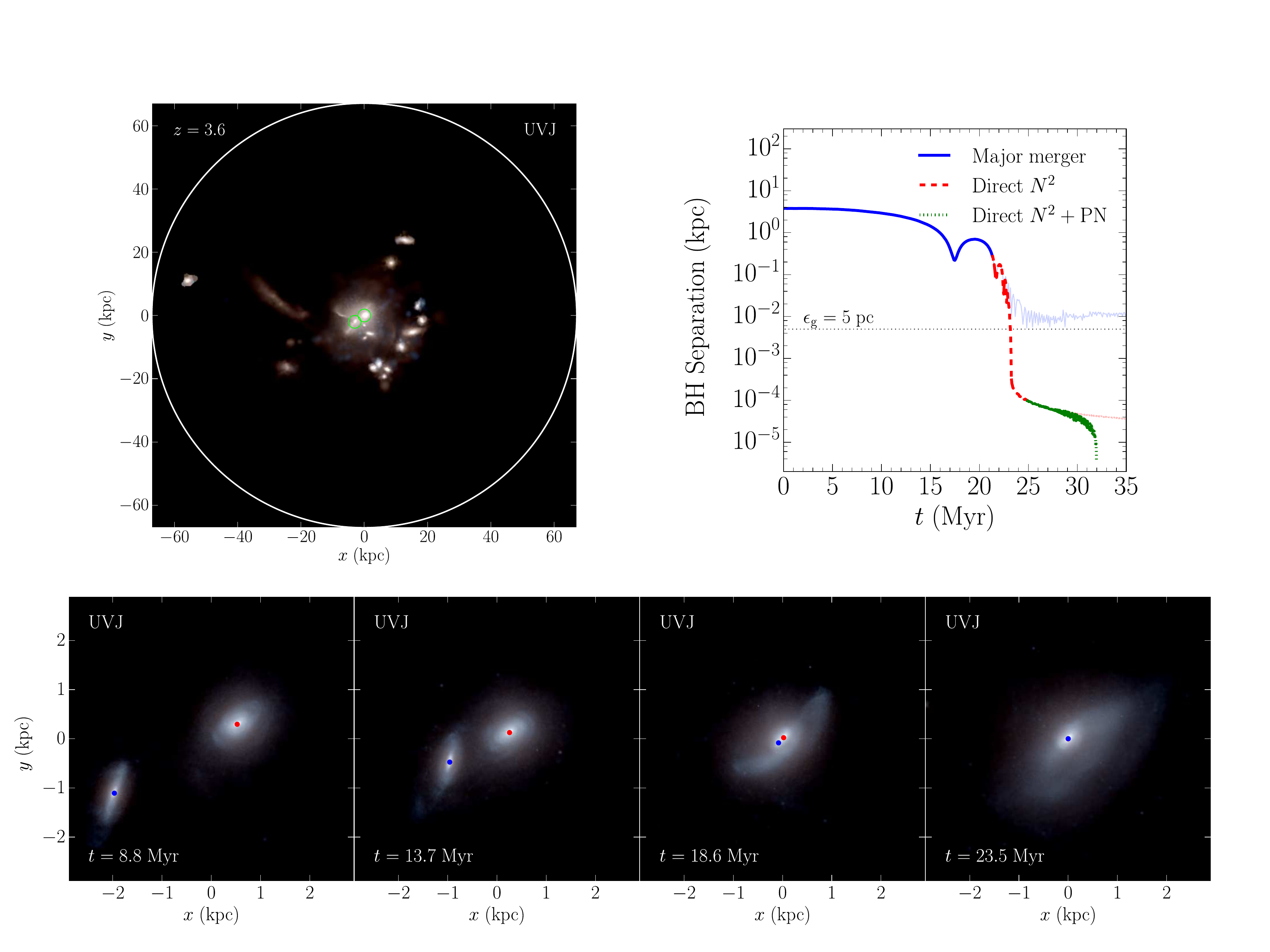}
\caption{Group environment of a galaxy merger simulated by \cite{Khan16}. The left  upper panel shows a mock UVJ map of the galaxy group at redshift $z=3.6$. The white
circle marks the virial radius of the group halo, while the green circles mark the galaxies in the verge of merging which host a massive black hole. The right upper panel shows the black hole separation versus time. Notice the huge dynamical range covered by this simulation, going from a few kpc  down to a separation of 5 pc (horizontal dotted line - pairing phase) corresponding to the force resolution of the large scale hydrodynamical simulation. 
Below 5pc the continued direct N-body simulation guarantees that repeated scatterings with stars drive the binary down to 0.1 pc (red dashed line - binary/hardening phase), and that Post Newtonian corrections 
in the dynamics (blu dotted line - gravitational driven phase )  drive the binary to coalescence within $\sim 30$ Myr after the formation of the oblate remnant, galaxy.  The lower panel shows, from left to right, the  time evolution of the galaxy merger depicted using mock UVJ photometric images of the merger, and the red and blue dots mark the position of the primary and secondary
black hole, respectively. Lengths are in physical coordinates.  We defer to \cite{Khan16} for details. Courtesy of F. Khan.}
 \label{BH-dynamics} 
\end{center} 
\end{figure} 

\noindent
(1) The {\it pairing phase}, when the black holes pair on galactic scales following the dynamics of the galaxies they inhabit
until they form a Keplerian binary (on pc scales) when the stellar/gas mass enclosed in their relative orbit is comparable to the sum of
the black hole masses. In this phase, the two galaxies first sink by dynamical friction against the dark matter background to form a new 
galaxy dragging the two black holes at the centre of the new system. Then, the black holes experience, as
individual massive particles, dynamical friction against the stars/gas and continue to spiral in and sink. \\ (2) {\it The  binary or hardening  phase}, when single stars
scattering off the black holes extract tiny amount of their orbital energy and angular momentum. If present in large numbers, the binary continues to contract.
 In gas-rich galaxies, torques from a circum-binary gaseous disc surrounding the binary can also  lead to hardening.\\
(3) The third phase of {\it gravitational wave driven inspiral} starts when the black holes get so close
(typically at around or below $\sim 10^{-3}$ pc)  
that they detach from their nearest environment, and gravitational waves dominate the loss of energy and angular momentum
driving the binary to coalescence. \\(4) Finally the new black hole that formed may experience  a {\it  recoiling phase} since 
gravitational waves carry away linear moment. Gravitational recoil velocities are between $\sim 300 \kms$ and $\sim 4000 \kms$ \cite{Lousto11}.
Thus the new black hole can either  oscillate
and sink back to the centre of the relic galaxy, or escape the galaxy.
Only state-of-the-art numerical simulations can describe this long journey that begins at 10 kpc scales and ends when the two black hole coalesce, typically on
scales of $10^{-6}$ pc. The {\it delay} between the galaxy merger and black hole merger varies from Myr to many Gyrs \cite{Colpi14}.
Figure \ref{BH-dynamics}  from \cite{Khan16} shows the three phases of a merger of two galaxies belonging to a group from a cosmological simulation (see the caption for details).

The merger of black holes in pristine dark matter halos is even more difficult to simulate as the dynamics is dominated by the gas and this requires use of  
self-consistent  high resolution hydrodynamical cosmological simulations with rich input physics  (chemistry network, cooling and radiative transport,  turbulence and magnetic field dissipation)  over a wide dynamical range. Preliminary studies indicate that the black hole dynamics is stochastic \cite{Fiacconi13}, implying a rather broad range of sinking timescales.

\begin{figure}[!t]
\begin{center}
\includegraphics[width=1.00\textwidth]{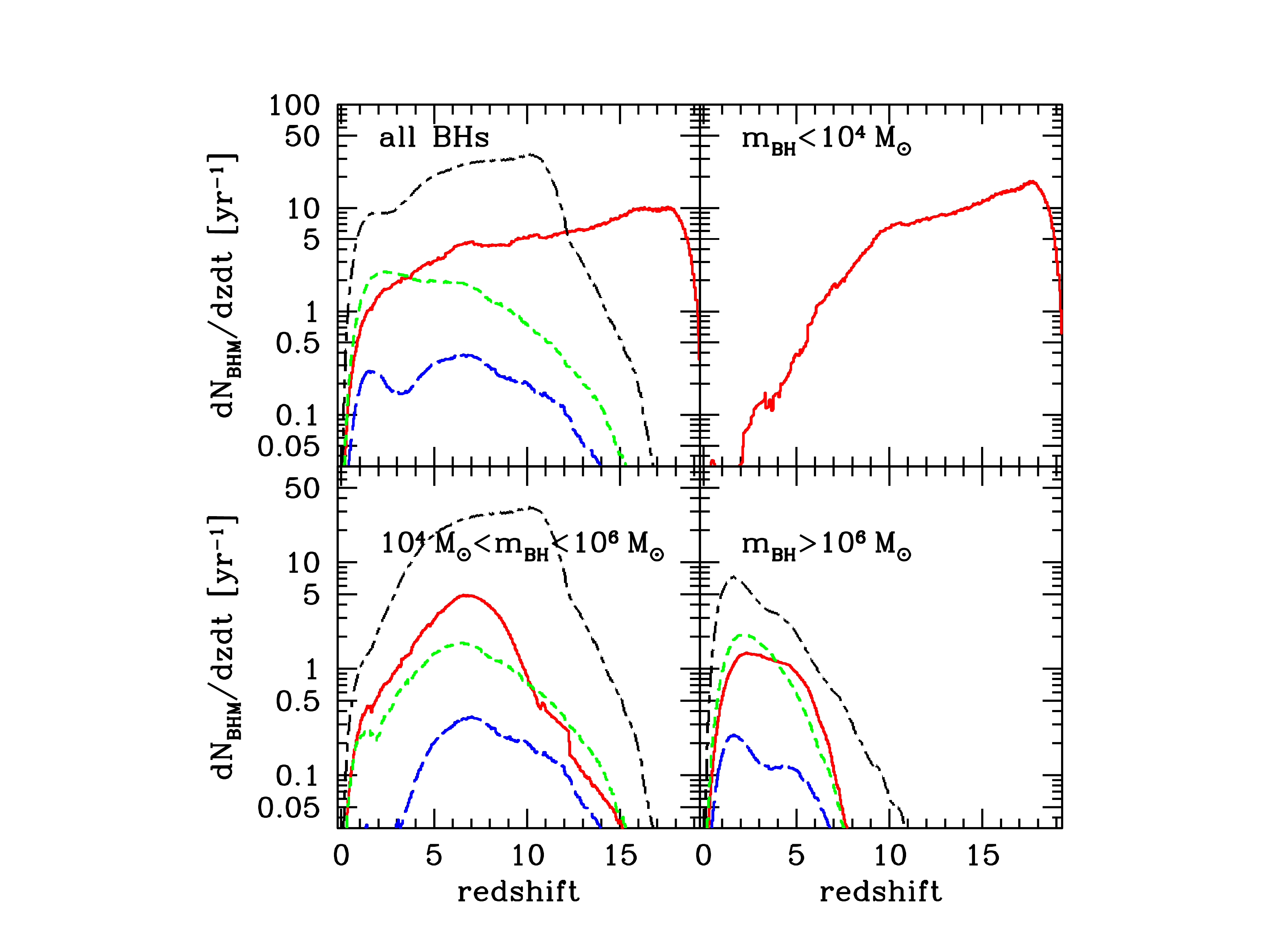}
\caption{Differential merger rates (per unit redshift and time expressed in yr) versus redshift for different black hole seed formation scenarios, from \cite{Sesana07}. 
Two main scenarios for black hole formation are considered: one where seeds are  remnant of Population III stars of $\sim 100\msun$, and one where seeds form with masses of $\sim 10^4\msun$. The black holes than continue to evolve due to accretion: see \cite{Sesana07} for details. 
Note how the lack of a physical understanding of the black hole seed formation mechanisms and their relation with the environment and cosmic epoch makes these rates highly uncertain.}
\label{rates-versus-z} 
\end{center} 
\end{figure}

\subsubsection{Reconstructing the cosmic evolution of massive black hole binary coalescences across the ages}

Given a mass distribution of black hole seeds, a cosmological model for the growth and assembly of dark matter halos, 
and an accretion recipe, one can infer the merger rates of massive black holes.
In Figure \ref{rates-versus-z} we show 
the merger rate per redshift bins of black holes as a function of $z$, for a variety of models of black hole seed formation, from Pop III stars to relic of supermassive stars collapsing as DCBH, computed using a Monte Carlo merger tree synthesis model within the extended Press and Schecter formalism for the assembly of galaxy halos \cite{Sesana07}.
The uncertainties are large with merger rate excursions of about two orders of magnitude, ranging from ten to several hundreds events per year.  
Each halo had experienced few to few hundred mergers in its past life, placing mergers among the critical key mechanisms driving galaxy evolution.

\section{The  sources of the gravitational wave universe}\label{sources}

Here we list the prospected sources of the gravitational wave universe, based on two criteria: the distinction between {\it high} and 
{\it low} frequency  sources, and between {\it short duration, transient} and  {\it continuous} sources. The concept of {\it backgrounds} is shortly introduced.



\subsection{The high frequency gravitational universe}\label{high-frequency-universe}

The sources of the {\it high frequency universe},  observed with  ground-based detectors 
 at frequencies between  $\sim10$ Hz  and 
to a few 1000 Hz ($\sim 1$ and $\sim 10^4$ Hz with the  Einstein Telescope under design), can  be grouped into four basic classes: 
compact binary coalescences, un-modelled bursts, continuous waves, and stochastic backgrounds\cite{Andersson13,Sathya09,SathyaET12}. These groups refer to different astrophysical settings and different  algorithms for their detection.

\begin{figure}[!t]
\begin{center}
\includegraphics[width=1.00\textwidth]{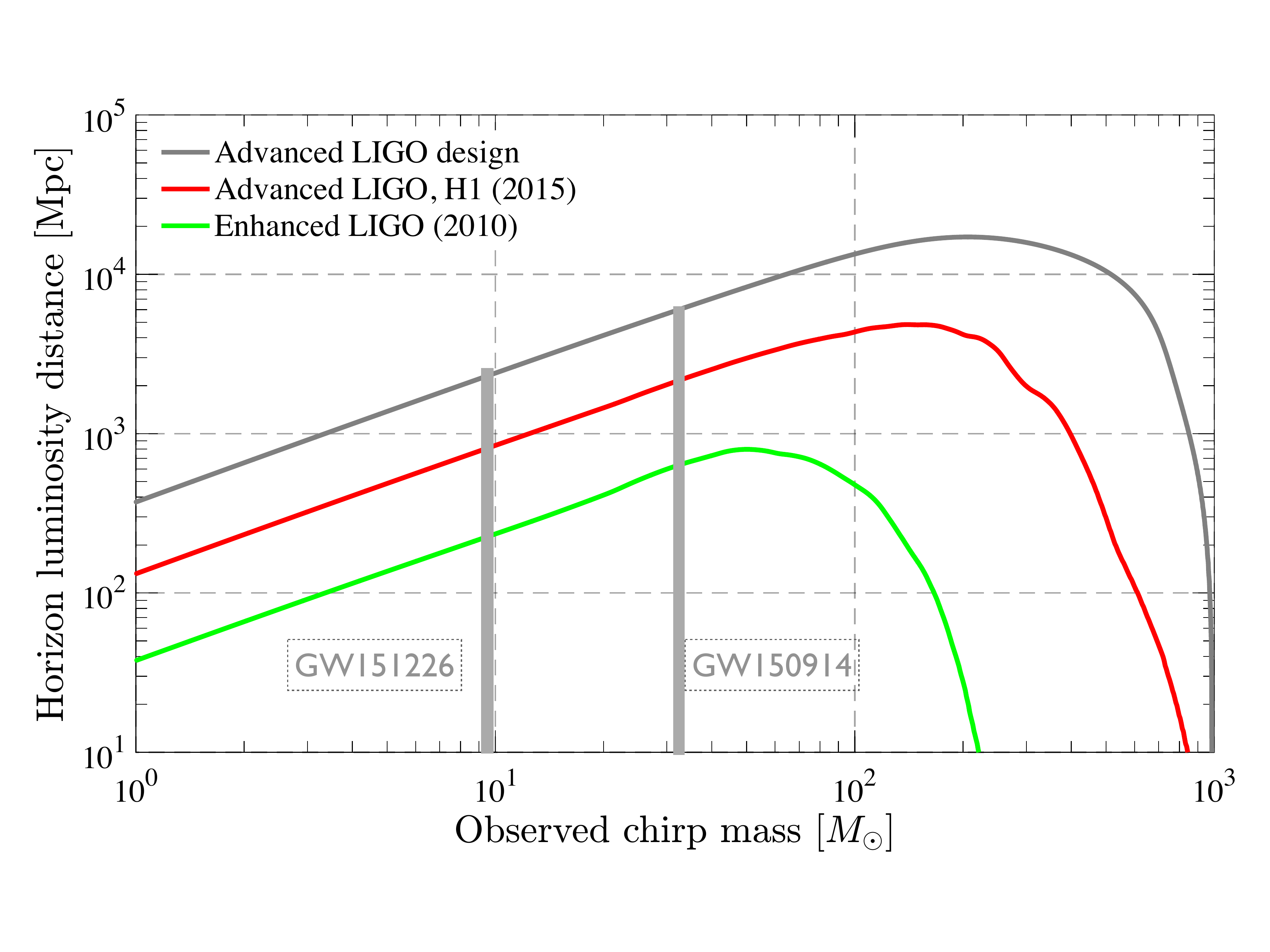}
\caption{Horizon Luminosity distance reach of CBCs as a function of the detector-frame chirp mass 
(we  defer to $\S 6$ of this Chapter for the definition of the chirp mass for binaries at cosmological distances). 
The Figure is adapted from \cite{Martynov16}.
The distance is computed for equal-mass binaries including the inspiral and merger signal. (NS,NS) binaries have observed chirp masses ${\cal M}_{\rm c}\simeq 1$,
while (BH*,BH*) binaries have ${\cal M}_{\rm c}$ which exceed unity. Light-grey lines refer to GW150914 and GW151226 which have detector-frame chirp masses of $\sim 30.2 \msun,$ and $ 9.7\msun$, respectively.
The range of chirp masses includes also the possibility of detecting intermediate mass black holes in binaries, if they exist with ${\cal M}_{\rm c}>100$.
The curves refer to three different stages of the 
Advanced LIGO design as indicated in the figure. A source at redshift $z=0.01$ (0.1, 1) has luminosity distance of 43 Mpc  (463 Mpc, 2.8 Gpc) according to the present cosmological model.
Courtesy of the LIGO Scientific Collaboration.}
\label{LIGO-reach-distance} 
\end{center} 
\end{figure}

\smallskip
\noindent
$\bullet$ {\underline {\bf Compact Binary Coalescences}} -  CBCs refer to binaries hosting the relics of massive stars and comprise (NS,NS),(NS,BH*) and (BH*,BH*) binaries.  CBCs are {\it transient}  sources. They are detectable  at the time of their coalescence, as they emit a sizeable fraction of their reduced-mass-energy, and have a modelled signal. 
(NS,NS) binaries are characterised by  
mass ratios $q\equiv m_2/m_1$ (with $m_2<m_1$) close to $\sim 1$, as observed in double neutron star binary systems, and a lower limit is $q\sim 1.4 \msun/M_{\rm max}^{\rm NS}\sim 0.46$.  The mass ratio of (NS,BH*) and  (BH*,BH*) binaries is less constrained, since we do not know the maximum mass of stellar black holes nor how they pair in binaries.  GW150914 and  GW151226 have mass ratio   $\sim 0.82$  and $\sim 0.53$, respectively \cite{Abbott-1,Abbott-GW151226}. 
GW150914  and GW151226 are expected to be the first two of a rich population of CBCs of different flavours that will be observed in the forthcoming  Advanced LIGO
and Virgo science runs. In $\S 6$ and $\S 7$ we describe in depth CBCs and their expected signal. 

\smallskip
\noindent
{\sl Horizon luminosity distance} - A key fact that makes binaries important sources is that the amplitude of the emission is calibrated just by a combination of the two masses
(to leading orders).  Given this,  their detectability can be expressed  in terms of the horizon luminosity distance $d_{\rm horizon}$ for a detector, defined as the distance  at which a detector measures a signal-to-noise ratio (SNR)  of 8 for an optimally oriented (face-on) and optimally located binary. Figure \ref{LIGO-reach-distance} indicates this distance reach for Advanced LIGO in  three  of its design
configurations. At present the distance reach for (NS,NS) binaries with Advanced LIGO and Virgo is $\sim 90$ Mpc, and $\sim 400$ Mpc for (BH*,BH*) binaries.
At design sensitivity Advanced LIGO can detect neutron star binaries out to a distance of $\sim 400$ Mpc scales, and black hole binaries as GW150914 out to $\simless 10$ Gpc, as shown in the figure. Black holes of intermediate mass can also be detectable, if they form in binaries over this mass range. We remark that all these binaries would not be detectable otherwise (neutron star binaries are observed as pulsars only in our Milky Way).
The third generation of Earth based  detectors as ET will be able to detect (NS,NS) out to redshift $z\sim 2$ corresponding to a distance of $1.6$ Gpc, and (BH*,BH*) as GW150914 out to a redshift $z\sim 5$ (47 Gpc) allowing ET to explore binary populations at cosmological distances, at the end of the cosmic dawn and
during high noon.

\begin{figure}[!t]
\begin{center}
\includegraphics[width=.99\textwidth]{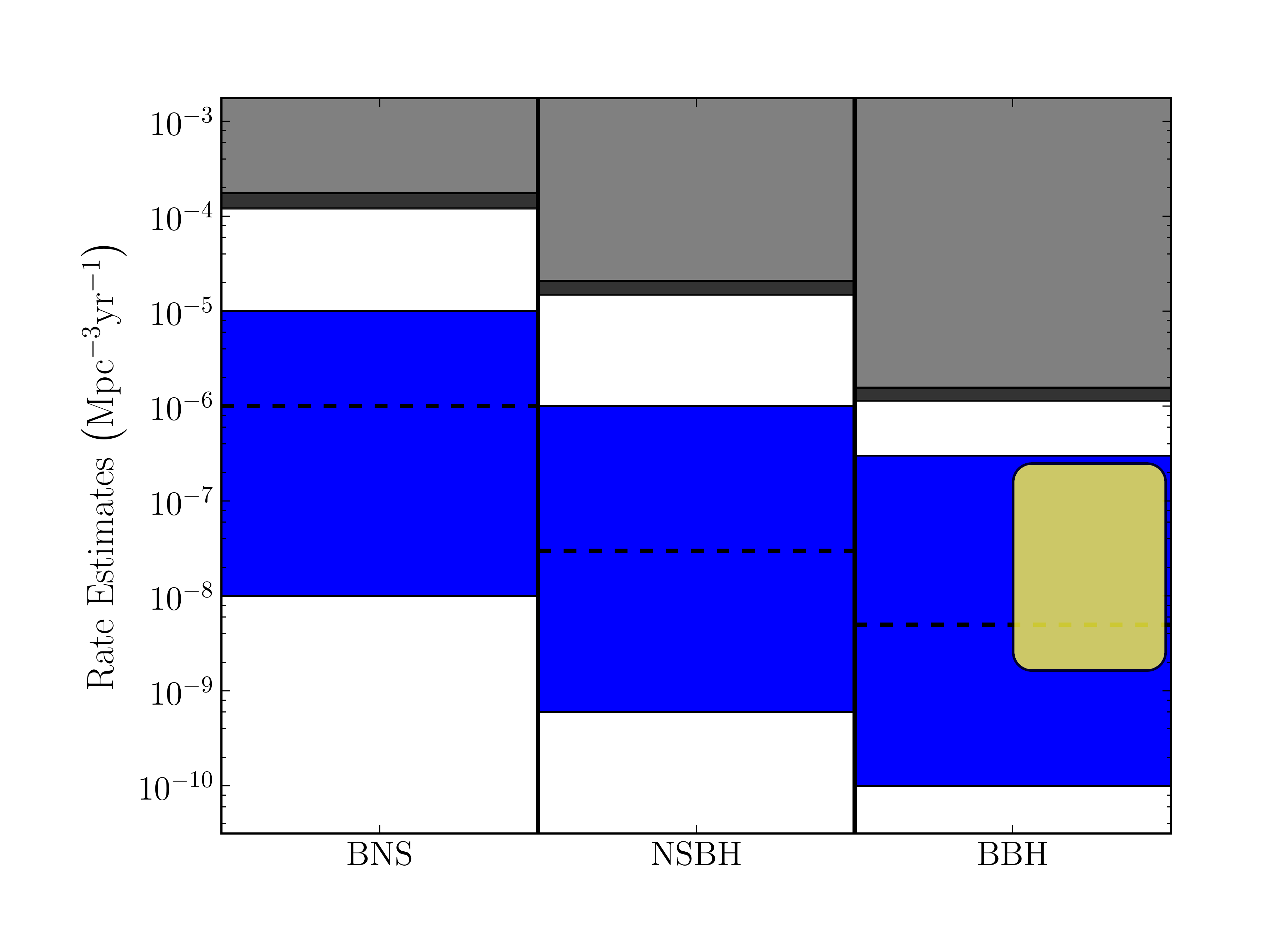}
\caption{Comparison of  merger rates  of 
(NS,NS), (NS,BH*) and (BH*,BH*) binaries adapted  from \cite{Abadie12}. The blue regions show the spread in the predicted rates with the dashed-black lines showing realistic estimates. A black hole mass of $10\msun$ was used for estimating the rates.   The  grey regions display the upper limits on the rates obtained in the S5-VSR1 analysis. Following the discovery of GW150914 and GW151226, conservative estimates now fall in the dark-yellow square (see \cite{Abbott-4-rates} and arXiv:1606.04856v1 for further details).
}
 \label{CBC-rates} 
\end{center} 
\end{figure}

\smallskip
\noindent
{\sl Expected coalescence rates} - Prior to the discovery of GW150914, the rate of CBCs relied entirely on theoretical population synthesis models
and dynamical models,  and for (NS,NS) binaries on constraints derived from electromagnetic observations \cite{Abadie10}.  The rates with their large uncertainties
are in Figure \ref{CBC-rates}.
 With the discovery of GW150914, the rates for (BH*,BH*) binaries now fall in the conservative range of 
  $9\times 10^{-9}$ and $250\times 10^{-9}$ Mpc$^{-3}$ yr$^{-1}$ (we defer to  \cite{Abbott-4-rates} and arXiv:1606.04856v1).

\smallskip
\noindent
$\bullet$  {\underline {\bf Unmodelled Bursts} refer to short-duration events caused by a sudden change of state in the source that do not have 
 a near-universal waveforms. Here we outline their key features.

{\it Core collapse supernovae} (CCSNae) and {\it hot remnants}  belong to this class \cite{Ott09-review,Fryer11-review}. 
In CCSNae, among the most powerful explosions in the electromagnetic universe, the available energy reservoir of $3\times 10^{53}$ erg is set by the difference 
in gravitational binding energy between the pre-collapse iron core and the collapsed
neutron star remnant. Much of this energy is initially
stored as heat in the proto-neutron star and most of it  ($99\%$) is released in the form of neutrinos, $1\%$ in kinetic energy of the explosion, $0.01\%$ is emitted in radiation across the electromagnetic spectrum, and an 
uncertain fraction is expected to be emitted in gravitational waves.
Electromagnetic observations of CCSNae yield secondary observables, such as progenitor type and mass, explosion morphology and energy, 
and ejecta composition. 
By contrast, gravitational waves, much like neutrinos, are emitted from the innermost
region (the core) of the CCSN and thus  convey primary, direct  live
information  on the dynamics of the core collapse and bounce.   They potentially inform us not only on
the general degree of asymmetry in the dynamics of the
CCSN, but also more directly on the explosion mechanism, the structural and compositional evolution
of the proto-neutron star, the rotation rate of the collapsed core, and the state of nuclear matter.

The violent dynamics in CCSNae and (possibly) in long GRBs (resulting from the collapse of rapidly rotating
low-metallicity massive stars, dubbed as collapsars) is expected to give rise, if aspherical,  to low amplitude bursts of 
gravitational waves with typical durations from milliseconds to seconds, over a wide frequency range, between 50-1000 Hz. 
The bursts  have no universal features as gravitational wave emission is influenced by the stochastic dynamics driven by the richness
in the input physics that accompany the infall of matter and its bounce. Many multi-dimensional processes may
emit gravitational waves during core collapse and the subsequent post-bounce
CCSN evolution. 
 In $\S 7$ we select a few mechanisms that lead to the emission of gravitational waves, and show the shape of the signal.

The proto-neutron star that forms at the end of a CCSN is a hot and rapidly evolving object.
After the first tenths of seconds of the remnantÕs life, the lepton pressure in the interior
decreases due to extensive neutrino losses, and the radius reduces to
about 20-30 km. The subsequent evolution is
Òquasi-stationaryÓ, and can be described by a sequence of equilibrium configurations.
 In these states the hot star can display a rich spectrum of non-radial normal modes, which can excite emission
 of gravitational waves in narrow intervals around the characteristic frequency of the mode, and extending 
 over times comparable to the damping timescale of the excited oscillation mode \cite{Andersson03-review}.
CCSNae should be  visible throughout the Milky Way with enhanced interferometric detector technology, while third generation observatories 
may be needed to explore events at a few  Mpc, out to which the integrated CCSNae rate is $\sim 0.5-1$  yr$^{-1}$. Detecting gravitational waves in coincidence
with optical, X-ray, $\gamma$-ray radiation or neutrinos could give unprecedented insight into stellar collapse.

 {\it Pulsar glitches} are also expected to fall in this category.  Glitches are enigmatic spin-up events seen in (mainly)
 relatively young neutron stars like Crab and Vela. 
 A glitch is a sudden increase (up to 1 part in $10^6$) in the rotational frequency of a pulsar. Following a glitch is a period of gradual recovery
 to a spin close to that observed before the glitch, due to braking provided by the emission of high energy particles and electromagnetic radiation. These gradual recovery periods have been observed to last from days to years. Currently, only multiple glitches of the Crab and Vela pulsars have been observed and studied extensively.  The energy of these events is $10^{42}$ erg, i.e. 
 of  the order of $10^{-12} \msun c^2$,
which set a benchmark energy level for the emission in gravitational waves by pulsars. 
These events are likely to be within reach of ET, but still too weak for Advanced LIGO and Virgo, and are observable only in the Galaxy.

 {\it Magnetar flares} could be important sources of gravitational waves. Magnetars are associated to
 the high energy phenomena known as Soft Gamma Repeaters and Anomalous X-ray Pulsars. These sources host slowly spinning, isolated neutron stars 
 endowed by ultra strong magnetic fields and whose emission is powered by the release of magnetic energy.  On December  2004 a giant flare has been observed in  SGR 1806-20 which
 released  $\sim 5\times 10^{46}$ erg in high energy radiation, implying an internal magnetic field strength of $10^{16}$ Gauss. 
 To explain this powerful emission, models require a substantial deformation of the neutron star in a direction non coincident with its spin axis. 
 The newborn fast spinning magnetar may radiate for a few weeks  gravitational waves at frequencies around a kHz, and may constitute a promising new class of gravitational wave emitters, visible once per year from galaxies in the Virgo cluster, out to a distance of 16 Mpc \cite{Stella05-magnetar}.

{\it Asteroseismiology of neutron stars}  is (at least in principle) a promising 
avenue for studying neutron star interiors \cite{Andersson03-review,Andersson-Ferrari11}.
 Neutron stars have a rich oscillation spectrum associated to non-radial normal modes with  frequencies in the kHz regime.  They can be excited 
 in different evolutionary phase, e.g. in rapidly and differentially rotating hot proto-neutron stars, or in  old neutron stars recycled in binaries whose accreting layers are sites of repeated nuclear explosions that produce X-ray flares.  In this last case,
 the rapid rise times of these instabilities may excite acoustic vibrations. If the rise time matches the period of a mode, than a substantial fraction of the energy released can 
 be channeled into mechanical vibrations and a large fraction of this energy could be carried away by gravitational waves, when other mode-damping mechanisms (e.g. viscosity) are less efficient.  
     
\smallskip
\noindent
$\bullet$ \underline {\bf Continuous wave source} are generally persistent sources which produce signals of  roughly constant amplitude and frequency, i.e.  
varying relatively slowly over the observation time. A number of mechanisms may cause the neutron star to emit a continuous signal.
These include deformations generated  either by strains in the star's crust or by  intense magnetic fields, precession, and long-lived oscillation modes of the fluid interior.
Target sources for this type of emission are the rotation powered neutron stars in the Milky Way \cite{Abbott10-pulsars}. 
 
  More than 2000 radio pulsars have been detected for which the sky location and frequency evolution have been accurately measured.  
 Among them, several tens  have spin frequencies greater than 20 Hz so that they are in the Advanced LIGO and Virgo bandwidth reach. 
 In the search of the gravitational wave signal, pulsars are assumed to be triaxial stars emitting gravitational waves at
precisely twice their observed spin frequencies (i.e. the emission mechanism is an $ l = m = 2$
quadrupole), with the wave phase-locked with the electromagnetic signal.  
No signal has been reported so far from targeted pulsars. 
  This null result can therefore be interpreted  as upper limit on the strength of the gravitational wave emission, and thus as upper limit on the level
  of asymmetry seeded in the star's equilibrium structure.
  Theoretical modelling of bumpy neutron stars has mainly focused on
establishing what the largest possible neutron star mountain would be \cite{Andersson-Ferrari11}.  Expressing
this in terms of a (quadrupole) ellipticity, detailed modelling of crustal strains suggest $\epsilon_{\rm crust}<2\times 10^{-5}(U_{\rm break}/0.1)$
  where $U_{\rm break}$ is the crustal breaking strain.
  State-of-the-art calculations indicate that solid
phases may also be present at high densities, allowing the construction of stars with
larger deformations. The magnetic field also tends to deform the star. For typical pulsar field strengths
the deformation is $\epsilon_B\sim 10^{-12} (B/10^{12}\, {\rm G})^2$, but it can be larger by a factor $\sim 10^3$ if the core is super-conducting 
 with critical field strength of $10^{15}$ G \cite{Andersson-Ferrari11}.
  
 An observational milestone was reached recently, when the LIGO and Virgo data  analysis from the
first nine months of the S5 science run was carried on to beat the Crab pulsar spin-down limit.
It was found
that no more than $2\% $ of the spin-down energy was being emitted in the gravitational wave channel. 
This  limit already indicates that  Crab is not a maximally strained quark star
for which larger ellipticities  are allowed \cite{Abbott08-crab,Abbott10-pulsars,Andersson-Ferrari11}.

\medskip
\noindent
$\bullet$ {\underline {\bf Stochastic backgrounds}} can arise from a large population of weak sources, so that there are many comparable strength signals with overlapping frequencies in each resolvable frequency bin. In the high frequency regime, the likely sources would be a population of inspiralling binaries at much greater distances than the resolvable CBCs. 
With the LIGO detection of  GW150914 there might exist a population of "heavy" binary black holes with mass above $\sim 30\msun$ contributing
to the stochastic gravitational wave background  at a level  higher than previously expected from CBCs \cite{Abbott-7-background}.

A {\it  stochastic background} can also arise from the {\it primordial gravitational waves} produced at the inflationary epoch. The standard cosmological model places this background at even lower levels than the expected foreground from unresolved binaries,  but alternative models can produce strong cosmological backgrounds in different frequency bands. Consequently, non-detections can place meaningful constraints on alternative cosmological models.

\subsection{The low frequency gravitational universe}
 \begin{figure}[!t]
\begin{center}
\includegraphics[width=.99\textwidth]{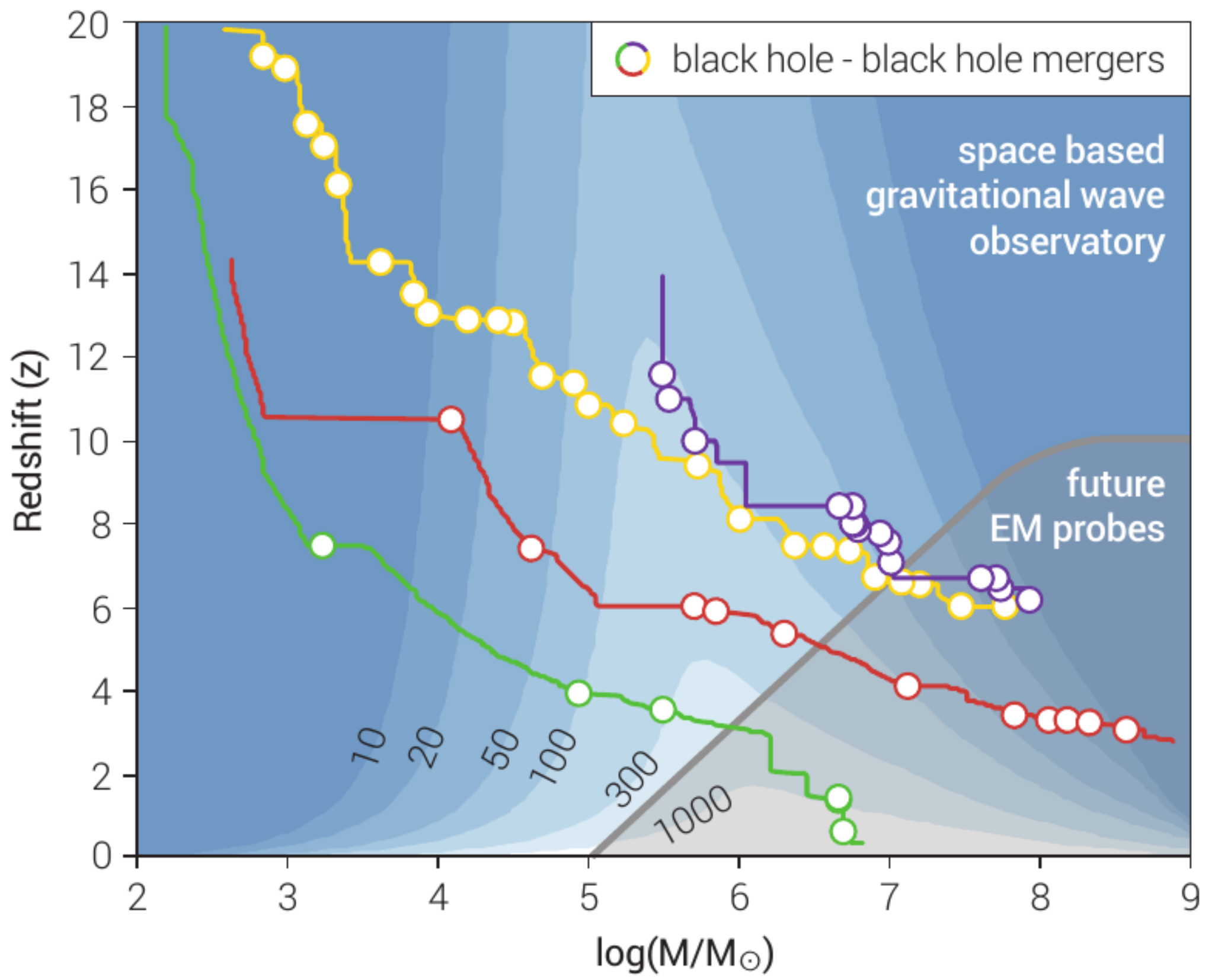}
\caption{Contour plot of constant sky-polarization averaged signal-to-noise ratio (SNR) of equal mass, non spinning massive black hole binary coalescences in the eLISA detector, plotted in the total rest-frame mass ($x$-axis) and redshift ($y$-axis) plane, from  \cite{whitepaper13}. Tracks represent selected evolutionary path of different MBHCs, obtained using semi-analytical population synthesis models. Circles mark MBHCs occurring along the way, whereas the bottom-right grey area identifies the portion of parameter space accessible to future electromagnetic observations of active supermassive black holes. The figure shows how complementary is the
parameter space explored by eLISA. In the overlap region there might also be the possibility of detecting precursors and or electromagnetic counterparts  of MBHCs.}
\label{fig_MBHB_Mzplot} 
\end{center} 

\end{figure}
The milli-Hz frequency range (between 0.1 mHz and 0.1Hz) will be probed by space based interferometers such as eLISA \cite{whitepaper13,GWnote13}\footnote{eLISA or 'evolving' LISA is, at the time of writing, a descoped version of the original LISA mission, to be flown by ESA in the early 2030s. For the purpose of our general discussion, there is no need to differenciate between the details of the two concepts. Detailed eLISA desings under consideration can be found in \cite{Klein16}.} and is usually referred to as the {\it low frequency universe}. This is expected to be by far the richest window in terms of number, loudness, distance reach and diversity of sources, including massive black hole coalescences (MBHCs), extreme mass ratio inspirals (EMRIs), galactic and extragalactic binaries of stellar mass compact objects, and more.

\smallskip
\noindent
$\bullet$ {\underline {\bf Massive black hole coalescences}} (MBHCs) are binaries resulting from the collision and merger of galaxies, and are detected
at the time of their coalescence.  
Figure \ref{fig_MBHB_Mzplot} shows that eLISA will observe signals coming from MBHCs in the mass range between  $10^4\msun$ and $10^7\msun$, with typical binary mass ratios  $0.1<q< 1$, out to redshift $z \sim 20$ (if they already exist) corresponding to a luminosity distance of $\sim 230$ Gpc and an age of the universe of 180 Myr. Overlaid to constant signal-to-noise ratio contours are mass-redshift evolutionary pathways ending with the formation of a supermassive black hole representing (i) an analogue of SgrA$^*$ (the black hole at our Galactic centre); (ii) a typical quasar at $z\approx2$; and (iii) two distant quasars at $z\approx 6$. White dots mark merger events and highlight the fact that any massive black hole we observe in bright  galaxies today has grown cutting through the eLISA sensitivity band. The forthcoming LISA-like observatory will therefore provide an highly complete census of MBHCs throughout the universe.

\smallskip
\noindent
{\sl Expected rates} - The expected rate of MBHCs is weakly constrained as it depends 
on the occupation fraction of (seed) black holes in haloes as a function of redshift, on their mass distribution (as depicted in Figure  \ref{rates-versus-z}), on their accretion 
history, and on the pairing and hardening efficiency inside the new galaxy that has formed \cite{Sesana07,SesanaGair11,Colpi14}. Cosmological
simulations of the galaxy assembly anchored to estimates of the local galaxy-merger-rate predict a few to few-hundred coalescences per year \cite{GWnote13}. 
Mergers are inevitable in a hierarchical universe, and whatever is the route to the massive black hole build-up, eLISA will provide a unique window to test MBHCs. 
MBHCs pinpoint places where galaxy mergers occur and in the eLISA band-width they inform us on the evolution of massive black holes in the low mass end
of their distribution extending down to the desert zone.

\smallskip
\noindent
{\sl Physics and astrophysics with precision gravitational wave measurements} -  In virtue of the extremely high signal-to-noise ratio  of most of the events, MBHC parameters will be extracted with exquisite precision \cite{GWnote13}. Individual redshifted masses can be measured with an error of $0.1\%-1\%$, on both components. Even more interestingly, the spins of two massive black holes  can be determined to an absolute uncertainty down to 0.01 in the best cases. This is a critical measurement, because the efficiency of accretion and mass growth of MBHs strongly depends on their spins  which are currently difficult  to determine through electromagnetic observations \cite{BertiVolonteri08} (see $\S 4.3$). 

The distinctive high signal-to-noise ratio of MBHCs will allow  black hole "spectroscopy" i.e., the direct measure of several frequencies and damping times associated to  the quasi-normal modes present in the ringdown signal of the  newborn massive black hole \cite{Berti15-testGR,BertiSesana16}. This will make it possible to carry on direct precision tests of the no-hair theorem.  Violations of general relativity predictions may indicate new physics or  the presence of exotic dark objects such as, e.g.  boson stars that carry a surface \cite{Colpi86}. The comparison between spectroscopy measurements 
from the LIGO-Virgo data  on (BH*,BH*) coalescences and those from LISA \cite{BertiSesana16} on MBHCs mapping the heaviest holes will be of enormous value: the proof 
of  the universality of black holes over a mass range of more than six orders of magnitude.   
  
 \begin{figure}[!t]
\begin{center}
\includegraphics[width=.99\textwidth]{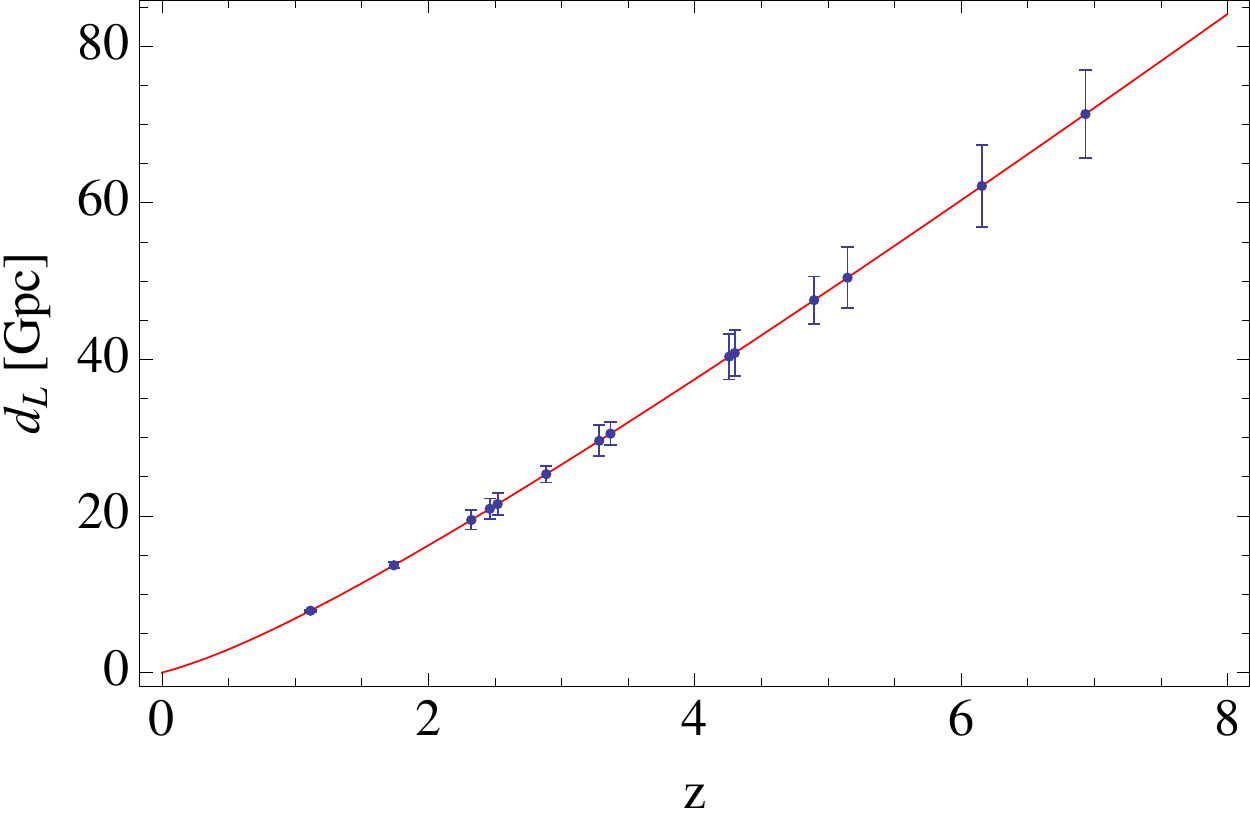}
\caption{ Simulated data points of "standard sirens" MBHCs in the luminosity distance versus redshift diagram \cite{Tamanini16}. 
Simultaneous electromagnetic counterparts  to the MBHCs are assumed, which provide an independent measure of the redshift of the source.  The data are from a simulated catalog of MBHCs extracted from a model of heavy seeds, and for a selected configuration of a LISA-like observatory.   Note that a LISA-like mission will be able to map the expansion of the universe out to $z>2$, which cannot be tested by Supernovae Type Ia or other current probes. Few standard sirens will be available at lower redshift, as MBHCs are rare events. See for details \cite{Tamanini16}. Courtesy of N. Tamanini.} 
\label{cosmography}
\end{center} 
\end{figure}

\smallskip
 \noindent
{\sl  Constraining the massive black hole cosmic history} -  While individual MBHC measurements will allow exquisite tests of general relativity and will probe several distinctive feature of massive black hole physics, information on the astrophysical evolution 
 is encoded in the statistical properties of the observed population. As first illustrated in \cite{SesanaGair11}, observations of multiple MBHCs can be combined together to learn about their formation and cosmic evolution. In particular the mass distribution of the ensemble of observed events encodes precious information about the nature of the first seeds, whereas the spin distribution will constrain the primary mode of accretion that grows them to become supermassive
\cite{BertiVolonteri08}.

\smallskip
\noindent
{\sl Cosmography} -
Another peculiar property of MBHCs is that their luminosity distance can be directly measured as it is encoded in the gravitational wave signal, and
its estimate does not involve cross-calibrations of successive distance indicators at different scales
(as the distance ladder in the electromagnetic universe) since the gravitational wave luminosity of MBHCs is determined by gravitational physics, only. 
Thus MBHCs are standard sirens (we defer to $\S 6.1$ for an exact definition). 
A LISA
like interferometer can provide the distance to the source to a stunning few percent accuracy. 
If an electromagnetic counterpart to the
MBHC event can be observed \cite{Schnittman12,Bogdanovic14}, it will make it possible to reconstruct  the luminosity distance versus redshift relation, as shown in Figure \ref{cosmography} offering the possibility of measuring the Hubble parameter at the level of $1-2\%$, and of inferring bounds on the dark matter and dark energy content of the universe \cite{Tamanini16}.

\smallskip
\noindent
$\bullet$ {\underline {\bf Extreme mass ratio inspirals} } (EMRIs) describe the inspiral and possibly the plunge of stellar mass compact objects
 into a massive black hole at the centre of a galaxy \cite{Pau07-EMRIs,GWnote13}.   EMRIs still fall in the class of "binaries" despite their small mass ratio $q\ll 10^{-3}$.

\smallskip
\noindent
 {\sl EMRI flavours and expected rate} -  Massive black holes in galactic nuclei are surrounded by a swarm of stars and compact objects. The densities can be as high as $10^{^8}$ stars pc$^{-3}$.  In such extreme environments, stars are easily deflected on very low angular momentum orbits,
 owing to repeated, distant stellar encounters and thus can enter the massive black hole sphere of influence.  The fate of main sequence stars on such
 "plunging"  orbits is to be tidally disrupted \cite{Rees88-tidal}. But, compact objects such as neutron stars, stellar  black holes and white dwarfs (for
 central black holes of  $M_{\rm BH}<10^5\msun$) can be captured in extremely eccentric orbits, with periastron of $\approx 10 R_{\rm G}$, avoiding
 disruption.  Their orbit will then slowly circularise because of gravitational wave emission and the slow inspiral can in principle lead to observable EMRI signals. 
 In general, stellar black holes are expected to dominate the observed rate for a LISA-like detector. This is because dynamical mass segregation tends to concentrate the heavier compact stars nearer the massive black hole \cite{Alexander06,Merritt11,Merritt13}, and because black hole EMRIs  have higher signal-to-noise ratio, and so can be seen out to a much larger distance, typically of  few Gpc ($z\simeq 1$). Their expected rate is uncertain due to the currently poor knowledge  on the low mass end of the massive black hole mass function in galaxies (at $M_{\rm BH}<10^6\msun$) and to the large uncertainties on the properties of typical compact object distributions in galactic nuclei. In general, a Milky Way type massive black hole is expected to form an EMRI every 10 Myr, implying a detection rate for a LISA-like mission in the figure of hundreds per year \cite{Pau-review12,GWnote13,Pau13-spin-EMRIs}. There is, however, at least a factor of 100 uncertainty on this number.

 \begin{figure}[!t]
\begin{center}
\includegraphics[width=.99\textwidth]{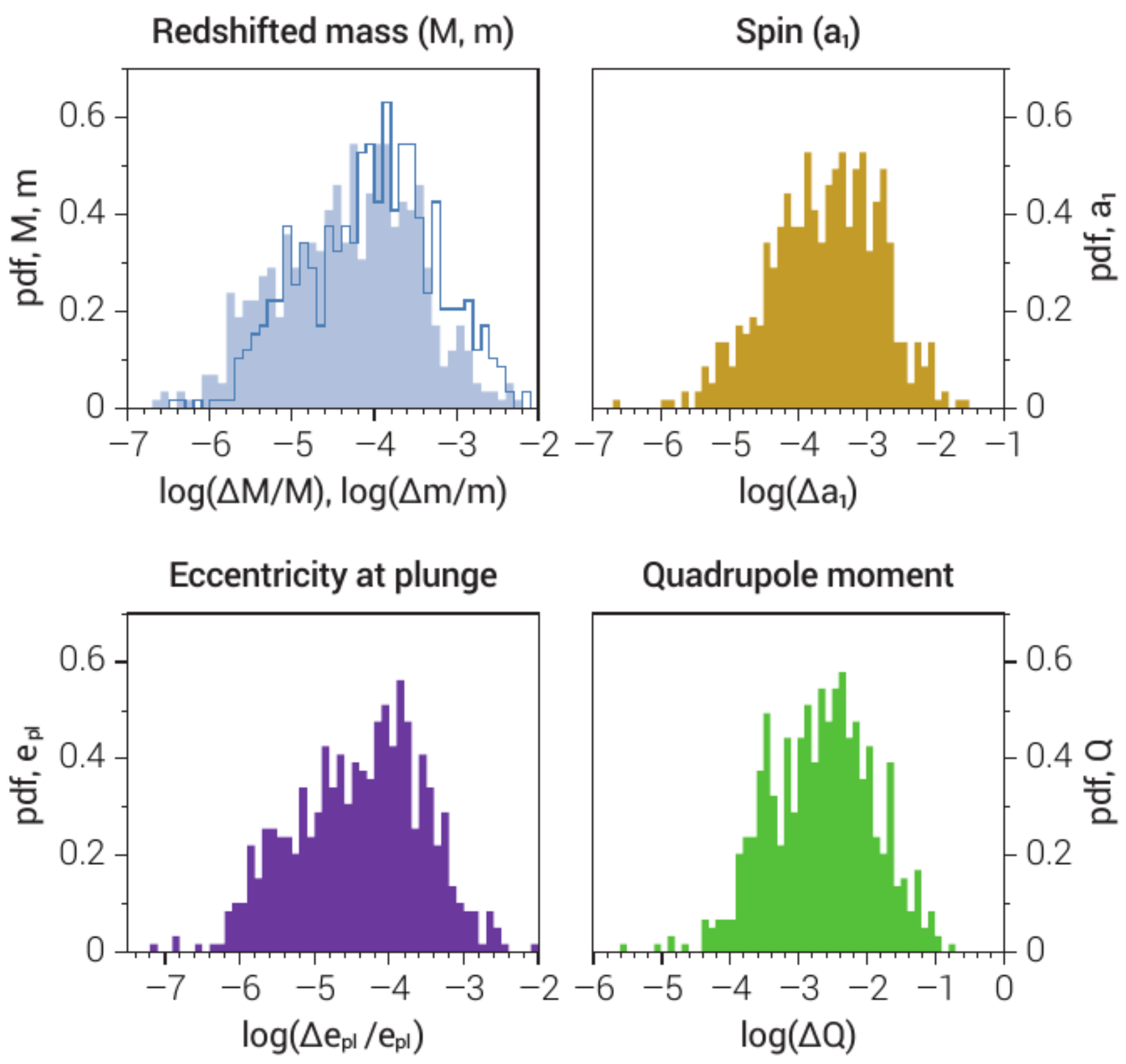}
\caption{Parameter estimation accuracy on a sample of EMRIs from  eLISA \cite{whitepaper13}.  In the top left panel the filled (empty) histograms are the distribution of mass determination precision for the central primary black hole (inspiralling compact object); the top right panel represents the distribution in the error determination of the spin magnitude of the central massive hole; the lower left panel represents the eccentricity error of the compact orbit at plunge; the lower right panel is the fractional precision at which the quadrupole moment of the primary black hole can be measured.}
\label{fig_EMRIPE} 
\end{center} 
\end{figure}

\smallskip
\noindent
{\sl Astrophysics and fundamental physics with EMRIs} -  High rates imply large astrophysical payouts following detection. The number and mass distribution of EMRIs will inform us about the  unconstrained low end of the mass function of massive black holes and on the dynamics of compact objects in the dense environment of galactic nuclei on scales that are impossible to probe otherwise. The requirement of matching hundreds of thousands of cycles to dig out the signal from the data stream, implies that detections will automatically come with exquisite parameter estimation \cite{GWnote13}. Figure \ref{fig_EMRIPE} shows that the mass of the two black holes and the spin of the massive black hole can be determined generally to better than a part in ten thousand, a precision that is unprecedented in astronomical measurements. This will make it possible to perform massive black hole population studies on a sample of relatively low redshift, quiescent black holes, complementary to the higher redshift, merging systems seen as MBHCs. EMRIs ensure that the inspiralling object essentially acts as a test particle in the background space-time of the central massive black hole. As such, the hundreds of thousands of wave cycles collected at the detector encode a very precise mapping of the stationary spacetime metric of the central massive black hole, providing the ultimate test of its Kerr nature, complementary to the ringdown one possible with MBHCs\cite{Berti15-testGR}. As shown in Figure \ref{fig_EMRIPE} deviations as small as  0.1\% from the Kerr mass-quadrupole moment will be detectable for typical EMRIs, pushing testing of spacetime metric to a whole new level.

\smallskip
\noindent
$\bullet$ {\underline {\bf Continuous sources}}  comprise double white dwarfs in binaries (WD,WD), and possibly (NS,NS) and (BH$^*$,BH$^*$), in the Milky Way emitting a nearly monochromatic signal, preferentially located at the low frequency end of the eLISA sensitivity interval \cite{GWnote13}. 
A number of (WD,WD) binaries are already known
to emit a nearly monochromatic signal in the eLISA band since they have been 
discovered in the electromagnetic window, and are known as {\it verification binaries.} 
The discovery of new (up to a few thousand for a two-year mission) ultra-compact binaries 
with orbital  periods below one hour and typically 5 to 10 minutes, determined from the periodicity of the gravitational wave, is one of the main objectives of a LISA-like mission.
For a number of systems it is possible to measure the
first time derivative of the frequency, and thus determine a combination of the masses of the two
component stars that can be used to distinguish white dwarf, neutron
star and black hole binaries. This will give  precious insight on the distribution of the binaries in their different arrangements and flavours, present 
in the thin and thick discs of our Galaxy as well as  in the halo and inside globular clusters.
 The highest signal-to-noise-ratio systems will allow us to study the complex
physics of white dwarf interactions in binaries and to establish how systems survive
as interacting binaries.  We recall that (WD,WD) binaries are considered to be potential progenitors of Type Ia supernovae \cite{Amaro-Seoane12,GWnote13}.

 \begin{figure}[!t]
\begin{center}
\includegraphics[width=.99\textwidth]{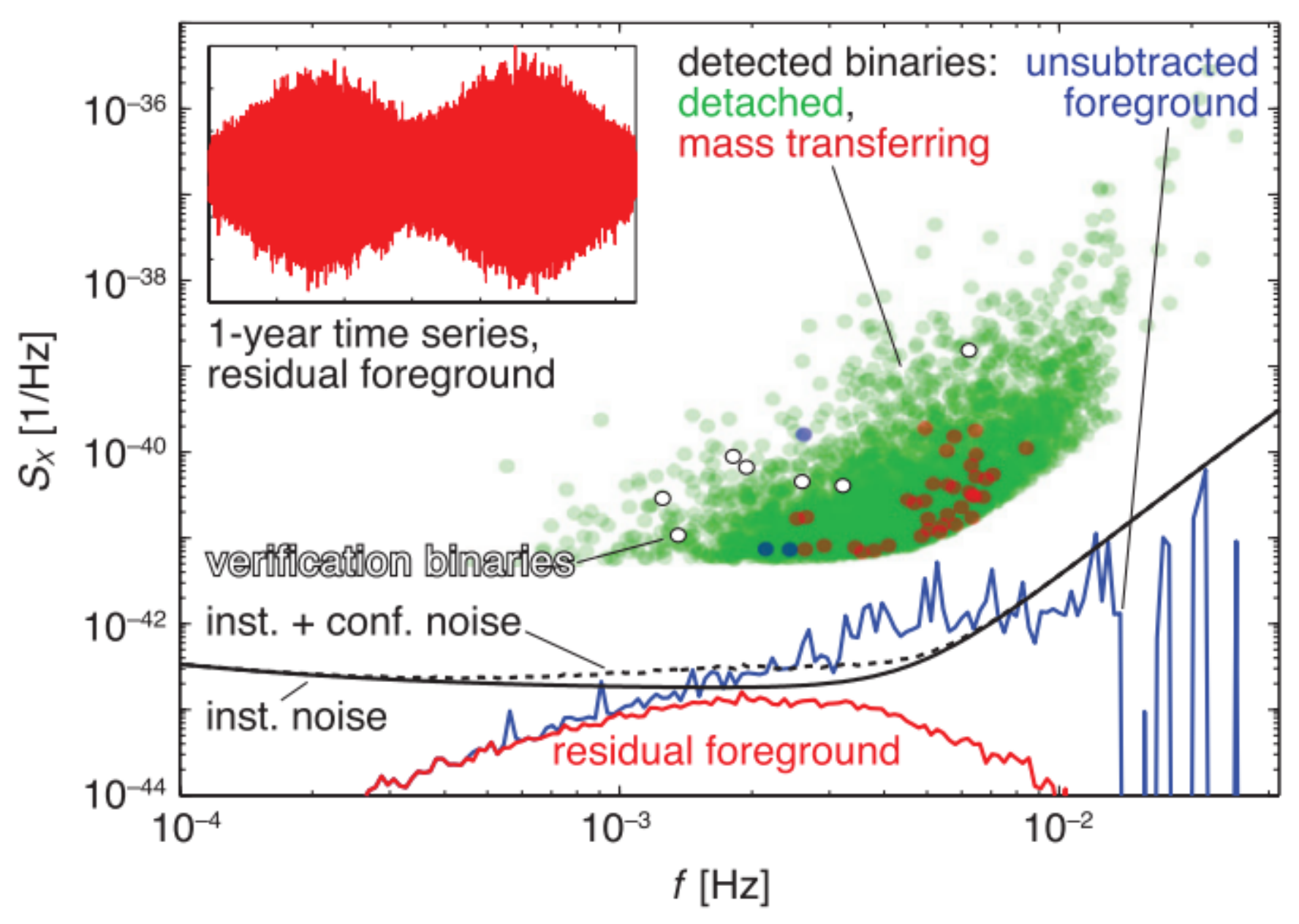}
\caption{Power spectral density in units of strain Hz$^{-1}$ of the gravitational wave foreground from galactic binaries, before (blue) and after (red) subtraction of resolvable systems, adapted from
\cite{Amaro-Seoane12}. Binaries are subtracted if they have a signal-to-noise ratio SNR$>7$ over two years of observations, and are plotted as green (detached binaries) and red/blue (mass-transferring binaries) dots. White dots identify few known verification binaries. The solid (dashed) black curves represent instrument noise without (with) the residual foreground confusion noise included. The small inset to the top left represents a time series realisation of the residual foreground, showing the typical annual modulation due to the detector's orbit around the Sun. Courtesy of M. Vallisneri.}
\label{fig_WDWD} 
\end{center} 
\end{figure}

\smallskip
\noindent
$\bullet$ {\underline {\bf Galactic foreground}} describes the signal coming from an unresolved population of million compact binaries emitting each a nearly monochromatic gravitational wave, which are 
confined in the thick disc of the Milky Way, preferentially (WD,WD) binaries  
which create a confusion-limited noise at frequencies below a few mHz, as illustrated in
Figure \ref{fig_WDWD}. Its average level is comparable to the instrument
noise, but due to its strong modulation during the year (by
more than a factor of two) it can be detected. The overall strength can be used to learn about the distribution of the
sources in the Milky Way. 

\smallskip
\noindent
$\bullet$ {\underline {\bf Cosmological background}} refers to the signal(s) coming from the primordial universe. The  frequency band of a LISA-like detector corresponds to 0.1 to 100 TeV energy scales in the early universe, at which new physics is expected to become visible. We defer to \cite{Caprini16} for a authoritative description of the potential sources of the primordial universe.

\subsection{The very low frequency gravitational universe}

\smallskip
Moving further down in frequency, we enter the {\it very low frequency universe} probed by Pulsar Timing Array  experiments EPTA \cite{Lentati15},  PPTA \cite{Shannon15},  NANOGrav \cite{Arzoumanian16}  and 
the International PTA (IPTA)\cite{Verbiest16}, which are especially sensitive in a window extending from $10^{-9}$ Hz to $10^{-7}$ Hz. 
 Arrays of  millisecond pulsars  can be used to detect correlated signals such  caused by passing gravitational waves.  The dominant contribution at these frequencies 
 is  expected to come from  supermassive black hole binaries 
(SMBHBs) in their slow inspiral phase,  month or years prior to merging \cite{Sesana13}.

\smallskip
\noindent
$\bullet$ {\underline {\bf Background from supermassive black hole binaries}}  refers to the incoherent superposition of signals coming from a large number of SMBHBs of $10^8-10^{10}\msun$ forming in massive galaxy mergers out to redshift $z\simless 1$ which gives rise to a confusion-limited foreground, and 
on top of which particularly bright or nearby sources might be individually resolved.
The main traits of the background are described  at the end of this Chapter, but we can anticipate the obvious payout of a PTA detection.

The background from SMBHBs informs us of the existence
of a vast population of sub-pc (to be precise, sub-0.01pc) SMBHBs expected to rise according to the current cosmological
model of galaxy assembly, and for which we have only indirect evidence \cite{Merritt06}.
Recently, a number of galaxies in the verge of merging has been discovered in large surveys, each galaxy harbouring an active supermassive black hole
\cite{Liu16}.
But,  these 
mergers are in their early stage of pairing, as galaxies are observed interacting on scales of several kpcs.  The detection of this foreground 
can provide a measure of the efficiency of  the pairing and hardening of these SMBHBs on pc and sub-pc scales. This will enable us to 
distinguish the role of  stellar and/or gas dynamics  in removing energy and angular
momentum from the binary (overcoming the last parsec problem, i.e. the possible stalling of the binary due its weak coupling with 
the environment \cite{Colpi14}.  In particular, 
from the shape  and amplitude of the signal we will learn whether binaries are eccentric or circular in their approach to coalescence, possibly
constraining the efficiency of the mutual coupling with stars or/and gas. Identification and sky localisation
of individual sources, will also open the possibility of identifying their electromagnetic counterpart,  making
multi-messenger studies of SMBHBs possible. In $\S 7.5$ we describe in more detail the background detectable by PTA.

\smallskip
\noindent
$\bullet$ {\underline {\bf The unknown}} is a universe hosting  totally unexpected sources over the whole multi-frequency gravitational wave sky. History shows that  every time a new window became accessible to electromagnetic observations we discovered sources that were never anticipated. 

\section{Binaries as key sources of the gravitational universe}

In this section, we introduce shortly key concepts required to identify the main traits of binaries as astrophysical sources of gravitational waves, and defer
to the book by M. Maggiore,  {\it Gravitational Waves}, and Poisson and Will, {\it Gravity}\cite{Poisson14}, for a comprehensive overview. 

In Newtonian gravity, two point masses in a binary move on circular or elliptical orbits around the common centre of mass. The motion is periodic  with constant Keplerian frequency
$f_{\rm K}=(GM/a^3)^{1/2}/(2\pi)$, where $a$ is the semi-major axis of the relative orbit and $M=m_1+m_2$ the total mass of the binary of components $m_1$ and $m_2$, respectively.
In general relativity, binary systems emit gravitational waves which radiate away orbital energy and angular momentum.
In the case of  circular binaries the gravitational wave, which tracks the large scale motion, is monochromatic with frequency  $f$ equal to $f=2f_{\rm K}$.
 To compensate the radiative energy losses, binaries back-react gradually hardening, i.e. decreasing their semi-major axis $a$ and increasing the orbital 
 frequency $f_{\rm K}$. The emission is weak initially and a  phase of nearly {\it adiabatic contraction}, lasting hundred to thousand million years, anticipates the
phase of {\it inspiral, merger and ringdown,}  which produce a detectable signal.  

The  {{\it inspiral}} refers to the phase 
 when the two binary components can still be considered as structureless  and their dynamics (both conservative and dissipative)  can be described by Post Newtonian (PN) theory.   In this phase, which is the longer lasting,  the signal, called {\it chirp}, has a characteristic shape, with both the amplitude and frequency of the wave slowly sweeping to higher values. This phase is crucial in obtaining first estimates of the binary system's parameters most of which can be extracted 
by matching the observed signal on general relativity predictions.
When the binary companions are spinning, the signal is modulated by spin-orbit and spin-spin couplings, and this modulation encodes in addition to the masses, orbit inclination, distance and sky location, also the spins of the two interacting bodies.

The {\it merger} refers to the phase of "very late inspiral"  and coalescence (no longer described within the PN formalism).  Moving at around one third of the speed of light, the two bodies experience extreme gravitational fields so that their dynamics and signal can be  described  only in the realm of Numerical Relativity (NR). 
The merger signal lasts for a shorter time (milliseconds for stellar origin black holes, minutes for massive black holes) compared to the inspiral, and in this phase finite-size effects become important for neutron star mergers, as the stars carry a surface.
NR simulations which account for the full non-linear structure of the Einstein's equation are highly successful in
tracing the dynamics and the gravitational wave radiation.

The {\it ringdown}  refers to the phase when the coalescence end-product relaxes to a new stationary equilibrium solution of the Einstein field equations: 
a new black hole for (BH*,BH*) and (NS,BH*) mergers or a hot hyper-massive or supra-massive neutron star or a black hole for the case of (NS,NS) mergers.  Likewise
MBHCs and EMRIs end with the formation of a new black hole.
The emitted radiation can be computed using Perturbation Theory and it consists of a superposition of quasi-normal modes of the compact object that forms. These modes carry a unique signature that depend only on the mass and spin in the case of black holes \cite{Berti15-testGR,BertiSesana16}. 

The merger and ringdown parts of the signal last for a short duration, yet they carry tremendous luminosity. Their inclusion in a matched filter search for binary systems dramatically increases the distance reach and the accuracy at which the masses and spins can be measured.
The access to the latest stages gives precious insight into the structure of neutron stars and the EoS at supra-nuclear densities; and in the case of black holes the possibility of testing
gravity in the genuinely strong-field dynamical sector, and possibly prove the "no-hair" conjecture \cite{Berti15-testGR,Yunes13-review}.
The emission of gravitational waves from a binary is a continuous process and phenomenological models for the merger dynamics have been developed, the most remarkable among the various approaches being the Effective One Body (EOB)  theory and the Phenom models which permits a continuous description of the three phases, as predicted by general relativity \cite{Buonanno99,DN16,Hannam14,Husa16}, including also tidal effects in the case of (NS,NS) coalescences \cite{Bernuzzi15}.

\subsection{Description of the inspiral}

Binaries are irreversibly driven to coalescence, and the reference frequency of the gravitational wave at the time of coalescence is 
\begin{equation}
f_{\rm coal}={1\over (\pi 6^{3/2})}{c^3\over GM}
\label{fcoal}
\end{equation} 
representing twice the Keplerian frequency of a test mass orbiting around a non-spinning binary black hole of mass $M$ (seen as single unit) 
at the innermost stable circular orbit $R_{\rm isco}.$ 
Neutron stars are so compact that their equilibrium radii are smaller than $R_{\rm isco}$ for many EoSs and typical masses \cite{Cook94}, so that $f_{\rm coal}$ represents a reference frequency for coalescing compact objects in general.

\medskip
\noindent
$\bullet$ {\underline{Radiated energy and angular momentum - back reaction}}\\
In the inspiral phase and to leading order,  the power radiated by a {\it circular binary} (averaged over a orbital period) 
is
\begin{equation}
 {\dot E}^{\rm circ}_{\rm gw} ={32\over 5}{c^5\over G}\left ({GM_{\rm c}\over c^3} \pi f\right )^{10/3}={32\over 5 \cdot 6^5}{c^5\over G}\nu^2 {\tilde f}^{10/3}
\label{dotEcirc}
\end{equation}
where $f=2f_{\rm K}$ is the frequency of the gravitational wave emitted, $\tilde {f}=f/f_{\rm coal}$ the unitless frequency, and 
\begin{equation}
M_{\rm c}\equiv{(m_1m_2)^{3/5}\over (m_1+m_2)^{1/5}}=\nu^{3/5}M=\mu^{3/5}M^{2/5}
\label{mchirp}
\end{equation}
 is the so called {\it chirp mass} expressed either in terms of  the {\it symmetric mass ratio}, $\nu=m_1m_2/M^2$ (equal to $1/4$ for an equal mass binary), or of the reduced mass $\mu=m_1m_2/M=\nu M$.
The luminosity ${\dot E}^{\rm circ}_{\rm gw}$  near coalescence (${\tilde f}\sim 1$) does not depend upon the mass $M$ of the coalescing objects,  but on the symmetric mass ratio $\nu$ only, approaching the value $\nu^2(c^5/G)\sim  \nu^2 ( 3.6\times 10^{59})\ergs.$ 
The independence on $M$ is just a consequence of the fact that [energy/time] is equivalent to [mass/time], and time is equivalent to mass in $G=c=1$ units.
For a short time lapse, this huge gravitational wave luminosity is far in excess of the electromagnetic luminosity of the entire universe (when $\nu>  0.01$).

We then remark that merging  black holes of stellar origin of $10\msun - 30\msun$ emit the same luminosity as merging black holes of $10^6\msun$ or $10^9\msun$, for a given $\nu,$  as the two fundamental constants $c$ and $G$ fix the scale uniquely.

 The orbital angular momentum ${\bf L}$ from a binary is radiated away at a orbit-averaged rate 
   \begin{equation}
 {\dot L}^{\rm circ,orb}_{\rm gw} ={32\over 5}M_{\rm c}c^2\left ({GM_{\rm c}\over c^3} \pi f\right )^{7/3}={32\over 5\sqrt{6^7}} \nu^2 Mc^2{\tilde f} ^{7/3}, 
\label{dotJbin}
 \end{equation}  
  in the direction of ${\bf L}$. 
 When the binary nears coalescence,  $ {\dot L}^{\rm circ,orb}_{\rm gw} \to [32/5 \sqrt{ 6^7}] \nu^2 Mc^2$
 whose value depends on $M$ and  $\nu$.
 
Binaries with non zero eccentricity $e$ and equal semi-major axis $a$, lose energy and angular momentum at a higher rate, as during closest approach when the mutual interaction is strongest, radiation is emitted more effectively. 
 The two rates are enhanced by a factor ${\cal E}(e)=(1+73e^2/24+37e^4/96)/ (1-e^2)^{7/2}$ in (\ref{dotEcirc}), and  ${\cal L}(e)=(1+7e^2/8)/(1-e^2)^2$ in 
 (\ref{dotJbin}), with respect to a circular binary. In the case of eccentric binaries, the signal carries a dependence on the eccentricity $e,$ and 
the emission spectrum is far richer than for a circular binary as more harmonics $nf$ of the fundamental frequency $f=2f_{\rm K}$ enter the expression, with 
$n>1$.
 
The emission of gravitational waves costs energy, and the source of radiation is the orbital energy of the binary, given 
by
$E_{\rm bin}=- (1/2)G\nu M^2/a=-(1/2)\nu M^2(GM)^{1/3}(\pi f)^{2/3}$ according to the virial theorem (computed to lowest order assuming Newtonian dynamics).  Likewise, angular momentum
is radiated away at the expense of the orbital angular momentum ${\bf L}=\nu M[GMa(1-e^2)]^{1/2} {\hat{\bf L}}, $ 
where ${\hat {\bf L}}$ denotes its direction. 

The inspiral can be represented as a sequence of quasi-closed orbits where  both the semi-major axis $a$ and eccentricity $e$ vary with  time.  During adiabatic contraction, and according to (\ref{dotEcirc}) and (\ref{dotJbin}), one can prove that 
energy is extracted  more rapidly than angular momentum, and binaries become more and more circular so that only little or null eccentricity is left at the time of coalescence.
(Notice however that this is not necessarily true for EMRIs as in their inspiral they can retain significant eccentricity up to 
the innermost circular orbit.)

The total energy of the binary decreases adiabatically at a rate equal to $ \dot E_{\rm bin}=-\dot E_{\rm gw}$. The binary hardens, the semi-major axis
decreases, and the gravitational wave frequency increases  at a rate 
 \begin{equation}
{\dot f}={96\over 5}\pi^{8/3}\left (  {GM_{\rm c}\over c^3} \right )^{5/3} f^{11/3}.
\label{fdotchirp}
 \end{equation}
 Equation (\ref{fdotchirp}) is derived setting  $ \dot E_{\rm bin}=(dE_{\rm bin}/df)(df/dt)$ with $dE_{\rm bin}/df$ inferred  using the
 expression of the binary's Newtonian energy $E_{\rm bin}$ given few lines above.
Equation (\ref{fdotchirp}) shows that to leading order  the frequency evolution of the gravitational wave emitted by a circular binary is determined uniquely by the chirp mass  $M_{\rm c}$. The evolution of $f$  is slow
initially and it progresses faster and faster with time, given the rapid dependence of ${\dot f}$ on the frequency itself.
The solution to (\ref {fdotchirp})
\begin{equation}
f(t)={5^{3/8}\over (256)^{3/8}\pi}  \left  ( {GM_{\rm c}\over c^3}\right )^{-5/8} (t_{\rm coal}-t)^{-3/8}
\label{ftcoal}
\end{equation}
 describes the rise in the frequency  $f$ of the gravitational wave emitted by the binary when chirping,  where $t_{\rm coal}$ gives the epoch of merger.  At  $t_{\rm coal}$, the frequency of the wave
formally diverges, but a non diverging cut-off frequency is found when the system evolves into the relativistic state and the two masses merge.

Figure \ref{GW150914} of $\S 2$ shows (bottom row) the spectacular {\it chirp} observed in GW150914 \cite{Abbott-1}, i.e. the increase in frequency during binary inspiral, and the convergence of a finite value at merger as a new black hole has formed.
 
According to (\ref{ftcoal}), a binary observed at a frequency $f$  takes a time to coalesce equal to
\begin{equation}
\tau_{\rm coal}^{\rm circ}(f)= {5\over 256\pi^{8/3}} {1\over \nu}\left (  { c^3\over GM} \right )^{5/3} {1\over f^{8/3}}\simeq  {7.4\over \nu} \left (  {\msun\over M} \right )^{5/3} \left ({1 \,{\rm Hz}\over f}\right )^{8/3}\,\rm days,
\label {tcoalf}
\end{equation}
which is a steep function of the frequency $f$. 

The late inspiral, merger and ringdown phases have a very short duration.   In terms of the dimensionless frequency ${\tilde f}\sim 1$ this time is 
\begin{equation}
\tau_{\rm coal}^{\rm circ}({\tilde f})={6480\over 256}{1\over \nu}{GM\over c^3}{1\over {\tilde f}^{8/3}}\simeq 1.25 \times 10^{-4} {1\over \nu}{M \over \msun}
{1\over {\tilde f}^{8/3}}\,\, \rm sec.
\label{tcoaltildef}
\end{equation}
Equation (\ref{tcoaltildef}) shows that inspiraling massive binaries weighing more than $10^6\msun$ 
are characterised by a longer duration signal than stellar origin binaries and that unequal mass binaries with same $M$ take much longer to coalesce
as $\tau_{\rm coal}^{\rm circ}\propto \nu^{-1}$.

To become observable sources of gravitational waves, binaries need to contract and merge on a timescale less than the  Hubble time of 13.6 Gyr.
Since one can relate $f$ to the Keplerian period, 
equation (\ref{tcoalf}) gives the characteristic time to coalescence of a generic binary as a function of the binary separation. Thus, 
 the typical semi-major axis that a binary should have in order to coalesce within  $\sim $ Gyr is 
\begin{equation}
a_{\rm gw}\simeq 2\times 10^{11} \left ({1\over \nu}\right )^{1/4} \left ({M\over \msun}\right )^{3/4} \left ({\tau^{\rm circ}_{\rm coal}\over \rm Gyr}\right )^{1/4}\,\rm cm,
\end{equation}
 corresponding to a few R$_\odot$ for stellar origin binaries, and to a $10^{-3}$ pc for  black hole binaries of $10^6\msun.$  These are remarkably small distances, so that
torques and dissipation processes of different origin need to be at work to "deposit" a binary in the domain of gravitational wave emission.

 Two further important quantities which describe circular binaries in their spiral-in phase are the energy spectrum $dE_{\rm gw}/df$ and the total energy radiated in gravitational waves 
 $E_{\rm gw}$. In the quadrupole approximation,
 \begin{equation}
 { dE_{\rm gw}\over df}={\pi^{2/3}\over 3G}(GM_{\rm c})^{5/3}f^{-1/3}
 \label{dedf}
 \end{equation}
and 
\begin{equation} E_{\rm gw}\sim{\pi^{2/3}\over 2G} (GM_{\rm c})^{5/3}f_{\rm max}^{2/3}
\label{egwinspiral}
 \end{equation}
 where $f_{\rm max}$ is the maximum frequency at which the inspiral is observed. 
If we extrapolate crudely  (\ref{egwinspiral}) up to $f_{\rm coal}$, the radiated energy  $ E_{\rm gw}\sim 0.08\, \nu M c^2$ 
depends on the reduced mass of the binary (a more accurate estimate of $E_{\rm gw}$ is given ahead  in this Chapter). 


\medskip
\noindent
$\bullet$ {\underline{Waveforms}}\\
Interferometers detect not the energy carried by the wave but the perturbation of spacetime itself, $h_{\mu\nu}$.
Chapter 1 provided the formalism to compute, to leading order, the two independent 
(traceless) polarisation states $h_+$ and $h_\times$ of a gravitational wave. For a circular binary at distance $r$  the two states read
\begin{equation}
h_+(t)={4\over r}\left (  {GM_{\rm c}\over c^2} \right )^{5/3}\left (  {\pi f(t_{\rm ret})\over c} \right ) ^{2/3}  {1+\cos^2(\iota)\over 2}\cos [\Phi_{\rm N}(t_{\rm ret})]
\label{h+bin}
\end{equation}

\begin{equation}
h_\times(t)={4\over r}\left (  {GM_{\rm c}\over c^2} \right )^{5/3} \left (  {\pi f(t_{\rm ret}) \over c} \right ) ^{2/3}  \cos(\iota)\sin [\Phi_{\rm N}(t_{\rm ret})]
\label{hxbin}
\end{equation}
where $\iota={\bf {\hat n}}\cdot {\bf {\hat L}}$ is the inclination angle between the line of sight ${\bf {\hat n}}$ and the unit vector parallel to the orbital angular momentum ${\bf { L}}$ of the binary.  $f(t_{\rm ret})$ is the instantaneous frequency
given by (\ref{ftcoal}) (evaluated at the retarded time $t_{\rm ret}=t-r/c$) and  
$\Phi_{\rm N}\equiv 2\pi\int f(t')dt'$  is the lowest order contribution to the orbital phase evolving according to 
  \begin{equation}
 \Phi_{\rm N}(t)=\Phi_0-2\left({5GM_{\rm c}\over c^3}\right )^{-5/8}(t_{\rm coal}-t)^{5/8},
 \label{phaset}
 \end{equation}
  where $\Phi_0$ is a constant  giving the orbital phase at the epoch of merger, when $f$ diverges nominally (finite size effect will impact
 on the waveform before this divergence is reached).
 It is worth noting that, at any given frequency,
both $h_+$ and $h_\times$ scale as  $\nu M^{5/3}$, i.e. faster than linear with respect to the total mass $M$ 
and linear in $\nu$, so that unequal mass binaries of total mass $M$ have weaker emission. 

Note further that the ratio of the two polarisation amplitudes $h_{+}/h_{\times}$ 
depends on the inclination angle $\iota$. When the binary is seen edge-on, i.e. $\iota=\pi/2$, $h_{\times}=0$
 and radiation has pure $+$ polarisation as, from the observer's view, the motion of the binary stars projected on 
the sky is purely linear. When $\iota=0,$  so that the binary is seen face-on,
 the stars execute a circular motion in the
sky, and both polarisation components have equal amplitude and are out of phase by
$\pi/2$, emitting a  circularly polarised wave. Thus, to leading order, the polarisation has a direct relationship to the motions of
the point masses projected on the observers sky plane.

A single detector can only  measure a linear combination of the polarisations, called {\it strain amplitude}, 
\begin{equation}
h(t)=F_+(\alpha,\delta,\psi)h_+  + F_\times(\alpha,\delta,\psi)h_\times
\label{h}
\end{equation}
where $F_+$ and $F_\times$ are the antenna patterns for a particular detector\cite{Sathya09}. The angles $\alpha$ and $\delta$ describe the binary's position on the sky,
and  $\psi$  the component of ${\bf{\hat L}}$ orthogonal to ${\bf{\hat n}}$. The angles $\iota$ and $\psi$ fully specify the orientation of
the binary's angular momentum ${\bf{\hat L}}$.  They  are constants for transient sources, but  they must be considered time-dependent for long lasting sources, when Doppler modulation of the signal due to the relative motion of the source and  detector can not be neglected.

It is expedient to write the response $h(t)=F(t)(\cos \xi h_+ + \sin\xi h_\times)$ with $F=(F^2_+ + F^2_\times)^{1/2}$ and $\tan \xi=F_\times/F_+$. In this way $F(t)$ is independent of the polarisation angle  and provides a measure of  the sensitivity of the detector to different locations in the sky. 

If the signal is extracted using {\it match
filtering techniques}, what is measured is the effective strain amplitude defined as 
\begin {equation}
h_{\rm eff} = {\cal N}_{\rm cycles}^{1/2} h, 
\label{heff}
\end{equation}
 where ${\cal N}_{\rm cycles} = fT$ represents the number of cycles the 
chirping binary covers over the observing (or emitting) time $T$.
 The number of cycles  spent by the source in the detector bandwidth $\Delta$ between [$f_{\rm min},f_{\rm max}$] is equal to ${\cal N}_{\rm cycles}=\int_{f_{\rm min}}^{f_{\rm max}} df (f/\dot f)$. From (\ref{fdotchirp}) we infer
 \begin{equation}
 {\cal N}_{\rm cycles}={1\over 32\pi^{8/3}}\left({GM_{\rm c}\over c^3}\right )^{-5/3} (f^{-5/3}_{\rm min}-f^{-5/3}_{\rm max}).
 \label{ncycles}
 \end{equation}
 Note that ${\cal N}_{\rm cycles}$ can be written in a compact form in terms of the dimensionless frequency ${\tilde f}$  and
 the in-band cycles can be estimated in terms of the minimum frequency of  the detector sensitivity $f_{\min}$  (since $f_{\rm min}\ll f_{\rm max}$)
  \begin{equation}
  {\cal N}_{\rm cycles}({\tilde f_{\rm min}})={6^{5/2}\over 32\pi}{1\over \nu}{\tilde f_{\rm min}}^{-5/3}.
   \label{cycleschirp}
   \end{equation}
   Note further that ${\cal N}_{\rm cycles}$ is proportional to $\nu^{-1}$ so that unequal mass binaries and in particular EMRIs remain 
   in band for a longer time covering many more cycles
   before plunging, which enhances the strength of their signal.  Typically, merging binaries are tracked for ${\cal N}_{\rm cycles}\sim O(10)-O(100),$
   while EMRIs for $O(10^{4-5}).$
 
Match filtering techniques,
can be very effective in extracting the signal from the noise, even when the typical amplitude is  a factor ${\cal N}^{1/2}_{\rm cycles}$ below the noise floor.  Since the  number of cycles spent in the interferometer bandwidth ${\cal N}_{\rm cycles}\propto f^{-5/3}$ decreases with 
increasing frequency more rapidly than
the rise of $h$ with $f$, the effective strain amplitude $h_{\rm eff}={\cal N}^{1/2}_{\rm cycles} h\propto f^{-1/6}$
decreases with $f$, even if the nominal instantaneous amplitude increases ($h\propto f^{2/3})$. 
\begin{figure}[!t]
\begin{center}
\includegraphics[width=.750\textwidth]{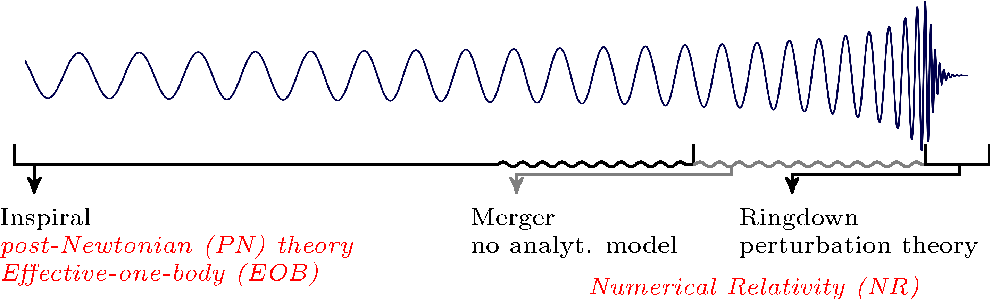}
\caption{Characteristic waveform $h(t)$  from coalescing non spinning  equal mass black holes, depicting the inspiral, merger and ringdown phases, from
\cite{Ohme12}. Wavy lines illustrate the regime close to merger where analytical methods
have to be bridged by NR.The different approximation schemes and their range of validity are indicated.Courtesy of F. Ohme.}
\label{BHsimple} 
\end{center} 
\end{figure}
Effective strain amplitudes from interesting astrophysical sources within reach have $h_{\rm eff}\approx 10^{-18}-10^{-22}.$

To have a description of the {\it late inspiral} linearised theory is insufficient.  As the two masses come closer, the relative orbital speed $(v_{\rm orb}/c)$ increases, and one should compute the orbit dynamics,  amplitude and phase of the gravitational waves in the so called PN  formalism \cite{Blanchet14,Poisson14},
which consists in an expansion in terms of  $(v_{\rm orb}/c)$, accurate down to relative binary separations of the order of 
$a_{\rm NR}\sim 10 GM/c^2$. Below $a_{\rm NR}$ instead, Numerical Relativity  (NR) gives the full and final description of the gravito-dynamics of the binary, from the late inspiral, final plunge and merger.
The Effective One Body\cite{Buonanno99,Ohme12,Buonanno14} method and the Phenom models \cite{Hannam14,Husa16} prove to be very effective in providing waveforms consistent with NR, across the three phases. 
 This allows to create template repositories that are 
used to model the observed signal, and extract the parameters at the source.
Figure \ref{BHsimple} illustrates in a rather simple way, the main features of the waveform from a non spinning, equal mass black hole binary. 
The key feature of black hole coalescence is that after merger the amplitude decays almost suddenly indicating the formation of a new event horizon.

\medskip
\noindent
$\bullet$ \underline{Binaries at cosmological distances in the PN frame work}\\
Coalescing binaries can be seen out to cosmological distances, and in this case the expansion of the universe during the propagation of the wave
from the source to the detector can not be neglected. This requires a re-exam of  (\ref{h+bin}) and (\ref{hxbin}).
It is easy to show that the frequency, mass and distance are affected by the underlying expansion of the universe in a rather simple way. If $z$ is the cosmological redshift of the source of gravitational waves, the frequency, mass and distance acquire the following corrections:
\begin{itemize}
\item the observed frequency $f_{\rm obs}$ is redshifted with respect to the frequency as measured in the source frame, $f_{\rm obs}=f/(1+z)$;
\item  the chirp mass in the source frame $M_{\rm c}$  is replaced by the redshifted chirp mass  ${\cal M}_{\rm c}=(1+z)M_{\rm c}$, i.e. the chirp mass in the observer-frame; 
\item  the source distance $r$ is replaced by the luminosity distance $d_L(z)=(1+z)a(t_0)r$,
where $a(t_0)$ is the scale factor today.
\end{itemize}

\noindent
According to the above scalings,  
the evolution of the waveform of a generic binary is  invariant under the change of $M_{\rm c}\to (1+z)M_{\rm c}$ and $f\to f/(1+z)$, 
so that the wave does not encode any information on the cosmological redshift of the source, and  the  two polarisation amplitudes are given by 
\begin{equation}
h_+(\tau^{\rm obs})={4\over d_L(z)}\left (  {G{\cal M}_{\rm c}(z)\over c^2} \right )^{5/3} \left (  {\pi f_{\rm obs}(\tau^{\rm obs})\over c} \right ) ^{2/3}  {1+\cos^2(\iota)\over 2}\cos[ \Phi(\tau^{\rm obs})]
\label{h+binchirpz}
\end{equation}
\begin{equation}
h_\times (\tau^{\rm obs})={4\over d_L(z)}\left (  {G{\cal M}_{\rm c}(z)\over c^2} \right )^{5/3} \left (  {\pi f_{\rm obs}(\tau^{\rm obs})\over c} \right ) ^{2/3} \cos(\iota)\cos[ \Phi(\tau^{\rm obs})]
\label{hxbinchirpz}
\end{equation}
where $\tau^{\rm obs}=(1+z)(t_{\rm coal}-t)$ is the time to coalescence measured  by the observer's clock  (note that $\tau^{\rm obs}_{\rm ret}=\tau^{\rm obs}$), and 
where the frequency and orbital phase $\Phi=\pi\int f_{\rm obs}(t')dt'$ are  computed solving for the equation 
\begin{equation}
{\dot f}_{\rm obs}={96\over 5}\pi ^{3/8} {\cal M}^{5/3}_{\rm c}(z) f^{11/3}_{\rm obs} \left  [1+{\cal D^{\rm PN}}\right ] \label{ftcoalchirpz}
\end{equation}
where $\cal D^{\rm PN}$ is the PN correction to the phase, up to the desired order. 
$\cal D^{\rm PN}$ depends on $t_{\rm coal}$ and $\Phi_0,$ and 
 can be expanded analytically in powers of the symmetric mass ratio $\nu$ and of ${\cal M}_{\rm c}(z) f_{\rm obs}$, both independent of redshift $z$ (we defer to \cite{Blanchet14} for the full  analytical expression of  $\cal D^{\rm PN}$ up to 3.5 PN order).  
 Again ${\dot f}_{\rm obs}$ is dominantly determined by the chirp mass, but PN corrections $\cal D^{\rm PN}$ depend on the symmetric mass ratio and on the two black hole spins (if different than zero). Therefore, these corrections help in breaking degeneracies in the waveforms and allow to measure individual black hole (redshifted) masses and spins.

\begin{figure}[!t]
\begin{center}
\includegraphics[width=.99\textwidth]{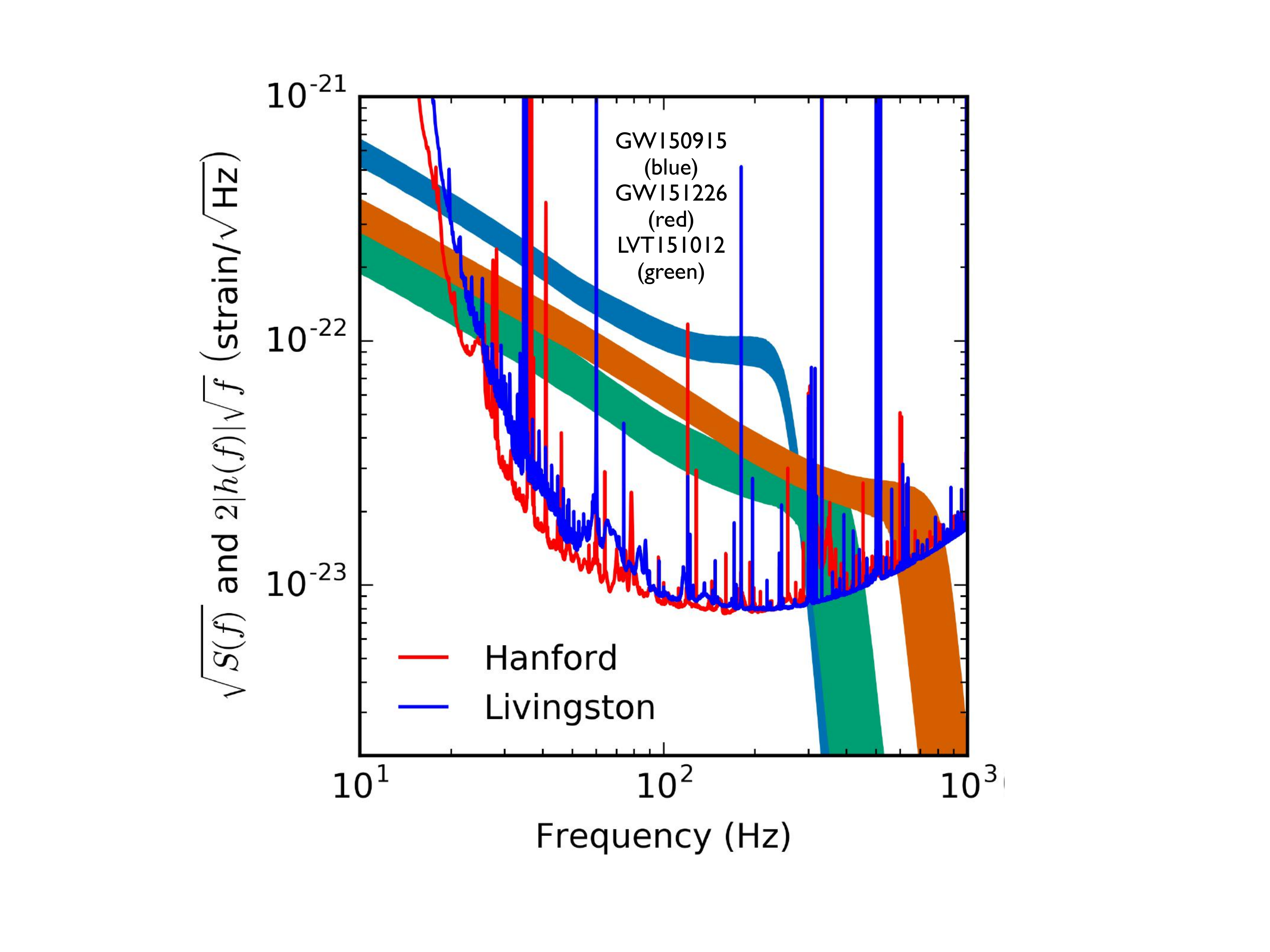}

\caption{The figure shows the spectral strain sensitivity of the total noise  $\sqrt {S_{\rm noise}}$ of the Hanford and Livingston detectors 
and the signals  of GW150914, GW151226 and of the trigger event  LVT151012 (which
has a $87\%$ probability of hosting a third coalescing  black hole binary) plotted as a function of frequency (in Hz).  The figure is adapted from \cite {Abbott-16-three-sources}.   Courtesy of the LIGO Scientific Collaboration and Virgo Collaboration.} 
  \label{fourier-domain}  
\end{center} 
\end{figure}
\medskip
\noindent
$\bullet$ \underline{Fourier domain}\\
In order to compare theoretical waveforms with experimental sensitivities, and to compute the signal-to-noise ratio it is necessary to describe the signal in the frequency domain. The Fourier transform of the two polarisation states of the wave are given by:

\begin{equation}
{\tilde h}_+(f)=A{c\over d_L}\left ({G{\cal M}_{\rm c}\over c^3}\right )^{5/6} e^{i\Phi^{\rm PN}_+(f)}{1\over f^{7/6}}\left ({1+\cos^2\iota\over 2}\right )
\end{equation}

\begin{equation}
{\tilde h}_\times(f)=A {c\over d_L}\left ({G{\cal M}_{\rm c}\over c^3}\right )^{5/6} e^{i\Phi^{\rm PN}_\times (f)}{1\over f^{7/6}}\cos \iota 
\end{equation}
with $A=\pi^{-2/3}(5/12)^{1/2}$. The phases $\Phi^{\rm PN}_+(f)=\Phi^{\rm PN}_\times(f) -\pi/2$ are as of today computed up to the 4th PN order, and can be expressed again in terms of $\nu$ and of ${\cal M}_{\rm c}(z) f$. 
 One can then compute the Fourier transform of the strain amplitude $h(t)$ (eq. \ref {h}), denoted as ${\tilde h}(f)$ (in units of Hz$^{-1}$) which includes the antenna pattern response
of the detector, and calculate the square of the signal-to-noise ratio S/N (or SNR) of a source as 

\begin{equation}
 \left ({{\rm S}\over {\rm N}}\right )^2=\int_0^\infty d\ln f\,{\vert 2 {\tilde h}(f)  \, \sqrt{f}\vert ^2\over S_{\rm noise}}
 \label{signal-to-noise ratio}
 \end{equation}

\noindent
 where $\sqrt{S_{\rm noise}}$ is the {\it spectral strain sensitivity} or {\it spectral amplitude} with dimension Hz$^{-1/2}$, and describes  the noise in the detector.
 Customarily, $\sqrt{S_{\rm noise}} $ and $2{\tilde h}(f) \sqrt{f}$ are plotted in the same diagram as
function of  frequency $f$, to describe  the sensitivity window of an interferometric experiment  jointly with the strength of the source signal. The signal from a coalescing binary sweeps through the frequency domain as illustrated in 
 Figure~\ref{fourier-domain}, where  for the first time we can plot the signals from the  three "real" sources (GW150914 and GW151226 mentioned above, plus a third candidate signal LVT151012) detected in the Hanford and Livingston interferometers above the sensitivity curves.

 \medskip
\noindent
$\bullet$ \underline{Chirping binaries as standard sirens}\\
Chirping binaries are {\it  standard sirens}, in that the measurement of their gravitational wave signal can determine the source absolute luminosity distance. This is the analogue of the {\it standard candles} of the electromagnetic universe, but differently from electromagnetic
observations, where one has to resort to cross-calibration of multiple distance indicators, for the gravitational wave sources  the only calibration is the assumption that general relativity describes the binary waveform.
We notice that the luminosity distance is a direct observable, if one can measure {\it both} polarisations $h_+$ and $h_{\times}$ 
separately as well as ${\dot f}_{\rm obs}$ directly from the observations. 
On one side the ratio of the two polarisations  $h_+/h_{\times}$ gives the inclination angle $\iota$ as both carry the same dependence on the chirp mass,
and by measuring the frequency derivative $\dot f_{\rm obs}$ imprinted in the phase of the chirping signal  one can infer the chirp mass (and at higher PN orders 
the individual masses). Then, all parameters in the expressions
of $h_+$ and $ h_{\times}$ are fixed except $d_L(z)$ which can be measured directly. This defines a standard siren. 
Once $d_L$ is determined, one can then infer the redshift of the source, given a cosmological model. But the measure of $d_L$ is slightly more complex.

A single detector can not measure $h_+$ and $h_\times$ independently, as it  measures a linear
combination of the two polarisation states as indicated in ({\ref{h}). Measuring the amplitude of the wave gives a combination of the angles $(\alpha,\delta,\psi,\iota)$ and $d_L$, even when
the redshifted chirp mass is known with sufficient precision.
Although ${\cal M}_{\rm c}$ decouples from the amplitude, the luminosity distance,
position, and orientation angles remain highly coupled and can be determined with overall fractional accuracy of 1/signal-to-noise.
One measures an effective luminosity distance $d_{L,\rm eff}=d_{L}/{\cal F}$ where ${\cal F}\equiv(F_+^2(1+\cos^2 \iota)^2+4F_\times^2\cos^2\iota)^{1/2}$. For non spinning binaries the signal is characterised by nine parameters: $M,\nu,t_{\rm coal},\Phi_0,\alpha,\delta,\psi,\iota, d_{\rm L}$.  When the phase is known to a high order PN, the masses $M$  and $\nu$, and the constant $\Phi_0,t_{\rm coal}$ can be determined. A network of three non-colocated interferometers can then measure three independent combinations of the polarisations and two time delays, and hence $d_{L}.$ 
In absence of non colocated multiple detectors one can exploit  the rotation of the interferometer around the Sun,  as this is the case for a space born
interferometer with the minimal number of links. 
With this procedure, the source position can  be inferred to
within a few/several square degrees in the best cases, giving information
about the source's distance and inclination \cite{Nissanke13}.   
A way to reduce these degeneracies
is to measure the event electromagnetically.  An electromagnetic signature, should it exist, 
can pin point  the event's position far more accurately than
gravitational waves alone. In this circumstance, the position angles decouple and multiple detectors
allow for a precise determination of the source distance \cite{Nissanke13}.
The remarkable fact is that when used in concomitance with an independent measurement of the redshift of the source, standard siren observations provide information on the luminosity distance - redshift relationship, hence an independent test of the current cosmological paradigm \cite{Schutz86-hubble}.  This is shown in Figure
\ref{cosmography} for the case of MBHCs.

\subsection{Spin effects in black hole binary inspirals}}\label{spin}
The formalism and the results described  so far  considered inspiralling  point particles characterised by their masses 
$m_1$ and $m_2$, only. But astrophysical objects are endowed by rotation, and it is necessary to evaluate the role of
their spins ${\bf S}_1$ and ${\bf S}_2$  in their dynamics \cite{Blanchet14}. Spins introduce spin-orbit and spin-spin couplings (i.e., the ``dragging of inertial frames'' by the bodies' spins)  affecting the binary dynamics, and consequently the phase and amplitude evolution of the gravitational signal during the inspiral, merger and ringdown.
In general, misaligned spins cause the orbital angular momentum to precess, modulating the emitted gravitational wave. Spin misalignment  increases the parameter space of templates needed to detect gravitational waves via matched filtering but also breaks degeneracies between estimated parameters in detected events. 


Spin induced effects  are important when the dimensionless spin parameter $s_{\rm spin}\equiv c\vert {\bf S}\vert/Gm^2$ of one or of the two binary components is fairly large. This is the case for black holes
which can be maximally spinning.  (We recall that massive black holes in galaxies can attain high values of their spin
due to accretion.) 
Neutron stars are unlikely to spin fast enough to drive interesting effects when they are in the detector band
during the inspiral phase.
To show this, consider the moment of inertia of a neutron star $I_{\rm NS}=(2/5) k_I\, M_{\rm NS}R_{\rm NS}^2$, where $k_I\sim 0.7-1$ measures the extent to which the  mass distribution inside the neutron  star is centrally condensed compared to a uniform sphere. For 
a neutron star with $M_{\rm NS}=1.4\msun$,  radius  $R_{\rm NS}=12$ km, and spin period $P_{\rm NS}$ of 10 ms, as observed in double neutron star binaries, the Kerr parameter 
\begin{equation}
s_{\rm spin}^{\rm NS}={c\over G}{I_{\rm NS}\over M^2_{\rm NS}}{2\pi\over P_{\rm NS}} 
\end{equation} is only $\sim 0.06 k_I$.  

\medskip
We focus here on spinning black holes only, and on
precession effects on the binary orbital angular momentum and spins.
There are astrophysical mechanisms \cite{Bardeen75,Perego09} that tend to align the black hole spins with the orbital angular momentum
during the early phases of binary evolution in gaseous circum-binary discs \cite{Dotti10,Bogdanovic07,Miller13}, but
 the alignment may be partial or may depend on the environment, so that spin-induced precession effects are astrophysically  important during the inspiral
 and merger of the two black holes\cite{Schnittman04,Kesden10,Gerosa15}.

If gravitational wave losses can be neglected, 
the total angular momentum  ${\bf J}$ of the binary, 
defined as  ${\bf J}={\bf S}_1+{\bf S_2}+ {\bf L}$,  is conserved together with  the modulus of $\bf L$ and  the moduli of the two  spins $s_{\rm spin,1 }$ and $s_{\rm spin, 2}$. Under these conditions the PN equations are  invariant under a re-scaling of the total mass $M$, leaving as free parameter only the symmetric mass ratio $\nu$, or equivalently $q=m_2/m_1<1$. 
%
%
 %

For a circular binary, to 2PN order in the spins, the evolution equations read \cite{Schnittman04,Kesden10}
\begin{equation}
{\dot {\bf S}}_1={\bf \Omega}_1\times {\bf S}_1,\qquad {\dot {\bf S}}_2= {\bf \Omega}_2\times {\bf S}_2,
\label{precessionspin}
\end{equation}
where 
\begin{equation}
{\bf \Omega}_1={G\over c^2 a^3}\left (2+{3\over 2}q -{3({\bf S}_2+q{\bf S_1})\cdot {\bf L}\over 2L^2}\right){\bf L} +{\bf S_2}
\label{omega1}
\end{equation}
\begin{equation}
{\bf \Omega}_2={G\over c^2 a^3}\left (2+{3\over 2}q -{3({\bf S}_1+q^{-1}{\bf S_2})\cdot {\bf L}\over 2L^2}\right){\bf L} +{\bf S_1}
\label{omega2}
\end{equation}
are the spin precession frequencies averaged over the orbital period, and
${\bf L}=\nu M(GMa)^{1/2}\, {\hat{\bf L}}$  the Newtonian angular momentum vector (with ${\hat{\bf L}}$ the unit vector indicating the direction). 
Equations (\ref{precessionspin},\ref{omega1},\ref{omega2})
imply that the direction of ${\hat {\bf L}}$  precesses according to 
 \begin{equation}
{\dot{\hat  {\bf L}}}=-\left ( \nu M(GM)^{2/3}/(\pi f)^{1/3}\right )^{-1}\left ({\dot {\bf S}}_1+{\dot {\bf S}}_2
\right ).
\label{precessionorb}
\end{equation}
Equation (\ref{precessionorb})
neglects the loss of orbital angular momentum by gravitational radiation and this is correct as long as the precession timescale 
\begin{equation}
\label{tau-prec}\tau_{\rm prec}\equiv\ 2\pi \Omega^{-1}\approx {2\pi c^2a^{5/2}\over G^{3/2}\nu M^{3/2}}
\end{equation}
is shorter than the timescale for gravitational wave emission $\tau^{\rm circ}_{\rm coal}$, that can be written in terms of the binary separation $a$ as
\begin{equation}
\tau^{\rm circ}_{\rm coal}={5\over 256}{c^5\over G^3}{a^4
\over \nu M^3}.
\end{equation}
%
Equations (\ref{precessionspin},\ref{omega1},\ref{omega2}) and ({\ref {precessionorb})
imply that highly spinning black holes
can change their spin orientation prior to merging. (Interestingly, it has been shown that the spin of a black hole may totally flip direction along the orbital angular momentum during the latest inspiral phase \cite{Lousto16flip}.)

A close inspection of the above equations
\cite{Schnittman04,Kesden10,Lousto14-spin}
reveals the existence of  subset  configurations  where all  the three vectors are locked in a plane as they jointly precess around ${\bf J}$ at the same rate.
These configurations are often referred to as "spin-orbit resonances" and are a consequence of the hierarchy  in the three timescales, the orbital time $\tau_{\rm orb}=2\pi f^{-1}_{\rm K}\ll \tau_{\rm prec}\ll \tau_{\rm coal}^{\rm circ}$.  
When the loss of energy and orbital angular momentum
through radiation reaction are included, systems can eventually be captured into these resonance orientations
\cite{Schnittman04}, thus reducing the range of precession frequencies. 

Studies by \cite{Kesden10} further indicate that
spin distributions that are initially partially aligned with the orbital angular momentum
can be distorted during the PN inspiral. Spin precession tends to align
(antialign) the binary black hole spins with each other if the spin of the more massive black hole
is initially partially aligned (antialigned) with the orbital angular momentum, thus increasing (decreasing) the average final spin. Spin precession is stronger for comparable-mass binaries and could produce significant spin alignment before merger for both supermassive and stellar origin black hole binaries \cite{Kesden10}. 
But in the case of quasi-circular binaries there are solutions which are unstable to large misalignments when the spin of the higher-mass black hole is aligned with the orbital angular momentum and the spin of the lower-mass black hole is antialigned.  This can occur in the strong field regime, near the merger \cite{Gerosa15}.
 It is therefore clear that the re-orientation of spins during the end of the inspiral and merger has important implication, and  it affects the final spin of the new black hole and the extent of the gravitational recoil described in the next paragraph.

\medskip

\subsection{Gravitational recoil}\label{recoil}
Merging black hole binaries radiate net linear momentum and the newly formed black hole receives a {\it  gravitational recoil}, acquired near the time of formation of the common horizon of the merging black holes.
In the framework of NR, it is now possible to obtain precise estimates
of ${\bf v}_{\rm recoil}$ as a function of the black hole parameters.

Gravitational recoil emerges when the two black holes are not symmetric. 
The asymmetry can be due to unequal masses, unequal spins, or a combination of the two. A non spinning black hole binary radiates net linear momentum if the component masses are not equal, and the maximum recoil  is of $\sim 175 \kms$  when the mass ratio is $q
\sim 0.195$ \cite{Lousto08-kick}. The complementary case, when the black holes have equal masses but unequal spins, first reported in \cite{Lousto08-kick}, leads to a maximum possible in-plane  recoil velocity of
$\sim 460 \kms$. This occurs when the spins have equal-amplitude, and are anti-parallel with respect to the orbital angular momentum direction.
But generic binaries with in-plane spin components may lead to much higher recoil velocities. 
Numerical relativity experiments find that the recoil normal to the orbital plane (due to spin components lying in the orbital plane) can be larger than the in-plane recoil originating from either the unequal-masses or the spin components normal to the orbital plane. Recoil velocities of nearly $4000\kms$ (known as super-kicks) arise
when equal-mass maximally spinning black holes merge with spins in the orbital plane equal in magnitude and opposite in direction.
 But even larger recoil velocities, of up to $\sim 5000\kms,$ emerge in "hang-up" binaries 
 with equal-mass, equal-spin magnitudes
having both spins forming an angle of $\sim 50^{\rm o}$ with the orbital angular 
momentum and in-plane components anti-aligned (see Fig. 1 \cite{Lousto11hangup}). 

Independent recoil velocity calculations have confirmed these behaviours, and empirical formulae have been derived to match the numerical results.
To illustrate the dependence of ${\bf v}_{\rm recoil}$ on the spins and mass ratio we report a handy formula for the recoil, referring to \cite{Lousto08-kick} for details:
\begin{equation}
{\bf v}_{\rm recoil}(\nu, {\bf S}_1,{\bf S}_2)=v_m\,{\bf e}_1+v_\perp(\cos\xi \,{\bf e}_1+\sin \xi\,{\bf e}_2)+v_{\parallel}{\hat {\bf L}}
\end{equation}
where $\perp$ and $\parallel$ refer to components perpendicular and parallel to the orbital angular momentum unit vector ${\hat {\bf L}}$,  $\xi$ measures the angle between the unequal mass and spin contribution to the recoil velocity in the orbital plane, 
\begin{equation}
v_m=A\nu^2{(1- q)\over (1+q)} [1+B\nu]
\end{equation}
is the in-plane component of the recoil velocity due to the asymmetry induced by the different masses carried by the black holes,  
\begin{equation}
v_\perp= H{ \nu^2\over (1+q)} [(1+B_H\nu)({\bf S}_2^{\parallel}-q {\bf S}_1^\parallel) + H_s{(1-q)\over (1_q)^2}({\bf S}_2^{\parallel}+q^2 {\bf S}_1^\parallel)]
\end{equation}
the additional  in-plane component of the recoil  related to both asymmetries in the spin and mass, and
finally 
\begin{equation}
v_\parallel= K{\nu^2\over (1+q)}[(1+B_K\nu)\vert {\bf S}_2^{\perp}-q {\bf S}_1^\perp\vert\cos\Theta_1 + K_s{(1-q)\over (1_q)^2}\vert {\bf S}_2^{\parallel}+q^2 {\bf S}_1^\parallel\vert \cos\Theta_2]
\end{equation}
the recoil velocity parallel to ${\hat {\bf L}}$.
It is useful to  notice how the perpendicular (parallel) components of the recoil couple with the parallel (perpendicular) components of the spin vectors.
In the above equations, $B,B_H,B_K,H_s,K_s$ are dimensionless constants (see \cite{Lousto08-kick}) derived from numerical relativity simulations, $\Theta_1$ and $\Theta_2$ are angles between the in-plane component of the total spin vector of the two black holes and the infall direction at merger, and ${\bf e}_1$ and ${\bf e}_2$ are unit vectors in the orbital plane and mutually orthogonal. The current best estimates for the dimensional constants are :  $A = 1.2 \times  10^4 \kms$, 
$H = (6.9\pm0.5)\times 10^3 \kms$ and $ K = (6.0\pm 0.1)\times 10^4\kms$ and  $\xi=145^\circ$. 

Recoil velocities have important implications in astrophysics, as the merger remnants may leave permanently or temporarily the sites inhabited by their
progenitor binaries. For stellar origin coalescences, this occurs when $v_{\rm recoil}$ exceeds the escape speed
form the parent star cluster, typically $\approx 50\kms$. For 
massive black holes in merging galaxies, the newly formed black hole leaves the galaxy if its recoil 
velocity exceeds $\approx 800 \kms$. For lower velocities, the kicked massive black hole can return
and sink back at the centre of the host, preserving a level of Brownian motion after having dissipated its kinetic energy via dynamical friction 
against stars or/and gas \cite{Gualandris08}. Off-set black holes can be seen in an active phase
since they may drag gas with them which can be accreted. Off-set AGN 
 clearly pinpoint, albeit indirectly, places where a merger has occurred, and a number of candidates have been observed, in large AGN surveys \cite{Komossa12}. 
 Super-kicks  naturally prompt the question of why supermassive black holes are often seen at the centres of their host galaxies. There are indications that super-kick configurations
 may not be the favourite arrangements as accretion torques tend to align the spins to the orbital angular momentum suppressing the in-plane components
 of the spins and reducing the extent of the recoil \cite{Dotti10,Bogdanovic07}. 
 But, hang-up configurations  may lead to non-negligible probabilities of  recoils of several  $1000 \kms$ in real astrophysical sources.  In general, spin-orbit resonances are likely to populate specific portions of the parameter space, and depending on which of the two black holes has
 more aligned spin with ${\bf L}$, kicks are either suppressed or enhanced. At present which of these scenarios is more common is observationally undetermined.

\subsection {Modelling the final spin and mass of coalescing black holes}\label{end-states}
A black hole binary coalescence is an "elementary" process that takes as input two black holes of initial mass $m_1$ and $m_2$ 
and spin vectors ${\bf S}_1$ and ${\bf S}_2$ (with ${\bf S}_{\rm i}=s_{\rm i}Gm^2_{\rm i}{\hat{\bf s}_{\rm i}}/c$, and $i=1,2$),  Newtonian energy $E_{\rm bin}$ and angular momentum $ L$  (the choice of ${\hat {\bf L}}$ is arbitrary and defines the $z$-axis) 
defined at a far-out initial separation  $a_{\rm in}$ and produces a new black hole of mass $M_{\rm fin}$, 
spin ${\bf S}_{\rm fin}=s_{\rm fin}GM^2_{\rm fin}{\hat{\bf s}_{\rm fin}}/c$ and  recoil speed  ${\bf v}_{\rm recoil}$.  
 The full mapping
between the initial and final states can be obtained by solving exactly the Einstein's field equations which incorporate the law of energy-momentum conservation. 
It is important to notice that a final state can correspond to different initial states, given that
the ten initial internal degrees of freedom are mapped into the seven of the new black hole that forms.

Astrophysical  black holes in binaries are expected to span a wide parameter space, too vast to be explored entirely.
At a lower level one can construct a mapping between 
a minimal set of initial conditions involving the mass ratio $\nu$ and spin vectors ${\bf S}_1$, ${\bf S}_2$, and   ${\bf S}_{\rm fin}$. 
In this way one can capture some important properties of the end-states, once the directions of the spins relative to the orbital angular momentum at a "sufficiently" far-out distance are specified. 
This mapping is degenerate as different initial states can lead to the same final state, also in this lowered parameter space \cite{Rezzolla09-final-states,Lousto14-spin,Rezzolla16}.
The binary's unit vector $\hat{\bf L}$ fixes a direction
in space and key angles are  $\cos \theta_1={\hat {\bf s}}_1\cdot{\hat{\bf L}}$, $\cos \theta_2={\hat {\bf s}}_2\cdot{\hat{\bf L}}$ , and $\cos \gamma\equiv {\hat{\bf s}_{\rm 1}}\cdot {\hat{\bf s}_{\rm 2}}.$ 

For generic binaries, it is possible to derive analytical fitting formulae, if one assumes to a first approximation that: (i) the mass-energy radiated in gravitational waves is negligible (i.e.,  $M_{\rm fin}=M$); (ii) the norm of the two spin vectors and of the vector ${\bf l}\equiv {\bf S}_{\rm fin} - ({\bf S}_1+{\bf S}_2)={\bf L}-{\bf J}_{\rm rad}$ 
(interpreted as being the residual orbital angular momentum contributing to ${\bf S}_{\rm fin}$)  do not
depend on the binary separation; (iii)
the final spin is parallel to the total angular momentum ${\bf J}$ defined at far distances (this amounts to assuming that, according to PN theory, the radiated angular momentum ${\bf J}_{\rm rad}\parallel {\bf J},$ and this is motivated by the fact that precession of ${\bf L}$ around ${\bf J}$ averages the gravitational wave emission orthogonal to ${\bf J}$);(iv)  the angles $\gamma$, $\theta_1$ and $\theta_2$ are locked in space; (v) when the initial spins are equal and opposite ($\cos\gamma=-1$), and the masses are equal, the final black hole has spin equal to that of a non spinning binary.  With these ansatz, a number of predictions can be made that are in close agreement with numerical simulations.
The first is that for equal mass binaries, aligned ($s_{\rm spin}>0$) or anti-aligned  ($s_{\rm spin}<0$) unequal spins\cite{Rezzolla16}, the final spin can be expressed as 
\begin{equation}
s_{\rm spin}=p_0+p_1(s_{\rm spin,1}+s_{\rm spin,2}) + p_2(s_{\rm spin,1}+s_{\rm spin,2})^2
\label{spin-final-equalmass}
\end{equation}
where the coefficients $p_0\simeq0.6869$, $p_1\simeq 0.1522$ and $p_2\simeq -0.0081$ are obtained from independent  fits to distinct data sets,
and where  $s_{\rm spin}$ takes a negative value when antialigned with respect to ${\hat {\bf L}}$.
Equation (\ref{spin-final-equalmass}) can be interpreted as power series of the initial spins and indicates that its zeroth-order term $p_0$ can be associated 
with the dimensionless orbital angular momentum not radiated in gravitational waves by a non spinning binary. This value is close to the most accurate measurement 
of the final spin of two non spinning black holes  $ s_{\rm fin}=0.68646\pm 0.00004$. 
The first order term $p_1$ can be seen as the contribution 
to the final spin from the spins of the two black holes and their spin-orbit coupling. The last, second order term $p_2$ accounts for the contribution of the spin-spin couplings.
Interestingly $s_{\rm spin}=p_0$ is the final spin also for equal-mass binaries with ${\bf S}_1=-{\bf S}_2$.

Recently, by combining information from the test-particle limit,
perturbative/self-force calculations, the PN dynamics, and an
extensive set of NR simulations collected from the literature,
Hofmann et al. \cite{Barausse-spin-16} have constructed a novel formula for the final spin from
the merger of quasi-circular black hole binaries with arbitrary mass
ratios and spins, and we defer the reader to \cite{Barausse-spin-16} for further details.

The above equations give a prescription to calculate the final spin of the new black hole. But how can the information on the spin be extracted
from the signal?
The measurements of the black hole individual spins from the signal is hampered by partial degeneracies in the 
phase evolution which is a function of
the black hole masses and the individual spin vectors. 
During the inspiral, for
a binary with spins aligned with ${\bf L}$, the spin influence on the evolution of the phase  arises in a weighted 
combinations  of the
spins projected on ${\bf {\hat L}}$
\begin{equation}\chi_{\rm eff}={c\over G}\left ({{\bf S}_1\over m_1}+{{\bf S}_2\over m_2}\right )\cdot {{\hat {\bf L}}\over M}
\label{chi-spin-parallel}
\end{equation}
which takes values between -1 (for  maximally rotating black holes when both have  spins antialigned with the respect to the orbital angular momentum) and +1 (aligned spins). Equation (\ref{chi-spin-parallel}) implies degeneracy in the estimate of the individual spins, in particular for equal mass binaries.

When the spins have also components lying in the orbital plane, their in-plane projections rotate within the orbital plane at different velocities and the
signal acquires further structure. 
In this case the mean influence of the four in-plane spin
components on the phasing  can be combined
into a single effective precession spin parameter 
 \begin {equation}
 \chi_{\rm p}={c\over (2+3q/2) Gm_1^2}{\rm max}[ (2+3q/2)S_{1\perp}, (2+3/2q)S_{2\perp}]
 \end{equation}
 where $ \chi_{\rm p}=1$ (0) corresponds to a binary with maximum (null) level of precession.

\medskip

For GW150914,  it has been possible
to measure the spin of the new black hole $s_{\rm fin}=0.67^{+0.05}_{-0.07}$ and $\chi_{\rm eff}=-0.06^{+0.17}_{-0.18}$,  and pose a limit on $\chi_{\rm p}<0.81$ (at 90\% probability level)  \cite{Abbott-1,LIGO16-spin}.
 The spins of the black holes prior to coalescence have been constrained to value 
$s_{\rm spin,1}=0.31^{+0.51}_{-0.27}$ and $s_{\rm sipn,2}=0.39 ^{+0.50}_{-0.34}$ \cite{LIGO16-spin}. 
For GW151226, the  weaker signal does not allow an estimate of $s_{\rm fin}$, but  from the longer-duration inspiral seen in GW151226 
it has been possible to pose  a lower
limit on $s_{\rm spin,1}>0.2$ at the 99\% credible level.  Only weak constraints have been placed on $\chi_{\rm p}$ suggesting that the data are no informative on
the level of precession in the binary prior to coalescence \cite{Abbott-GW151226}.

\medskip
As far as the radiated energy ${E^{\rm rad}_{\rm gw}}$ and final mass $M_{\rm fin}$ are concerned,  the combined approach that uses fitting formulae from the PN expansion calculation calibrated
with NR experiments has led to predictions on the final mass \cite{Rezzolla16}. 
Two regimes are described here as they guide intuition: (i) the test particle limit (in which $m_2\to 0$), and (ii) the case of equal-mass binaries with spins aligned or anti-aligned with respect to the orbital angular momentum ${\bf L}$.
 
 In the {\it test particle} limit, the energy radiated by $m_2$ during the inspiral onto the central black hole of mass $m_1$ (with $m_1\gg m_2$), from large distances down to $R_{\rm isco}$ is
  \begin{equation}
 {E^{\rm rad}_{\rm gw}}=[1-{\tilde  E}_{\rm isco}(s_{\rm spin})]\nu M c^2 +O(\nu)
 \end{equation}
 where  ${\tilde E}_{\rm isco}(s_{\rm spin})$ the binding energy per unit mass at the innermost stable circular orbit 
  \begin{equation} 
 {\tilde E}_{\rm isco}(s_{\rm spin})=\left [1-2/(3 R^{\rm spin}_{\rm isco, \pm})\right ]^{1/2}
 \end {equation}
 where $R^{\rm spin}_{\rm isco, \pm}$ is the coordinate radius of the innermost stable circular orbit in the Kerr metric, equal to  $6GM/c^2$ for a non-spinning black hole, and $GM/c^2$ ($9GM/c^2$)
 for a maximally rotating black hole and a test particle on a co-rotating (counter-rotating) orbit. For $s_{\rm spin}=0$, ${\tilde E}_{\rm isco}=\sqrt{8/9}$ while for $s_{\rm spin}=1$, ${\tilde E}_{\rm isco}=\sqrt{1/3}$  ($\sqrt{25/27}$) for prograde (retrograde) orbits. The maximum radiated energy is equal to $ {E^{\rm rad}_{\rm gw}}=0.423 \nu M c^2$, for
 a co-rotating orbit around a maximally rotating Kerr black hole, neglecting the plunge phase and high order corrections.

\begin{figure}[!t]
\begin{center}
\includegraphics[width=.9900\textwidth]{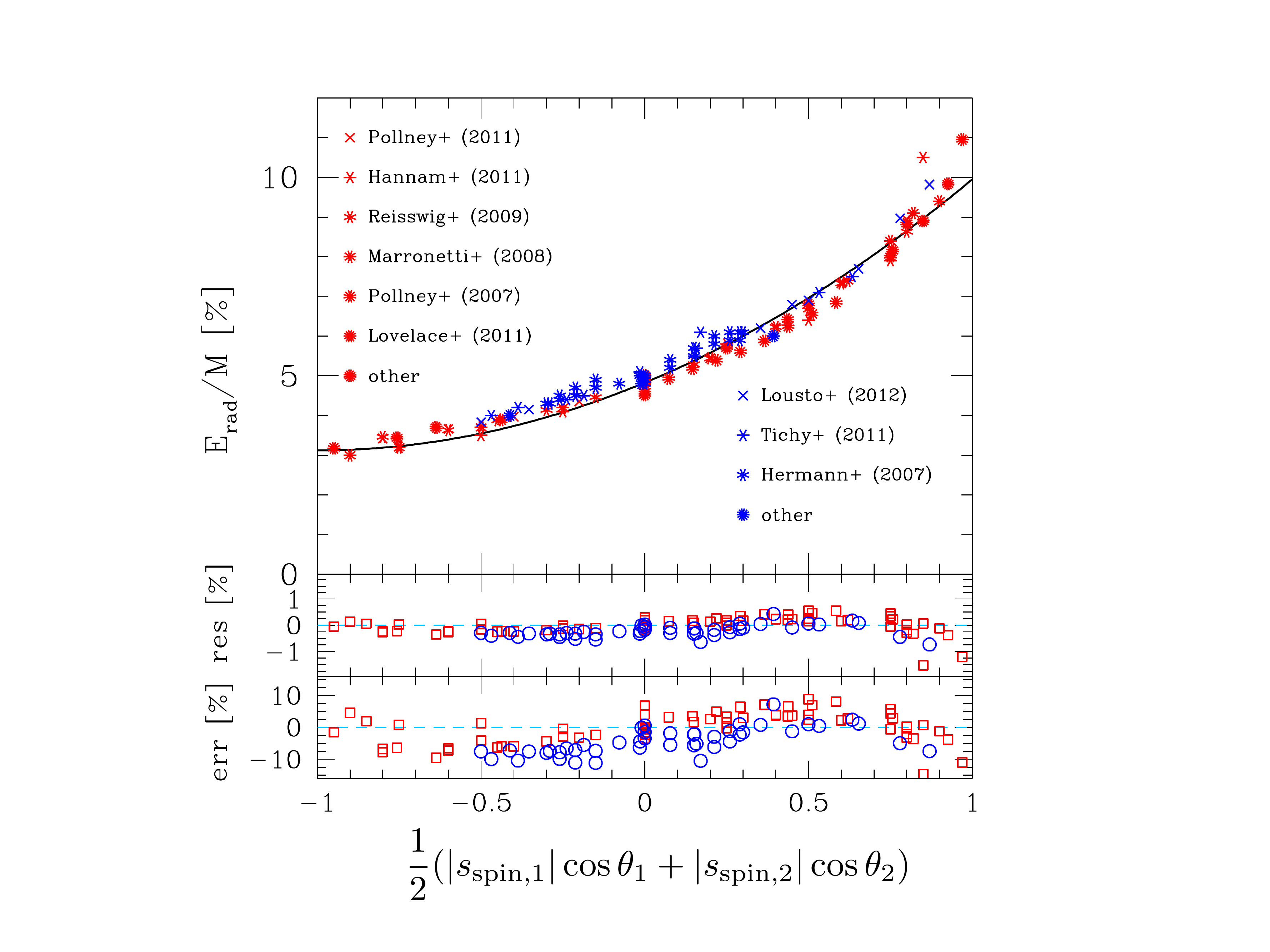}
\caption{Radiated energy, $E_{\rm gw}^{\rm rad}$ per unit mass  as a function of the total
spin of the system projected along the orbital angular momentum, for  different sets of published NR simulations 
of equal mass binaries, both with
aligned/antialigned spins (in red) and for misaligned spins (in blue).The black solid line is
the fit whose analytical expression  can be found in \cite{Barausse-mass-12}. Bottom panels give the residuals
of the NR data from the fitting expression and the corresponding error
relative to  $E_{\rm gw}^{\rm rad}$.  Courtesy by E. Barausse.
\label{Radiated-mass}}
\end{center} 
\end{figure}

The case of equal-mass binaries and both spins aligned or anti-aligned with respect to the orbital angular momentum, has been explored numerically \cite{Reisswig10,Lousto14-spin}.
The energy emitted by these binaries during the inspiral (from infinite separation), merger and ringdown (computed by combining PN calculation to numerical relativity results) can be described by a polynomial fit\cite{Rezzolla16}
\begin{equation}
{E^{\rm rad}_{\rm gw}}=[w_{\rm ns}+w_{\rm  so}(s_{\rm spin,1}+s_{\rm spin,2}) +w_{\rm ss}(s_{\rm spin,1}+s_{\rm spin,2})^2]Mc^2
\label{e-rad-eq}
\end{equation}
where  again $s_{\rm spin}$ takes negative values when the spin is anti-aligned with respect to ${\bf {\hat L}}.$ The fitting coefficients 
$w_{\rm ns}=0.04827\pm0.00039$,  $ w_{\rm so}=0.01707\pm0.00032,$ and $w_{\rm ss}=w_{\rm so}/4,$ can be interpreted as the non-spinning contribution to the radiated energy (which is the largest one and which accounts for $\sim 50\%$ of the largest possible mass loss occurring when $s_{\rm spin,1}=s_{\rm spin,2}=+1$),
the spin-orbit contribution (which is $\simless 30\%$ of the largest possible loss) and spin-spin contribution (which is $\simless 20\%$ of the largest possible loss), respectively.  Equation (\ref{e-rad-eq}) reproduces the data available to within $\sim 0.005M c^2$ except for almost extremal spins where higher order terms may be needed.
This result teaches us that spin effects
 can either amplify or reduce  the amount of energy radiated away so that binaries with given $M$ and $\nu$ do not radiate away the same ${E^{\rm rad}_{\rm gw}}$.

In Figure \ref{Radiated-mass}  we show the results of NR simulations for equal mass binaries \cite{Barausse-mass-12}.
As illustrated in Figure \ref{Radiated-mass}, equal-mass, maximally spinning black holes
with spins aligned with the orbital angular momentum are the most efficient emitters of gravitational waves, with ${E^{\rm rad}_{\rm gw}}=0.0995 Mc^2.$ 
Also equal-mass non spinning binaries
lose  a considerable fraction of their mass via radiation having ${E^{\rm rad}_{\rm gw}}\sim 0.048Mc^2$, while maximally spinning binaries
with spins parallel but anti-aligned with the orbital angular
momentum have ${E^{\rm rad}_{\rm gw}}\sim 0.037Mc^2$ \cite{Reisswig10}.  
More complex fitting formulae have been derived \cite{Lousto14-spin} for generic binaries.

 %
\begin{figure}[!t]
\begin{center}
\includegraphics[width=.450\textwidth]{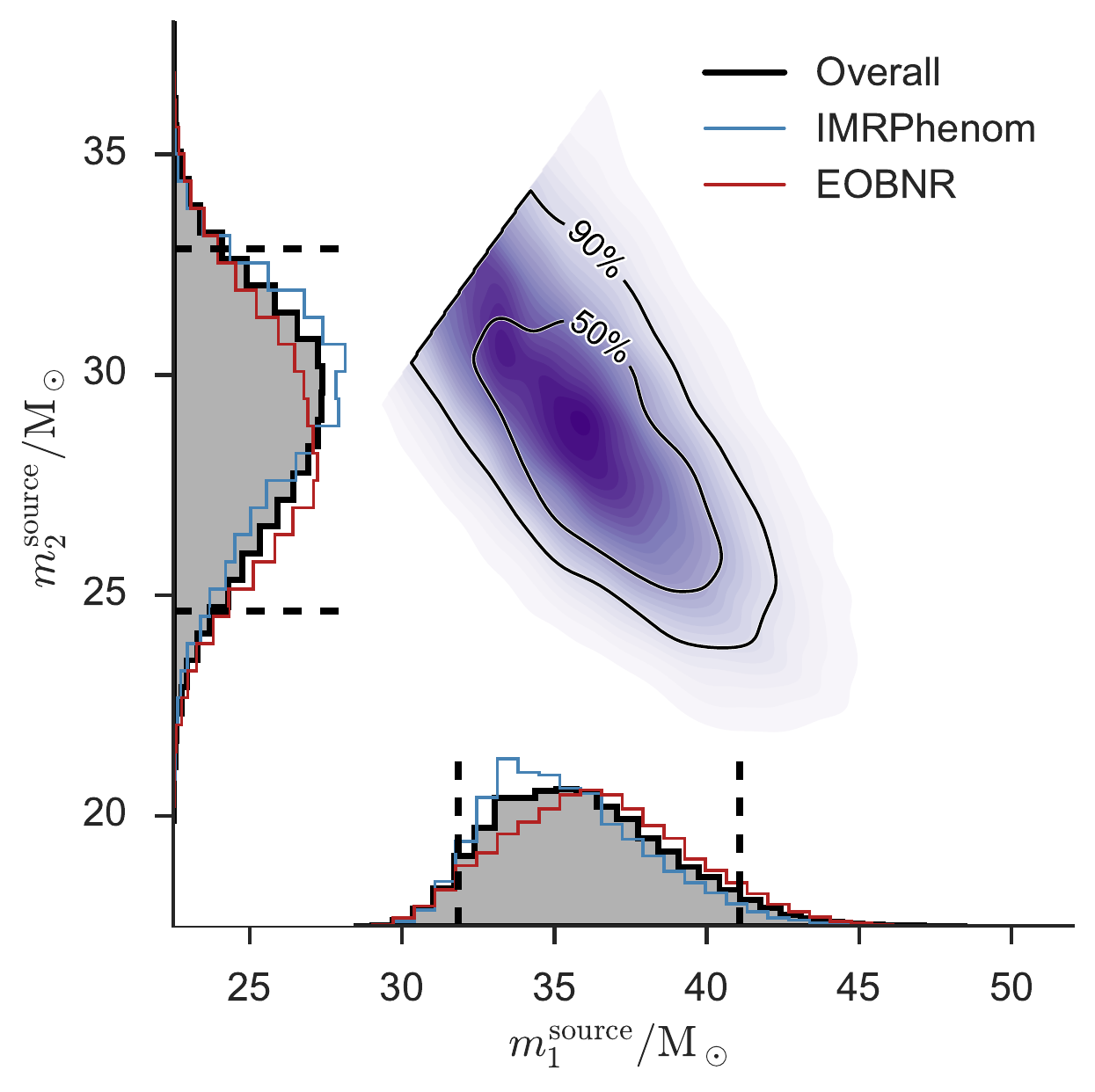}
\includegraphics[width=.450\textwidth]{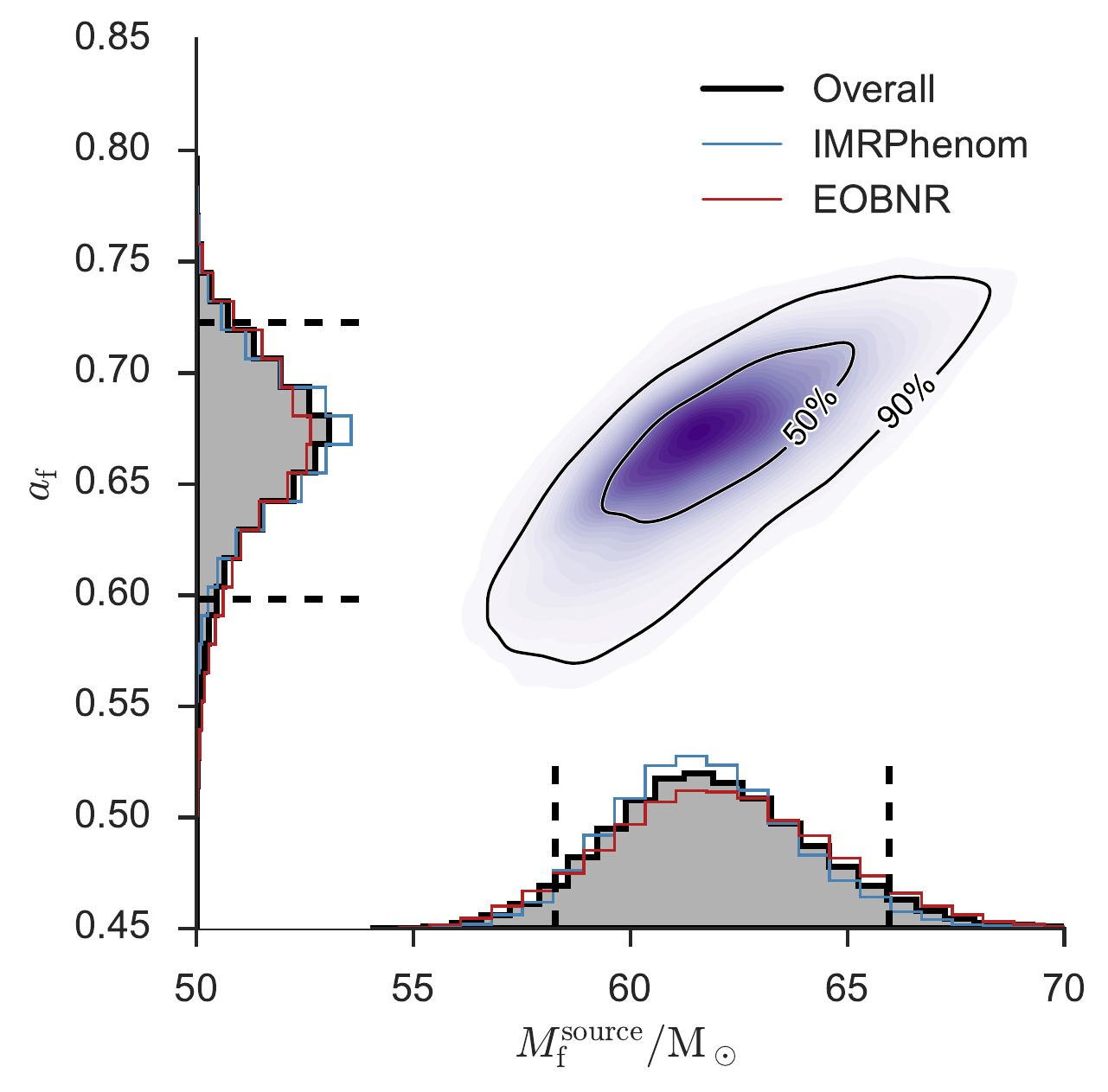}
\caption{Probability Distribution Functions  (PDFs) for the source-frame component masses $m_1$ and $m_1$ (left), and for the mass $M_{\rm fin}$ and spin 
$s_{\rm fin}$ (right) of the remnant
black hole in GW150914 \cite{Abbott-1,LIGO16-spin}.  In the 1-dimensional marginalised distributions, the figure shows  the overall (solid black), IMRPhenom (blue) and EOBNR (red) PDFs, where the acronym IMRPhenom  and EOBNR refer to the two model waveforms for spinning (non-precessing) binaries used to analyse the signal. 
Figure adapted from Abbott et al.  arXiv:1602.03840v1.  For detail we defer to \cite{LIGO16-spin}. The
dashed vertical lines mark the 90\% credible interval for the overall
PDF. The 2-dimensional plot shows the contours of the 50\%
and 90\% credible regions plotted over a colour-coded posterior
density function.   Courtesy of the
LIGO Scientific Collaboration and Virgo Collaboration.
\label{Abbott-GW150914}  }
\end{center} 
\end{figure}

Figure \ref{Abbott-GW150914} shows the Probability Distribution Functions  (PDFs) for the source-frame masses of the two black holes in GW150914, and for the mass and spin of the remnant black hole \cite{Abbott-1,LIGO16-spin}.  From these values one can infer a radiated energy of $3.0^{+0.5}_{-0.5}\msun c^2$, the majority of
which emitted at  frequencies in the LIGO sensitive band. For the second source, GW151226 we defer to \cite{Abbott-GW151226}.

\section{Waveforms and the Laws of Nature}

In this section we give a brief overview on the properties of the gravitational wave signals, which inform us on the nature of the sources and on their 
physical properties.

\begin{figure}[!t]
\begin{center}
\includegraphics[width=.850\textwidth]{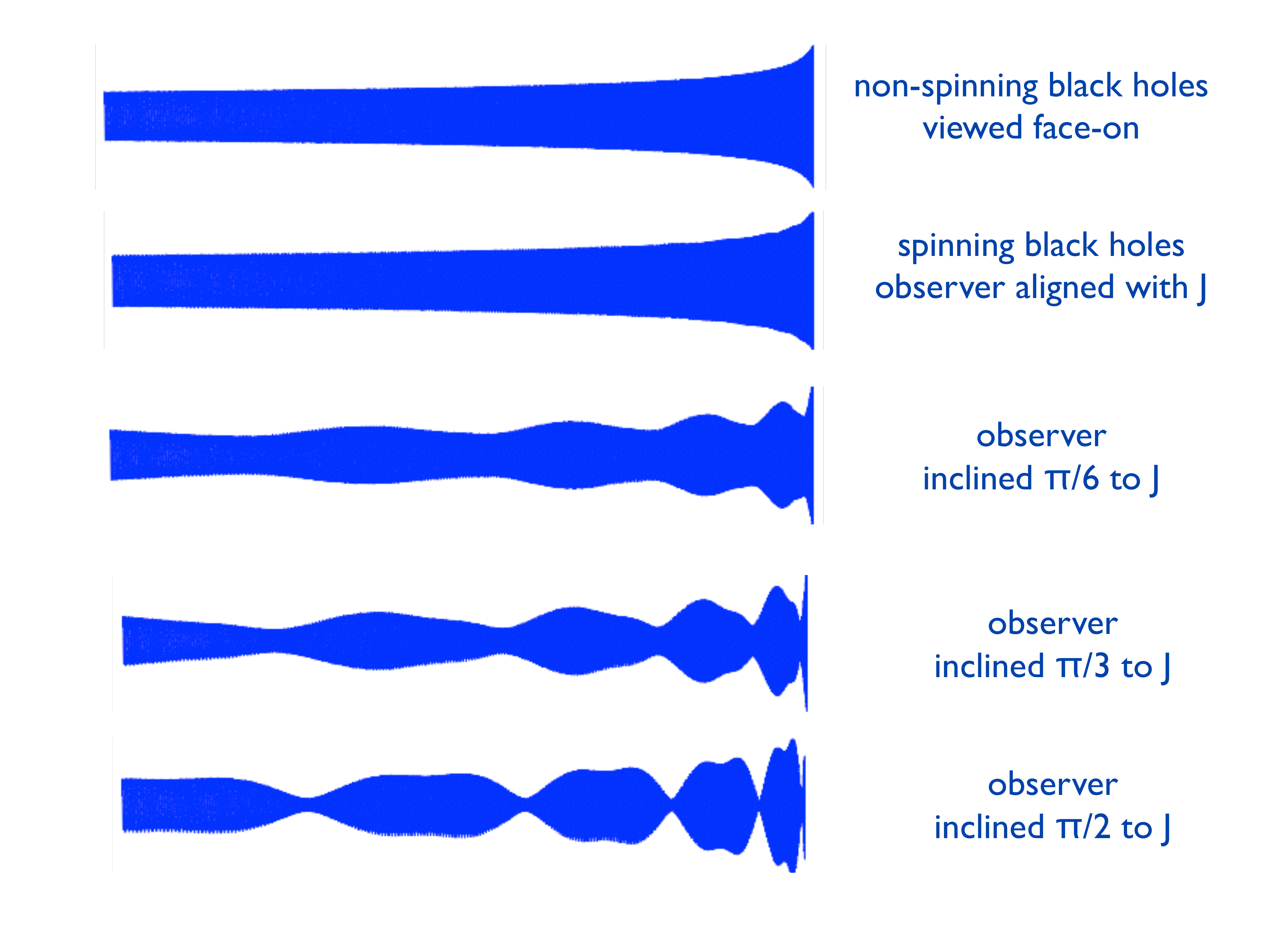}
\caption{Waveforms from a black hole binary with mass ratio $q=1/3$. The upper panel refers to a non-spinning binary
viewed face-on, i.e. with optimal orientation corresponding to the highest signal. The remaining panels refer to a spinning binary with in-plane spin $s_{\rm spin,1}=0.75$, and $s_{\rm spin,2}=0$, 
 viewed at different angles with respect to the total angular momentum ${\bf J}$.
The morphology  changes with the viewing angle, and the modulation by precession is absent when the binary is seen
along ${\bf J}$.  Adapted from M. Hannam talk, Hannover 24 May 2016 : https://gw150914.aei.mpg.de/program/mark-hannams-talk }
\label{BHwaveform-spin-morphology}
\end{center} 

\end{figure}

\subsection {Black hole binary coalescences}\label{bhb}
Black hole binaries (BH*,BH*) of stellar origin, or massive  (MBH,MBH) have universal signals as black holes are geometrical objects, according to general relativity. 
Spin-precession colours the waveform in various ways, changing the morphology.
This is illustrated in Figure \ref{BHwaveform-spin-morphology} where we show the spin-induced modulation on the gravitational wave amplitude, during the inspiral phase, as view from different orientations with respect to the observer's  viewing angle.

For an accurate measure of the physical parameters and in order to break degeneracies, one needs to know the full waveform and detect
a source with a high signal-to-noise ratio.

  The whole waveform (including the ringdown) contains precious information on the mass and spin of the new black hole. We recall that if  the object is truly a black hole
as predicated by general relativity, then the
mass and spin obtained from the different oscillation modes present in the ringdown signal should all be consistent within the measurement errors \cite{Berti15-testGR, Yunes13-review}.
Inconsistencies in the values of the mass and spin inferred in this way would be an indication of
the failure of general relativity or that the radiation was emitted from an object that is not a black hole. 
Test of consistency with the early inspiral and merging phase will also be critical, providing extremely valuable information, both on the nature
of the interacting bodies and on the properties of gravity, as fundamental interaction.

\begin{figure}[!t]
\begin{center}
\includegraphics[width=.990\textwidth]{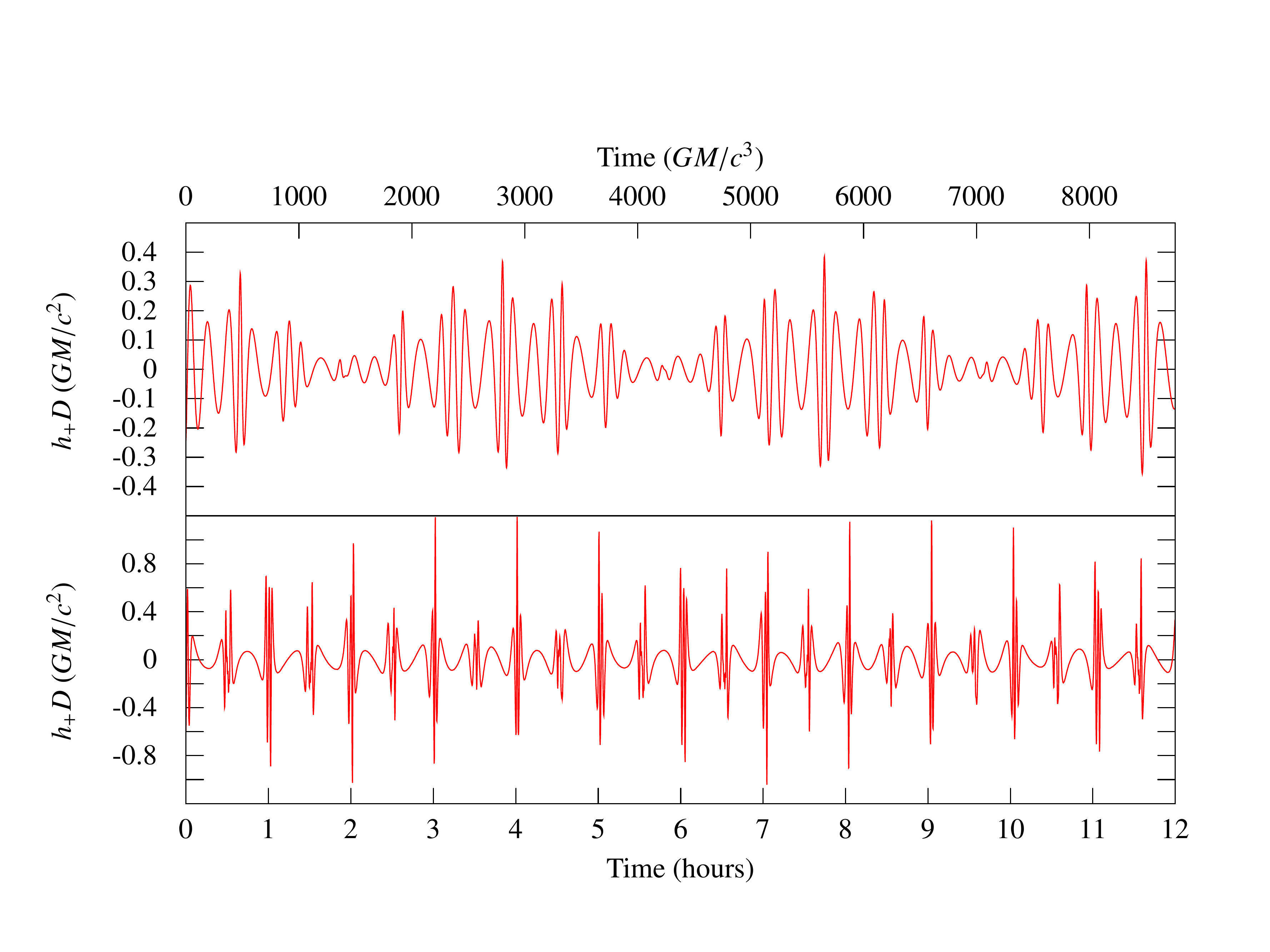}
\caption{Segments of generic EMRI waveforms \cite{GWnote13}. These are the plus-polarised waves
produced by a test mass orbiting a $10^6\msun$ spinning black hole  with $s_{\rm spin, 1}=0.9$, at distance D from the observer. Top panel: Slightly eccentric and inclined retrograde orbit modestly far from the horizon. Bottom panel: Highly eccentric and inclined prograde orbit much closer to the horizon. The amplitude
modulation visible in the top panel is mostly due to Lense-Thirring precession of the orbital plane. The more eccentric orbit in the bottom panel produces sharp spikes at each pericentre passage.}
\label{EMRI-waveform}
\end{center} 
\end{figure}

\subsection {Extreme Mass Ratio Inspirals}\label{emris}
EMRI are expected to be very clean astrophysical systems, except perhaps in the few percent of galaxies containing
accreting massive black holes, where interactions with the accretion disc could possibly affect the EMRI dynamics and waveform. EMRIs  trace geodesics in the spacetime of the massive black hole. Over a typical eLISA
observation time (months to years), EMRI orbits are highly relativistic (radii smaller than $10 \,R_{\rm S}$) and display
extreme forms of periastron and orbital plane precession due to the dragging of inertial frames by the massive black holes
spin, as illustrated in Figure \ref{EMRI-waveform}.

Given the large amount of gravitational wave cycles collected in a typical EMRI observation (${\cal N}_{\rm cycle}\sim 10^5$), a fit of the
observed gravitational waves to theoretically calculated templates will be very sensitive to small changes in the physical
parameters of the system. eLISA should be able to determine the mass of the
massive black hole to fractional accuracy of about $10^{-3}-10^{-4}$ for gravitational wave signals with an SNR of 20 and the spin with $10^{-3}$
in its absolute value. 

This level of precision suggests that we can use  EMRIs for highly precise observational tests of the {\it Kerr-ness} of
the central massive object \cite{Berti15-testGR,Gair13,Yunes13-review}. That is, if we do not assume that the larger object is a black hole, we can use gravitational
waves from an EMRI to map the spacetime of that object. The spacetime outside a stationary axisymmetric object is
fully determined by its mass moments $M_l$ and current multipole moments $S_l$. Since these moments fully characterise the
spacetime, the orbits of the smaller object and the gravitational waves it emits are determined by the multipolar structure of
the spacetime. Extracting the moments from the EMRI waves is analogous to geodesy. Black hole geodesy, also known as holiodesy, is very
powerful because Kerr black holes have a very special multipolar structure. In units with $G=c=1$, a Kerr black hole has multipole moments given by
\begin{equation}
M_l+iS_l=(is_{\rm spin})^l M_{\rm BH}^{l+1},
\end{equation}
where $M_{\rm BH}$ is the mass of the large black hole. 
Thus, $M_0=M_{\rm BH}$, $S_1=s_{\rm spin} M^2_{\rm BH}$, and $M_2=-s_{\rm spin}^2 M_{\rm BH}^3$, and similarly for all other multipole moments; they are all completely
determined by the first two moments, the black hole mass and spin. This is equivalent of stating the black hole 'no-hair'
theorem: the properties of a black hole are entirely determined by its mass and spin.  The mass moment $M_2$ will be measured with extreme accuracy by eLISA, for a signal-to-noise-ratio larger than 30.

Any inconsistency with the Kerr multipole structure could signal a failure of general relativity, the discovery of a new type of
compact object, or a surprisingly strong perturbation from some other material or object. EMRI signals can be used to
distinguish definitively between a central massive black hole and a boson star \cite{Colpi86}. In the black hole case
the gravitational wave signal "shuts off" shortly after the inspiraling body reaches the last stable orbit (and then plunges
through the event horizon), while for a massive boson star, the signal does not fade, and its frequency derivative changes
sign, as the body enters the boson star and spirals toward its centre \cite{Berti15-testGR,Gair13}.

 \subsection {Neutron star binary coalescences}\label{ns-coalescence}

Neutron stars  have a surface and are deformable bodies.  In the latest moments of the inspiral of a (NS,BH*) or (NS,NS) binary,  the tidal interaction 
becomes important.  The neutron star acquires a tidal deformation, and this affects the external gravitational field and the relative orbital motion.
Thus the deformability of each star leaves a subtle signature in the amplitude and phasing of the gravitational
wave.

The effect of the tidal interaction on the orbital motion 
and gravitational wave signal is measured by a quantity
known as {\it tidal Love number}, defined for each body in the binary. 
To guide intuition, let us treat the first body (either one of the two NS or BH*)  as a point mass
sourcing the external monopole potential  $U_{\rm ext}$, and 
focus on the tidal deformation on the second body excited by $U_{\rm ext}$.  
The deformed mass distribution can be described in terms of a quadrupolar deformation $Q_{ab}$ which turns out to be proportional to the 
external tidal field, $T_{ab} = -\partial_a\partial_b U_{\rm ext}.$
Dimensional analysis requires $Q_{ab}= -2/(3G)\kappa R^5 T_{ab}$, where $\kappa$ is the tidal Love number for the quadrupolar deformation, and $R$ the stellar radius.  The resulting external  gravitational potential, relative to the centre of mass of the deformed body, is 
\begin{equation}
U={Gm_2\over r}-{1\over 2}\left [1+2\kappa \left ({R\over r}\right )^5\right ]T_{ab}x^ax^b,
\end{equation}
the first term representing the monopole contribution, the second the gravitational potential by the point mass sourcing the tide, and the third the correction to
the gravity's field induced by the distorted mass distribution.

\begin{figure}[!t]
\begin{center}
\includegraphics[width=.70\textwidth]{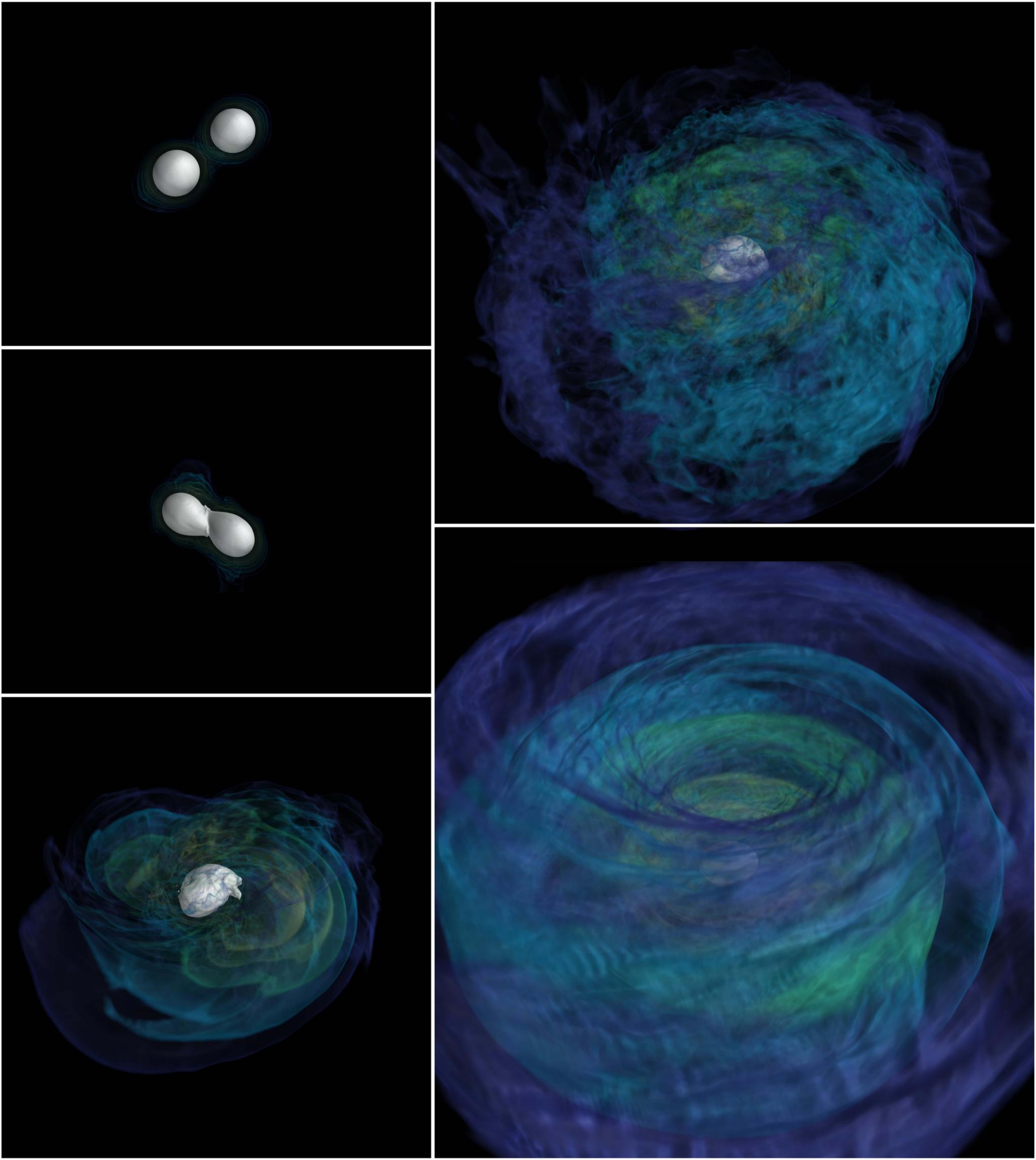}
\caption{The figure depicts a merger of an equal-mass (NS,NS) binary \cite{Giacomazzo16}. The two neutron stars (drawn in white)
spiral-in (top left panel), touch (middle left), and merge 
into a hyper-massive neutron star while ejecting debris (bottom left).
Most of the ejecta form a torus orbiting around the remnant (right
panels), while the rest escapes. Courtesy of B. Giacomazzo. 
\label{NS-movie} }
\end{center} 
\end{figure}

The formalism has been extended within the framework of general relativity \cite{Poisson09,Nagar12,Bernuzzi15} and the quadrupolar Love number $\kappa$  has been computed
for neutron stars with different EoS:  $\kappa=0.11$ for the lowest ${\cal C}=0.139$, and $\kappa=0.0647$ for the highest ${\cal C}=0.1924.$
Interestingly enough, non-spinning black holes have zero Love numbers. 
Both the compactness ${\cal C}$ and Love number $\kappa$ enter the evolution of the orbital phase, and with Advanced LIGO and Virgo it will be possible
to measure the degree of tidal polarisability, defined as $2/(3G)\kappa R^5,$ for neutron stars when the signal-to-noise ratio is sufficiently high ($\sim 16$),
opening the prospects of extracting information on the nuclear EoS from a coherent
analysis of an ensemble of gravitational wave observations of separate binary merger events \cite{Nagar12,Bernuzzi15}.

\begin{figure}[!t]
\begin{center}
\includegraphics[width=0.990\textwidth]{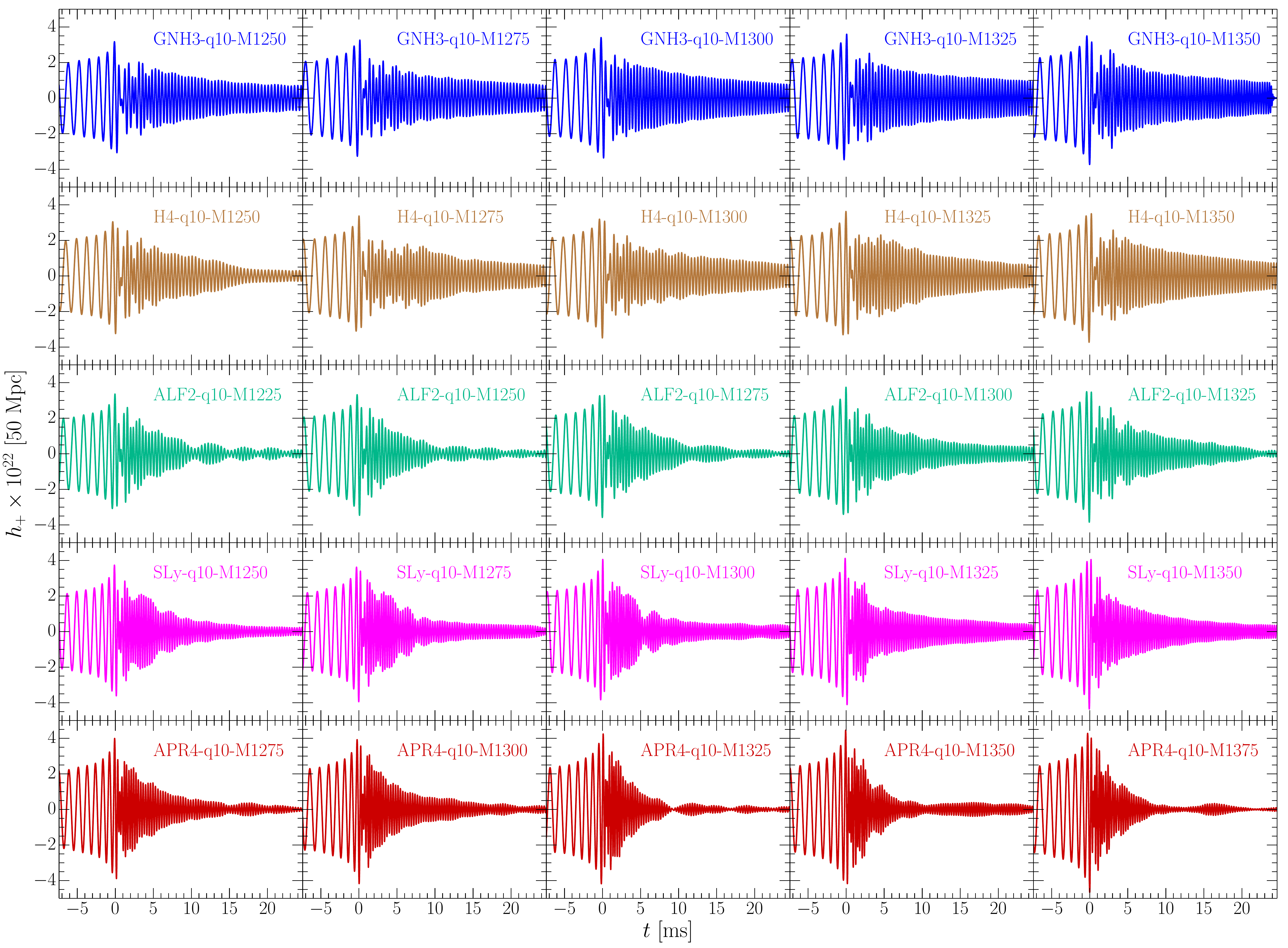}
\caption{ 
Gravitational wave signals $h_+(t)\times 10^{22}$ for sources at 50 Mpc, 
from a large suite of equal-mass (NS,NS) coalescences and for a variety of nucleonic EoS, from \cite{Baiotti-review16,Takami14}. Courtesy of L. Rezzolla. 
\label{NS-waveforms-time-domain} 
}
\end{center} 
\end{figure}

After being tidally deformed and shock heated at impact, the two neutron stars merge, and depending on the total mass of the binary, they may form 
a black hole or a hyper-massive neutron star which may later collapse into a black hole. The merger and ringdown phases can only be followed numerically, and  Figure \ref{NS-movie} 
shows the coalescence of 
two equal-mass (non magnetised) neutron stars forming a hyper-massive neutron star \cite{Giacomazzo16}.  During the late inspiral, the two stars develop
strong deformations due to the tides that they mutually exert, and after merging the highly  non-symmetric hyper-massive star is surrounded by tidal debris.

Binary neutron star mergers have been simulated, within the context of NR,
in order to explore a  variety of EoS \cite{Baiotti-review16} and to extract 
and study the morphology of the gravitational wave signal, and in particular its evolution in the frequency 
domain, in search of characteristic frequencies that may inform us bout the degree of compactness of the two 
merging neutron stars\cite{Bernuzzi15,Bernuzzi15-spectrum}.
\begin{figure}[!t]
\begin{center}
\includegraphics[width=0.80\textwidth]{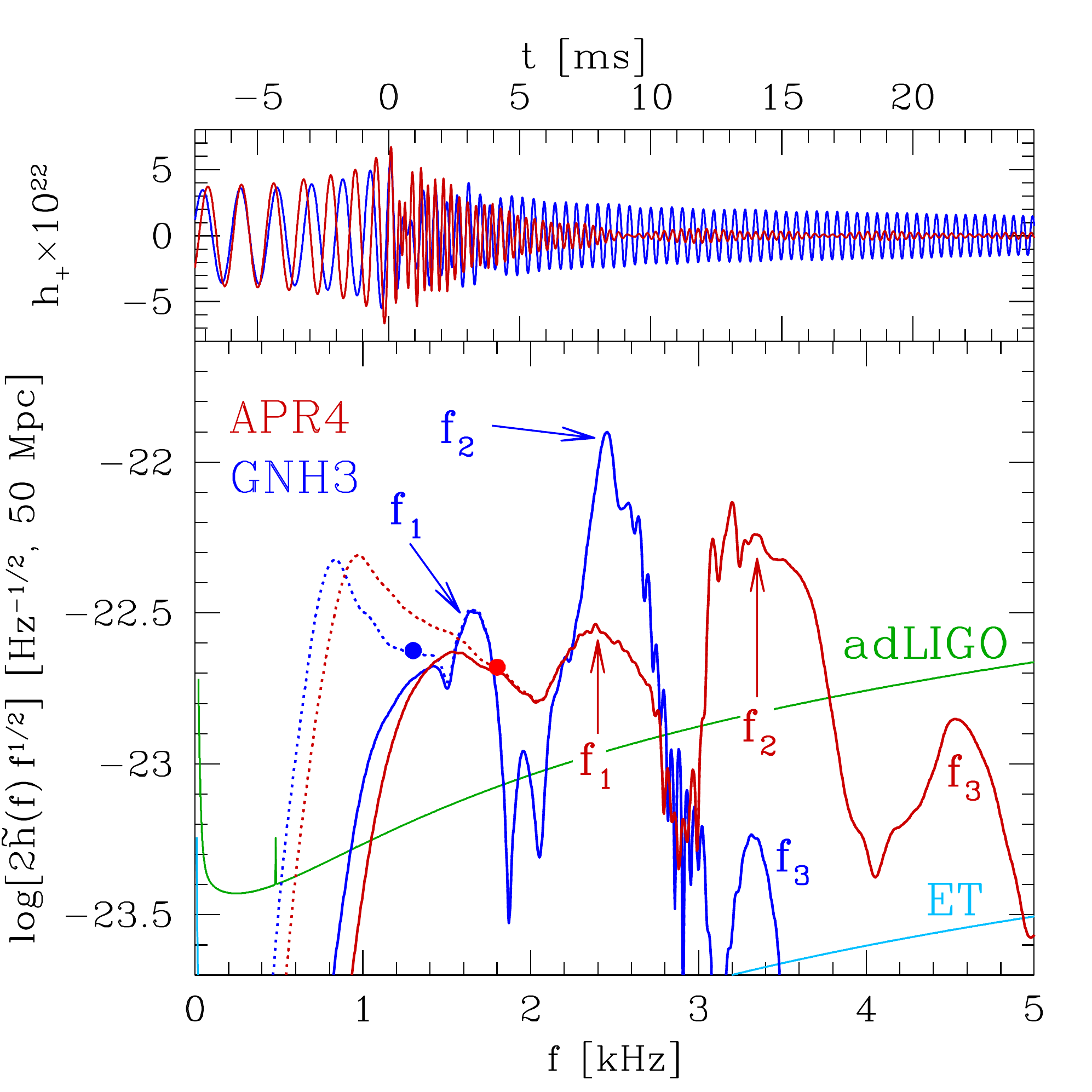}
\caption{This figure from \cite{Takami14} shows  in the top panel the evolution of $h_+$   from the coalescence of 
 equal-mass (NS,NS) binaries at a  distance of 50 Mpc  (the two curves  refer to two selected EoS  (APR4 dark-red and GNH3 EoSs blue lines).  The  bottom panel shows the wave spectral density  $2{\tilde h}(f)\sqrt{f}$  versus frequency $f$   windowed after the merger, for the two EoSs. The sensitivity curves of Advanced LIGO (green line) and ET (light-blue line) are also
plotted. The dotted lines show the spectral density in the inspiral, while the circles mark the contact frequency
$f_{\rm cont}={\cal C}^{3/2}/(2\pi \bar{M})$,
where ${\cal {C}} =\bar{M}/\bar{R}$ is the average compactness, with $\bar {M}$ the 
initial gravitational mass of the two stars,
$\bar{R}= (R_1+R_2)/2$, and $R_{1,2}$ are the radii of the
non-rotating stars associated with each binary. The frequency $f_1$ appears to correlate closely with the compactness $\cal C$ of the stars. 
See \cite{Takami14} for details. Courtesy of L. Rezzolla.  
\label{NS-waveforms-fourier-domain}
}
\end{center} 
\end{figure}

Figure \ref{NS-waveforms-time-domain} collects an inventory of gravitational wave signals from equal-mass binaries simulated with 
different EoS, shown in different colours. Each
column refers to a given initial gravitational mass and we defer to \cite{Baiotti-review16,Takami14}  for details. Figure 
\ref{NS-waveforms-time-domain} shows that the 
gravitational wave signal from (NS,NS) mergers does not show a rapid decay (as in the case of a (BH*,BH*) coalescence) as the hyper-massive neutron star that forms undergoes violent oscillation modes prior to collapse into a black hole.

In the Fourier domain, each signal carries precious information on the compactness $\cal C$ of the neutron stars.
This is illustrated in Figure \ref{NS-waveforms-fourier-domain} where the spectral density $2{\tilde h}(f)\sqrt{f}$ is plotted against the
frequency $f$ of the wave, for two selected models \cite{Takami14}. 
The low frequency dotted line corresponds to the inspiral phase which is then followed by
a sequence of three prominent peaks with different amplitudes. Those are related to the oscillations of the
hyper-massive neutron star, and would be absent or much smaller if a black hole forms promptly. 
In this latter case, the gravitational-wave signal would terminate abruptly with
a cut-off corresponding to the fundamental quasi-normal-mode frequency of
the black hole. The behaviour summarised in Figure \ref{NS-waveforms-fourier-domain}
 has been
investigated by a number of authors over the last decade.
As of today, there is a widespread consensus that: (i) the post-merger
gravitational wave signal possesses spectral features that are robust and
that emerge irrespective of the EoS or the mass ratio; (ii) the
frequencies of the peaks, referred to as "contact frequencies" in the post-merger signal can be used to obtain
important information on the stellar properties (i.e. mass and radius) and
hence represent a very good proxy to deduce the EoS.

\subsection{Electromagnetic counterparts of {\bf CBCs}}

\begin{figure}[!t]
\begin{center}
\includegraphics[width=1.00\textwidth]{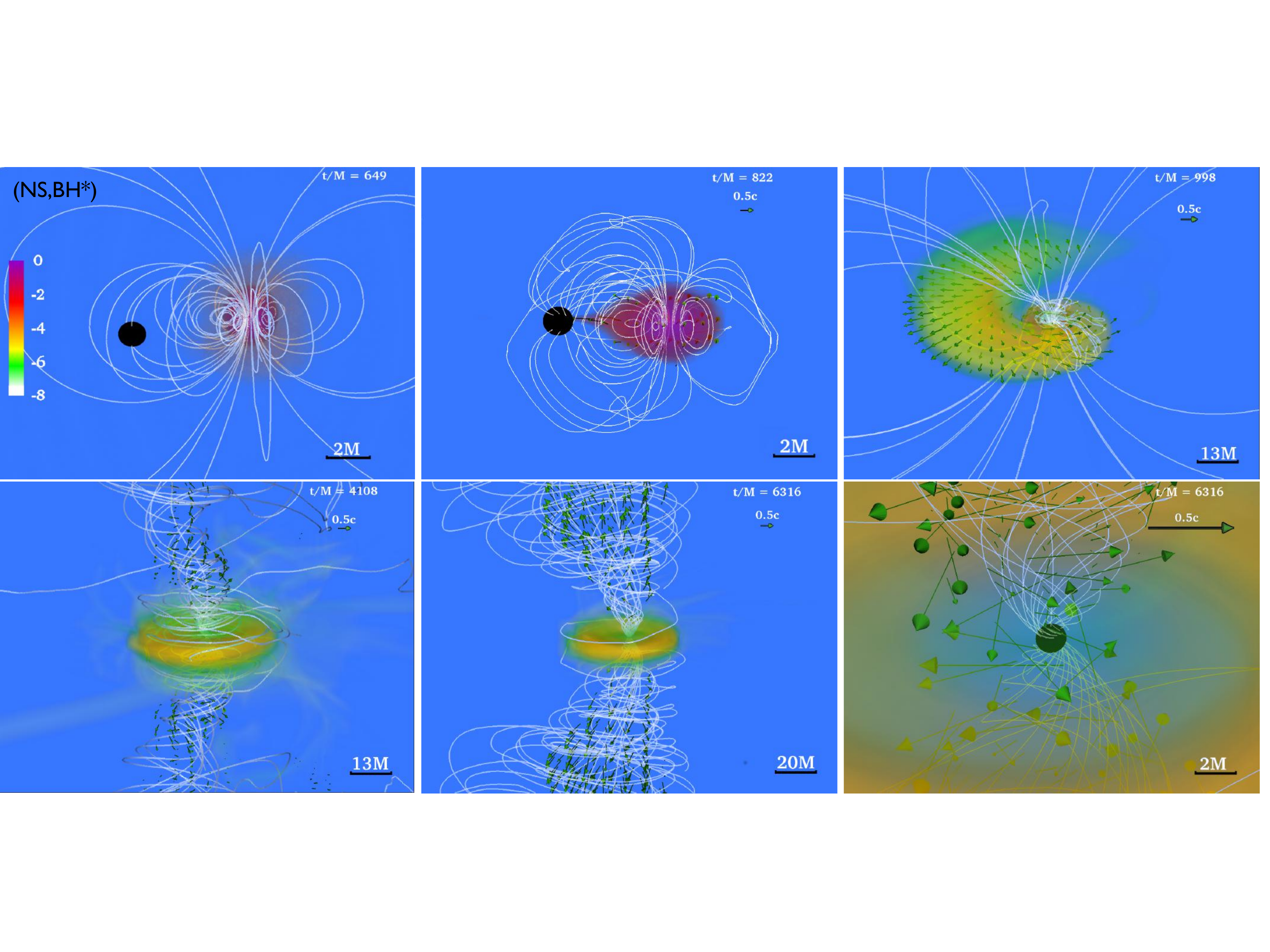}
\caption{Magnetohydrodynamical simulation in general relativity of the merger of a (BH*,NS) binary with 3 to 1 ratio.  Figure adapted from\cite{Shapiro15}. The black hole is 
rotating with spin 
$s_{\rm spin}=0.75$ aligned with the orbital angular momentum.  About two orbits prior to merger, a dynamically weak interior 
dipole  magnetic field is seeded into the neutron star.  The figure shows snapshots of the rest-mass density in log-scale, 
in units of  $8.92\times 10^{14}(1.4\msun / M_{\rm NS})^2\,\rm gr \,cm^{-3}$, at different  times $t/ \rm M$. 
Here, time is in units of $2.5\times 10^{-3} (M_{\rm NS}/1.4\msun) \rm ms= 7.58 (M_{\rm NS}/1.4\msun) km.$
Arrows indicate plasma velocities and white lines show the magnetic field lines. Bottom panels highlight the
system after an incipient jet is launched.  The scale-size in each panel is indicated on the right bottom corner in units of M. Courtesy of S. Shapiro.}
\label{BH*-NSmerger}
\end{center} 
\end{figure}

Very recent magneto-hydrodynamical simulations in general relativity of (NS,NS) mergers with an initially high, but dynamically weak magnetic field, 
have proven that at the end of the merger an incipient jet forms around the remnant \cite{Shapiro16-ns-ns}.
 This occurs following the delayed collapse of a hyper-massive neutron star into a black hole, 60 ms after the merger. The region above the black hole poles becomes strongly magnetised, and a collimated, mildly relativistic outflow is launched. 
 Figure \ref{BH*-NSmerger} gives a glimpse into the merger of a (BH*,NS) binary with mass ratio 3:1.
In this simulation \cite{Shapiro15}, the mass ratio is not extreme and the neutron star is tidally disrupted  as a whole, so that the 
black hole ends being enshrouded by a torus of debris representing the leftover of the star that is torn apart
by the intense tidal field of the black hole.  A weak magnetic dipole field was seeded in the neutron star extending to the exterior,
which is responsible for the launch of an incipient jet  which could account for  the energetics of a short GRB.

One of the current challenges of Advanced LIGO and Virgo is to detect the electromagnetic counterparts of  (NS,NS) and/or (NS,BH*)
merger events related to outflows
produced during the merger and post-merger phases, as illustrated in Figure \ref{EM-NSmerger}. This would have a tremendous impact as it will prove the origin of short 
GRBs as coalescing binaries \cite{Berger14}, and will open the possibility of
identifying the host galaxy and thus the redshift of the source opening the possibility of using (NS,NS) and (NS,BH*) mergers as standard sirens \cite{Schutz86-hubble}.

\begin{figure}[!t]
\begin{center}
\includegraphics[width=1.00\textwidth]{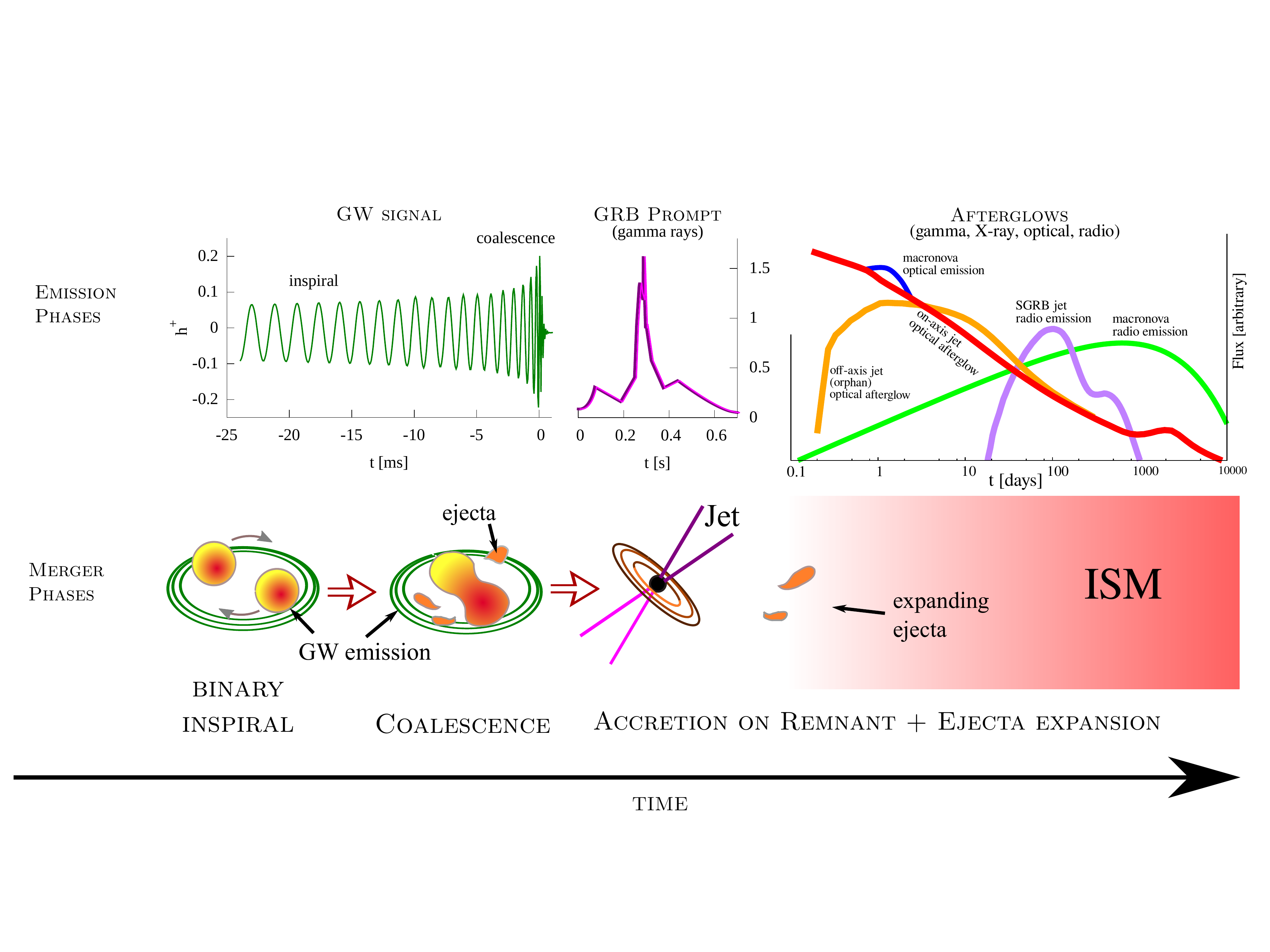}
\caption{Electromagnetic follow up of a (NS,NS) merger event.  See text for the description of the figure.} 
\label{EM-NSmerger}
\end{center}  
\end{figure}

In more detail, the tidal interaction
between the two neutron stars during the merger phase very likely leads to nearly
isotropic ejection of neutron rich material. Due to the subsequent
expansion, such material undergoes rapid $r$-process nucleosynthesis,
followed by radioactive decay of unstable isotopes, thus producing a
short-lived optical/near-IR transient, known as {\it macronova} (blue solid line in Figure \ref{EM-NSmerger})
lasting hours to days after the merger \cite{li98}. The same
ejecta may also produce a long-lived radio transient (green solid line),
lasting months to years, due to the interaction with the surrounding interstellar medium \cite{Nakar11}. In the post-merger phase, rapid accretion of a
centrifugally supported disc on the compact remnant powers a collimated
relativistic jet of semi-aperture $\theta_{\rm jet}$, which produces a
short GRB\cite{Berger14}. Prompt collimated emission at
gamma-ray energies (violet solid line in the top-middle plot) is followed
by extended, lower energy emission (afterglow), due to the interaction of
ejecta with the interstellar medium. Owing to relativistic beaming, the
gamma-ray emission is visible to observers with viewing angle within the
narrow cone of the jetted emission. Optical afterglow emission (red and
orange solid lines) is detectable by observers at angles  $\lesssim 2\theta_{\rm jet}$. Radio afterglow emission (purple solid line) is
observable from all viewing angles once the jet decelerates to mildly
relativistic velocities on a timescale of weeks to months (up to years,
depending on the inter-stellar medium density and jet energy).

\subsection {Core-collapse supernovae}\label{CCS}

Electromagnetic  observations suggest that many, if not
most, core collapse supernova  explosions exhibit asymmetric features, and  this is also suggested by results of multidimensional
 simulations.  Thus, core collapse supernovae (CCSNe) are likely to be accompanied by emission of gravitational waves,  associated  to 
the quadrupole mass-energy dynamics \cite{Ott09-review}. 
Spherical symmetry could be broken by stellar rotation, convection in the proto-neutron
star and in the region behind the shock, and
by the standing accretion shock instability. State-of-the-art models predict strains 
 $h\sim 10^{-23}-10^{-20}$ for core collapse events at 10 kpc with frequencies of $\sim 1$ to 1000 Hz, and 
total emitted energies in gravitational waves of $10^{42}-10^{47}$ erg, corresponding to $10^{-12}-10^{-7} \msun c^2$.
 
\begin{figure}[!t]
\begin{center}
\includegraphics[width=.999\textwidth]{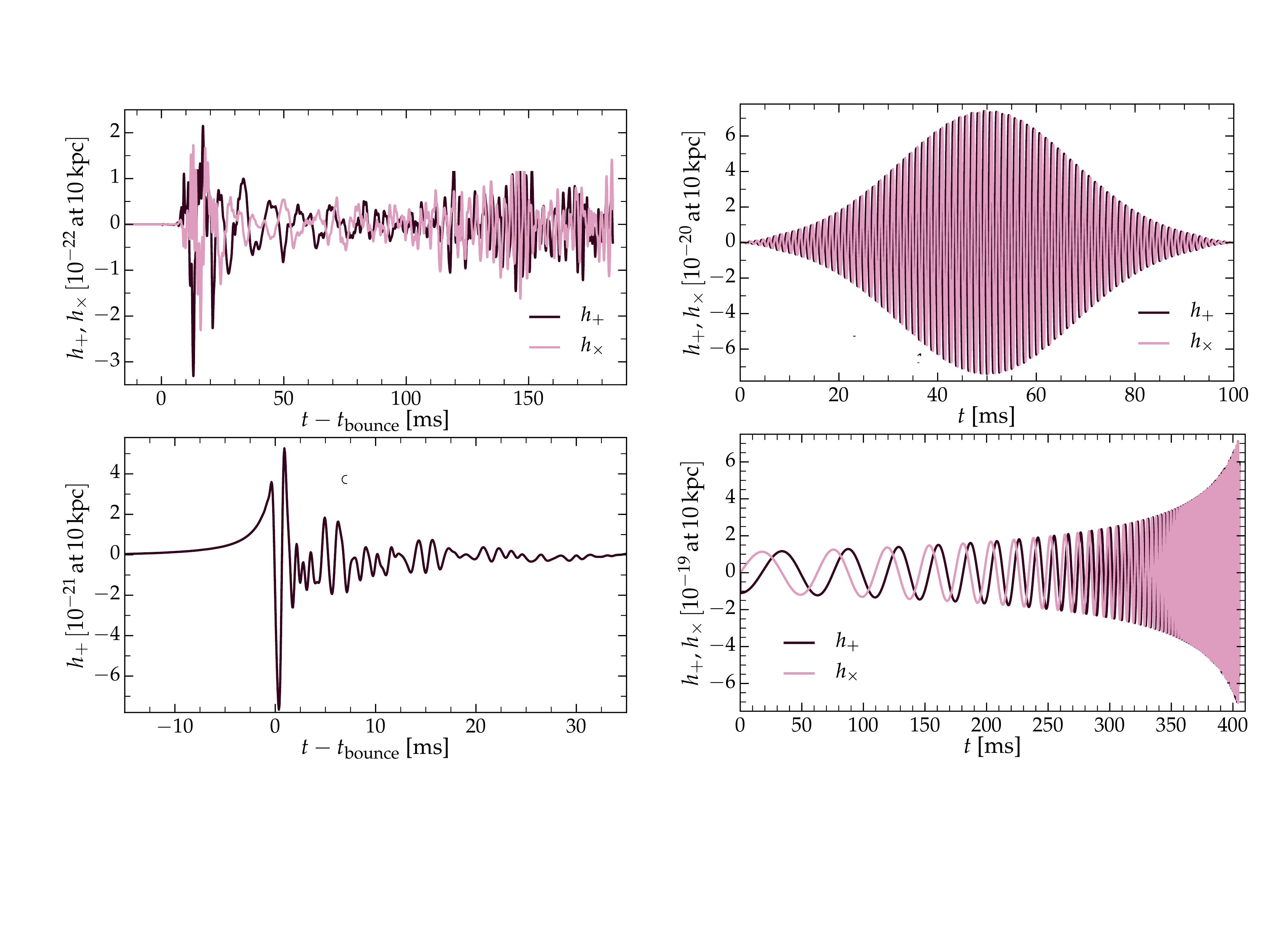}
\caption{Inventory of waveforms from CCNSe, adapted from \cite{Gossan16}.
Left panels show the strain for a neutrino (top)  and rotation (bottom) driven
core collapse as seen from an equatorial observed at a distance of 10 kpc. 
Right panels show the strain from a bar-mode  (top) and disc
fragmentation instability (bottom), as seen by a
polar observer at 10 kpc, calculated  to leading (quadrupole) order from simple analytical models of the instabilities. 
Courtesy of S.E. Gossan.}
\label{CCSNe}
\end{center} 
 
\end{figure}

There is a garden-variety of signals expected from CCSNe and in Figure \ref{CCSNe} we show, as an illustration,  the waveform expected from
four representative models \cite{Gossan16}.
In the left panel,  the strain from a neutrino (upper left) and rotation  (lower left) driven CCSN are seen by an equatorial observer at 10 kpc.
The wave from the neutrino-driven deformation carries two polarisation states, and is characterised by a long-lived signal with complex
time-domain structure. The rotation-driven instability leads to oblateness  and to an $l=2$, $m=0$ quadrupole deformation of  the inner quasi-homologously
collapsing core. The extreme accelerations experienced by
the inner core at bounce lead to a large spike in the
 gravitational wave signal, followed by ring-down of the proto-neutron
star as it settles to its new equilibrium state.  In contrast, the right panels of Figure \ref{CCSNe} show the signal  
from a differentially rapidly rotating  proto-neutron star unstable to a bar-mode instability $m=2$ (upper right panel), modelled in the quadrupole approximation as a sine-Gaussian
morphology \cite{Gossan16}.  If the CCSN mechanism fails to re-energize the stalled
shock, the proto-neutron star will collapse  to a black hole, due to fall-back on a timescale set by accretion. Provided matter has sufficient angular momentum,
a massive self-gravitating accretion disc/torus may form
around the nascent stellar mass black hole. The inner regions of the disc are geometrically thin
due to efficient neutrino cooling, but outer regions are
thick and may be gravitationally unstable to fragmentation
at large radii. The predicted signal shown in the lower right panel comes instead from orbiting fragments around the newly formed black hole.

\subsection {Signal from a cosmological background of supermassive black holes}\label{SMBH}
We are now in the position to characterise, after $\S 5$,  the cosmological background of gravitational waves resulting from the incoherent superposition of SMBHBs 
still away from coalescence, and  to
possibly identify individual sources.

Following \cite{Sesana08-Vecchio}, the characteristic amplitude 
of this cosmic background can be written as
\begin{equation}
h_c^2(f) =\int_0^{\infty} 
dz\int_0^{\infty}d{ M}_{\rm c}\, \frac{d^3N}{dzd{ M}_{\rm c} \,d{\rm ln}f_r}\,
h^2(f_r),
\label{hch2}
\end{equation}
where $f_r=(1+z)f$, $d^3N/dzd{M}_{\rm c} d{\rm ln}f_r$ is the comoving number of binaries  with chirp mass and redshift in the range $[{M}_{\rm c},{M}_{\rm c}+d{ M}_{\rm c}]$ and $[z, z+dz]$, emitting in a given logarithmic frequency interval
$f_r$. In the approximation of quasi circular binaries, $h(f_r)$ is the inclination--polarisation averaged strain given by 
\begin{equation}
h(f_r)=\left ({\frac{8\pi^{2/3}}{10^{1/2}}}\right) {1 \over  d_L} \left ({G{ M}_{\rm c}\over c^2}\right )^{5/3}\left ({\pi f_r \over c}\right )^{2/3},
\label{haverage}
\end{equation}
where $d_L$ its luminosity distance to the source \cite{Sesana08-Vecchio}. It is straightforward to show that for SMBHBs driven by 
gravitational wave emission, equation (\ref{hch2}) results in a spectrum of the form $h_c=A(f/f_0)^{-2/3}$. The normalisation $A$ encodes information about the cosmic population of SMBHBs and directly depends on the rate of massive galaxy mergers, on the intrinsic relations between the properties of the galaxies and the mass of the SMBHs they host, and on the pairing efficiency of SMBHs following galaxy mergers. Figure \ref{fig_PTAlimits} shows examples of the uncertainty in the expected signal level, assuming two specific $M_{\rm BH}$-galaxy correlations from the literature. The figure also shows up-to-date sensitivities from the three major PTAs, which started to pierce into the range predicted by SMBHB population models. A detection of the signal is expected within the next decade, also aided by the advent of new observational facilities such as MeerKAT, FAST and SKA \cite{Janssen15}.
 \begin{figure}[!t]
\begin{center}
\includegraphics[width=.99\textwidth]{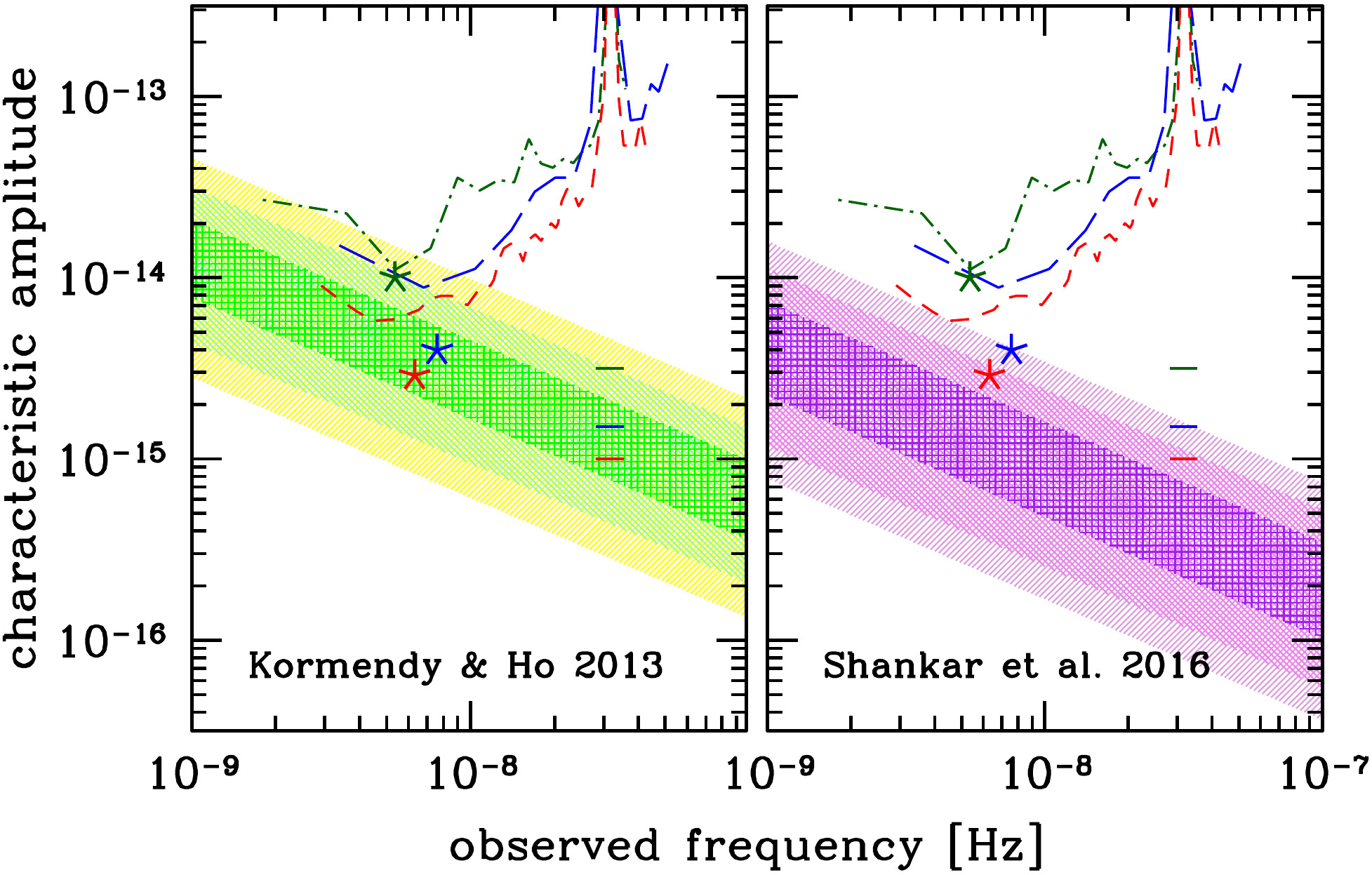}
\caption{Characteristic amplitude of the gravitational wave background assuming specific $M_{\rm BH}$-galaxy correlations as indicated in \cite{Sesana13}. In each panel, shaded areas define the 68\% 95\% and 99.7\% confidence intervals of the signal amplitude. Jagged curves represent current PTA sensitivities: EPTA (dot-dashed green), NANOGrav (long-dashed blue), and PPTA (short-dashed red). Stars represent the integrated upper limits to an $f^{-2/3}$ background coming from each sensitivity curve, and the horizontal ticks are their extrapolation at $f=1$yr$^{-1}$.}
\label{fig_PTAlimits} 
\end{center} 
\end{figure}

 \begin{figure}[!t]
\begin{center}
\includegraphics[width=.40\textwidth]{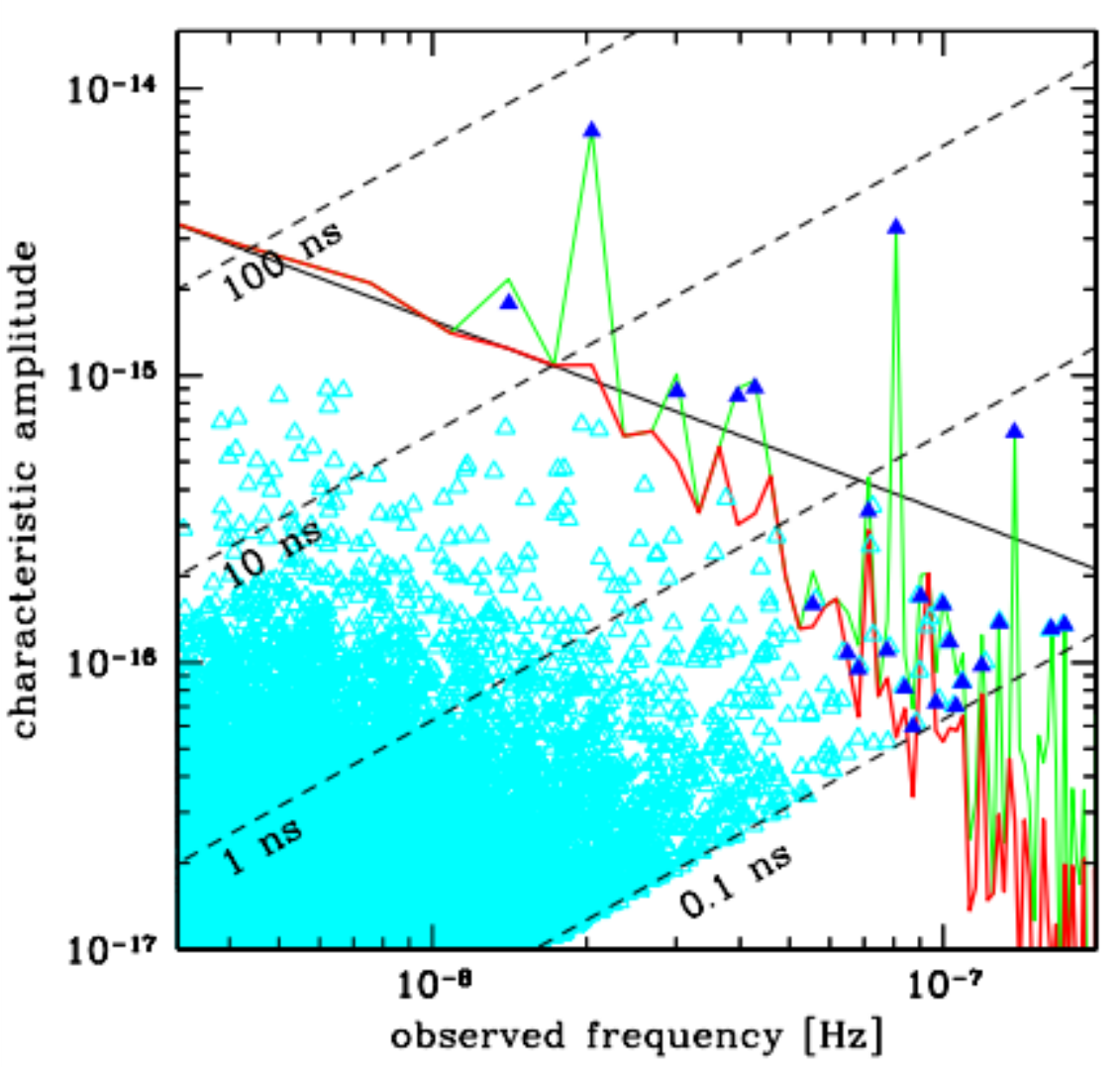}
\includegraphics[width=.50\textwidth]{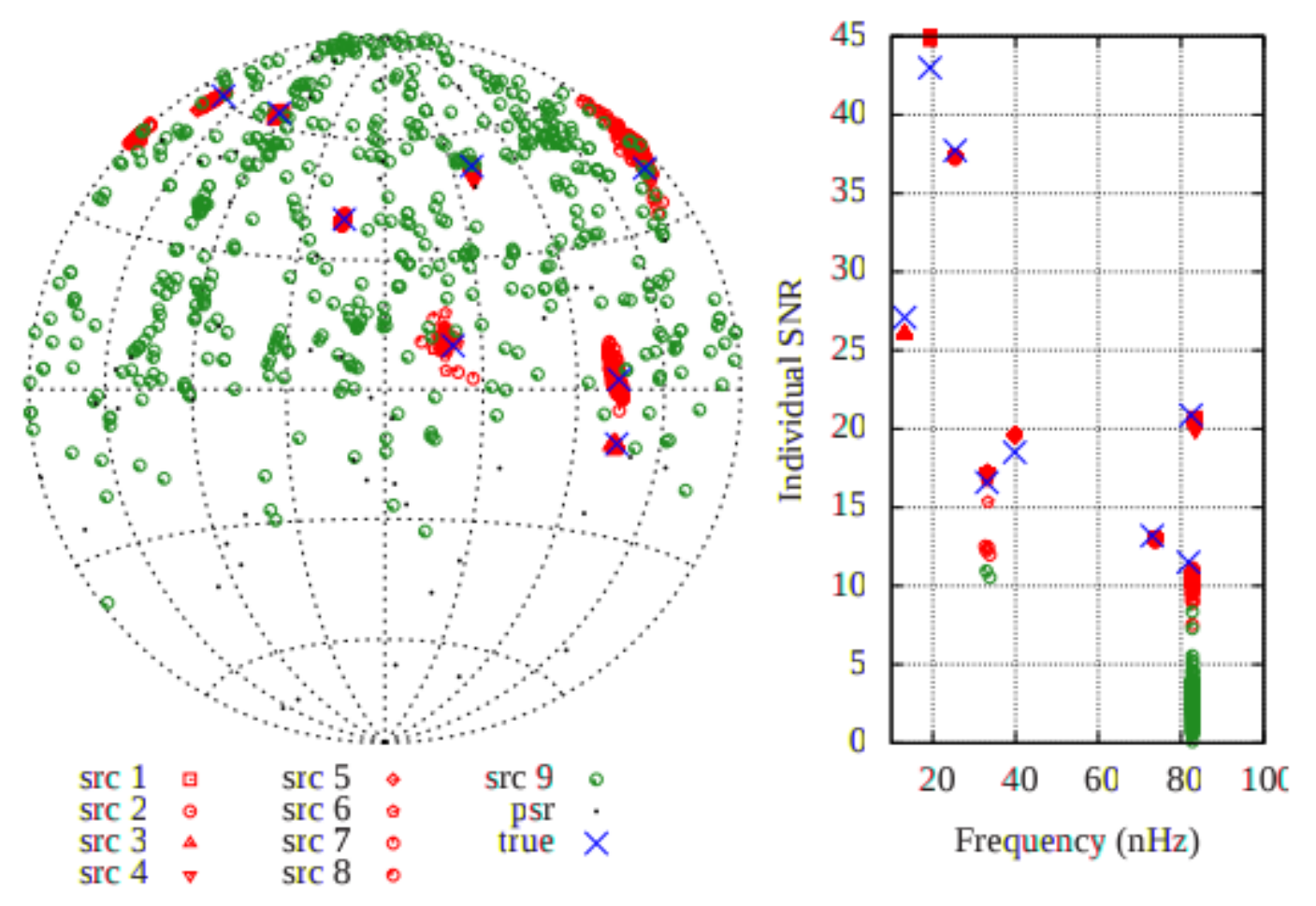}
\caption{The left panel shows characteristic amplitude versus frequency of a Monte Carlo realisation of the gravitational wave signal from SMBHBs. Each cyan point represents the emission of an individual binary and  blue triangles are individually resolvable sources. The green (red) line is the collective signal including (subtracting) resolvable sources. The black solid line is the expected analytical $f^{-2/3}$ power law, and the black dashed lines represent different timing residual levels. Adapted from \cite{Sesana13}.
The right panel, adapted from \cite{Petiteau13}  represents an exercise of individual SMBHB recovery in simulated PTA data. The blue crosses represent injected sources, whereas the clusters of red points are the signal recovered by a Multi-Modal genetic algorithm running on the synthetic dataset. This shows PTA potential of correctly identifying multiple SMBHBs.}
\label{fig_PTAGWB} 

\end{center} 
\end{figure}

If $A$ encodes information about the abundance and mass distribution of SMBHBs in the universe, the shape of the spectrum critically depends on the dynamical properties of SMBHBs on their path to final coalescence \cite{Sesana13}. The $f^{-2/3}$ spectrum is in fact valid only in the circular gravitational wave  driven approximation; coupling to the stellar and gaseous environment, necessary to shrink the binaries at sub-pc scales, as well as significant eccentricities will modify the shape of the spectrum at the low end, possibly causing a turnover. Observation and characterisation of the background shape will therefore provide unique insights on the population and dynamics of sub-parsec SMBHBs \cite{Arzoumanian16}.

 As far as resolvable SMBHBs are concerned, the $f^{-2/3}$ power-law shown in Figure \ref{fig_PTAlimits} is an approximation to a much more complex reality. Especially at $f>10$nHz, the signal enters in a low-number statistic regime, and is typically dominated by few sources. This is elucidated in the left panel of Figure \ref{fig_PTAGWB}, in which the blue triangles highlight SMBHBs that rise above the confusion noise generated by the overall population, and therefore can potentially be detected individually. These are, however, very low frequency signals involving SMBHBs far from coalescence. As such, they are essentially monochromatic and only few cycles can be observed, making a reliable estimate of basic parameters such as the source chirp mass extremely difficult. However PTAs can reconstruct the source localisation in the sky within tens of deg$^2$ thanks to triangulation. This will make it possible to follow-up the interesting regions in the sky in search for electromagnetic periodic signals, likely associated to the presence of a circum-binary massive disc which envelops the SMBHB.  
 This possibility will open new avenues in multi-messenger astronomy as shown in the right panel of Figure \ref{fig_PTAGWB}. A counterpart identification would, for example, enable the measurement of the redshift (and thus distance, assuming a cosmological model) to the source, thus allowing a precise measurement of the SMBHB chirp mass. Although the first PTA detection is expected to involve the unresolved gravitational wave background, several SMBHBs will eventually be individually resolved, providing another formidable tool to study the astrophysics of supermassive black holes.

 \begin{figure}[!t]
\begin{center}
\includegraphics[width=.99\textwidth]{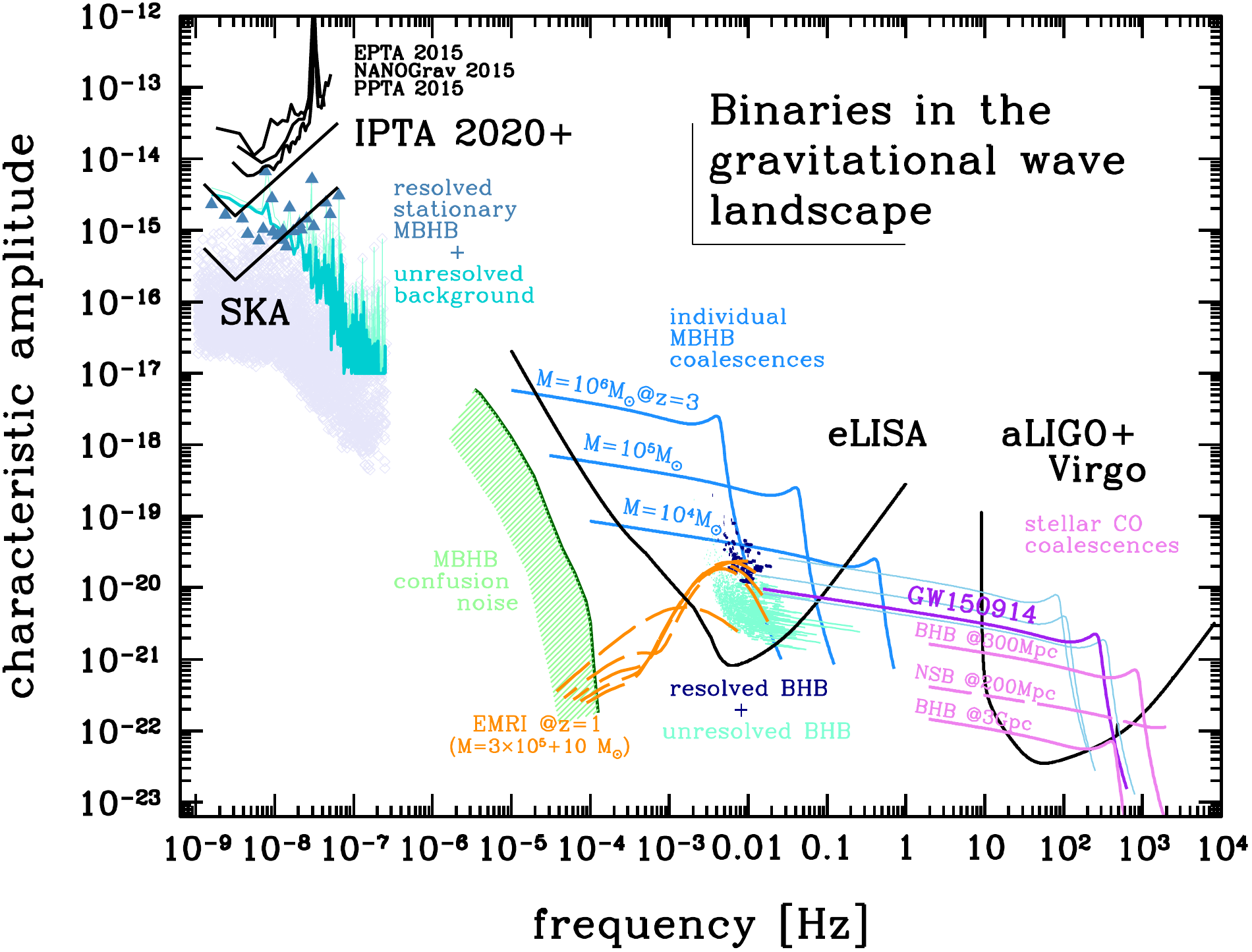}
\caption{Binaries in the gravitational wave universe. The Figure shows the dimensionless characteristic strain amplitude in Fourier space versus frequency,  from 
black hole binaries of all flavours, and of neutron star binaries, as described in this Chapter.
Plotted are also the sensitivity curves of PTA, eLISA,  and Advanced LIGO and Virgo.  In eLISA and LIGO-Virgo, merging  binaries above the sensitivity curve can be
detected with a signal-to-noise ratio that can be computed from (\ref{signal-to-noise ratio}), and sweep across the
sensitivity band increasing their frequency up to coalescence. 
At the lowest frequencies $\sim 10^{-8}$ Hz of PTA, inspiralling supermassive black hole binaries of $10^{9-10}\msun$  give rise to a stochastic background contaminated by 
individual, loud  sources.  Around $\sim $ mHz frequencies of eLISA coalescing binary black holes weighing $10^{4-7}\msun$ are the main sources. They sweep across the band months before merging. Together with EMRIs, they stand out  from an hypothetical background of binaries  like GW150914  and of Galactic WD-WD binaries (signal not shown to keep the figure readable).  At the highest frequencies accessible with Advanced LIGO-Virgo, coalescing stellar origin black holes and neutron stars 
 are the main sources and sweep across the band in minutes to a fraction of a second.  GW150914 sweeps first in the eLISA band and emerges again in the LIGO and Virgo sensitivity range at the time of coalescence (see Figure \ref{fig_LIGOLISA}). }
\label{granfinale} 
\end{center} 
\end{figure}

\section{Conclusion}

This Chapter is a first attempt to describe the {\it multi-band gravitational wave universe} in a unified way. {\it Binaries} of
all favours have been our main actors, and in Figure \ref{granfinale} 
we show their signals sweeping across the different frequency intervals.
We have demonstrated that the information gathered in the gravitational waves emitted by these sources is immense as will make it possible to answer to the deep and urgent questions posed in the beginning of this Chapter: on the Laws of Nature, on the intimate link between stellar and galaxy formation and evolution processes, and on the  geometry of the universe itself.\footnote{We regret for not having described the 
cosmic background(s) from the very early universe, for length limitations. We defer to  \cite{Caprini16} for a review.}

A new, golden era of exploration of the universe has started. Detecting the vibrations of spacetime 
will let us explore the cosmos from a new perspective.  This is thanks to the major advances we witnessed in experimental physics: 
laser interferometry proves to be successful and powerful in detecting the minuscule perturbations of spacetime produced by cosmic sources.
The discovery of GW150914 and GW151226 represents the culmination of an experimental research able to achieve
 strain noise sensitivities at the level of $10^{-23}\,\rm Hz^{-1/2}$ around 200 Hz \cite{Abbott-1,Abbott-5-sensitivity}.
 
On the 8th of June 2016, ESA announced  to the public the first results of the LISA Pathfinder in-flight experiment at the Sun-Earth Lagrangian point, designed to measure
 the level of residual acceleration noise
on two test masses free-falling with respect to a local inertial frame. The results \cite{Armano16} beautifully demonstrate that the two 
masses can be put and remain in "free fall" with a relative acceleration noise characterised by a power spectral density
of $5.2 \pm  0.1$  fm s$^{-2}$ Hz$^{-1/2}$ corresponding to $0.56\pm 0.01\times 10^{-15}$ {\it g} Hz$^{-1/2} $ with $g$ the standard acceleration gravity on Earth \cite{Armano16}. 
The result of the LISA Pathfinder experiment shows that the technology on board meets the requirement for a space based gravitational wave
observatory with a sensitivity close to what was originally foreseen for LISA.  With an expected launch of a LISA-like Observatory by 2030, the gravitational
wave universe will be scrutinised across all accessible frequency bands from nano-Hz, to mHz and kHz. 

 \begin{figure}[!t]
\begin{center}
\includegraphics[width=.99\textwidth]{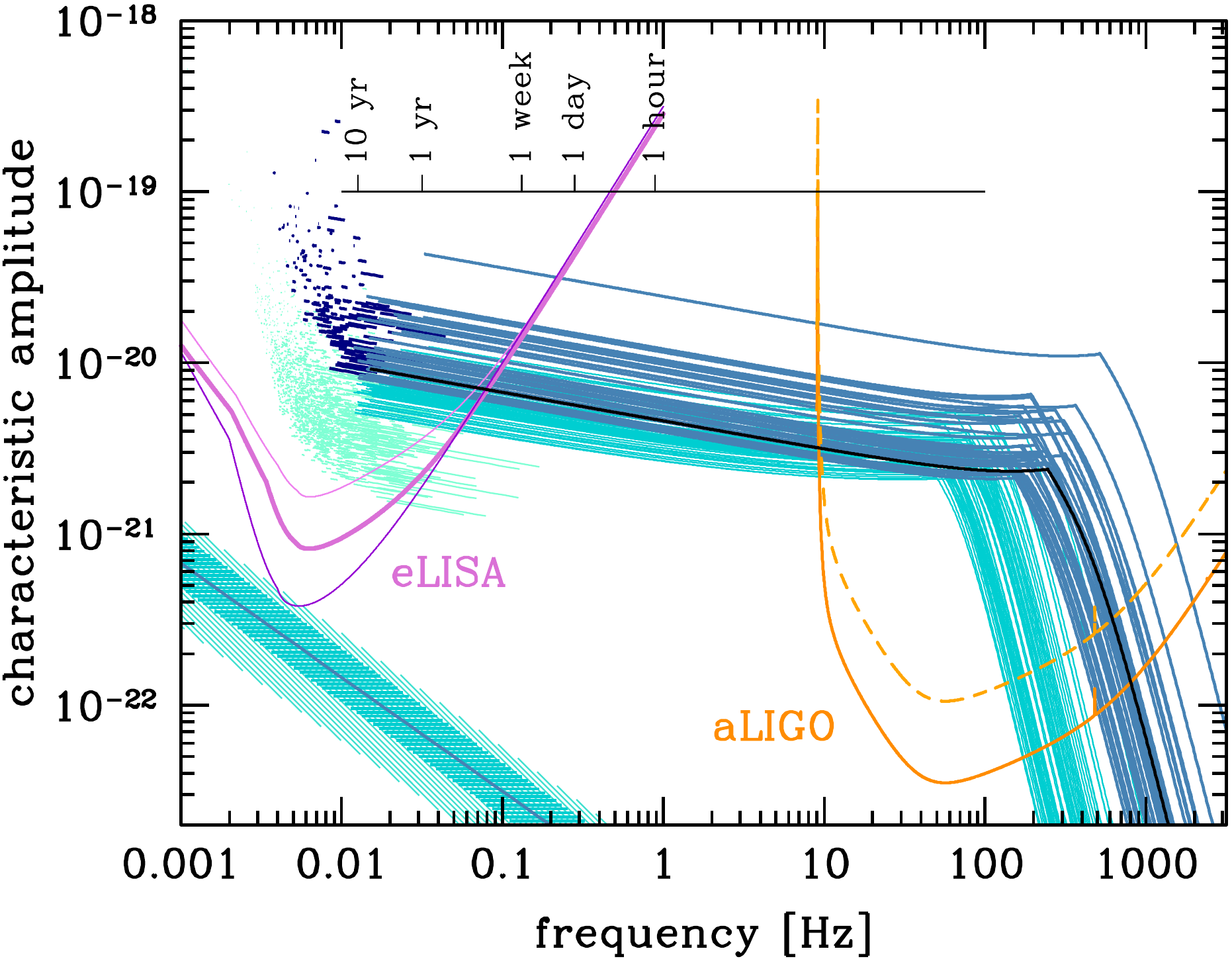}
\caption{Multi-band gravitational wave astronomy of (BH$^*$,BH$^*$) binaries, adapted from \cite{Sesana16-multiband}. 
Plotted is the dimensionless characteristic amplitude versus frequency as in Figure \ref{granfinale}.  The violet lines are the sensitivity curves of three eLISA configurations; from top to bottom N2A1, N2A2, N2A5 \cite{Klein16}. The orange lines are the current (dashed) and design (solid) Advanced LIGO sensitivity curves \cite{Abbott-5-sensitivity}. Blue lines represent tracks of a sample of (BH$^*$,BH$^*$) binaries. The light turquoise lines are systems with signal-to-noise ration (SNR) 
between 1 and 5 in the eLISA band.  The light and dark blue curves crossing  the Advanced LIGO band are sources with SNR $>5$ and SNR $>8$ respectively in eLISA; the dark blue ticks are binaries with SNR$>8$ in eLISA not crossing the Advanced LIGO band within five years. The characteristic amplitude track completed by GW150914 is shown as a black solid line, with the top label indicating its frequency progression in the last 10 years before coalescence.}
\label{fig_LIGOLISA} 
\end{center} 
\end{figure}

The high, low and very low frequency gravitational universes which encompass the multi-band universe should not be perceived as disconnected.
 They are instead inter-winded, and Figure \ref{fig_LIGOLISA} 
shows beautifully how profound is the level of reciprocity.  The detection of GW150914 has and will  have profound implications for the science  in the mHz regime. It has in fact became clear that, besides the vast population of galactic compact binaries, the future LISA-like interferometer in space might detect up to $O(10^{3})$ "heavy" 
GW150914 like (BH$^*$,BH$^*$) binaries out to $z\approx 0.5$ \cite{Sesana16-multiband}. GW150914 itself, five years prior to coalescence, was emitting gravitational waves at about $15$ mHz, accumulating a signal-to-noise ratio of $\approx 10$, sweeping across the eLISA window as shown in Figure \ref{fig_LIGOLISA}. Although rates are still uncertain and will be constrained by Advanced LIGO  and Virgo in their forthcoming runs, the detection of several such "heavy" (BH$^*$,BH$^*$) binaries in both mHz and kHz bands  will open the era of {\it multi-band "correlated" gravitational wave astronomy}, with profound implications for tests of general relativity and multi-messenger astronomy.

\bibliographystyle{ws-rv-van}
\bibliography{Bibliouniverse}

\end{document}